\documentclass[aps,prb,reprint,superscriptaddress,longbibliography]{revtex4-2}

\usepackage[T1]{fontenc}
\usepackage[utf8]{inputenc}
\usepackage{graphicx}
\usepackage{amsmath,amssymb}
\usepackage{bm}
\usepackage{xcolor}
\usepackage{url}
\usepackage{hyperref}

\DeclareUnicodeCharacter{2248}{\AA}

\newcommand{\CeCo}{CeCoGe$_3$}

\begin{document}

\title{Tunable magnetotransport through kinetically hindered first-order phase transitions in an antiferromagnetic metal}

\author{Jaime M. Moya}
\affiliation{Department of Chemistry, Princeton University, Princeton, NJ 08544, USA}
\author{Scott B. Lee}
\affiliation{Department of Chemistry, Princeton University, Princeton, NJ 08544, USA}
\author{Sudipta Chatterjee}
\affiliation{Department of Chemistry, Princeton University, Princeton, NJ 08544, USA}
\author{Nitish Mathur}
\affiliation{Department of Chemistry, Princeton University, Princeton, NJ 08544, USA}
\author{Grigorii Skorupskii}
\affiliation{Department of Chemistry, Princeton University, Princeton, NJ 08544, USA}
\author{Connor J. Pollak}
\affiliation{Department of Chemistry, Princeton University, Princeton, NJ 08544, USA}
\author{Leslie M. Schoop}
\email{lschoop@princeton.edu}
\affiliation{Department of Chemistry, Princeton University, Princeton, NJ 08544, USA}

\date{\today}

\begin{abstract}
Controllable multilevel resistance states are of interest for memory technologies like neuromorphic computing, but robust materials platforms toward such behavior remain limited. Here, we show that the non-centrosymmetric antiferromagnetic metal CeCoGe$_3$ suggests one such route through a kinetically hindered first-order magnetic transition. Cooling through the kinetically hindered first-order transition in an applied magnetic field produces a magnetic glass state in which high- and low-temperature magnetic phases coexist. The relative fraction of these phases can be controlled by the applied field in which the sample is cooled, and the electrical resistance is directly sensitive to that fraction. As a result, it is demonstrated that CeCoGe$_3$ supports stable multilevel resistive states. These results identify kinetically hindered first-order phase transitions as a promising route towards controllable multilevel magnetoresistive states.

%at 126 words. Max word count is 200 so we are clear.

%Disorder-broadened first-order phase transitions can host unusual functional responses, particularly when slow kinetics prevent complete transformation into the equilibrium ground state.
%Here, we show that CeCoGe$_3$ exhibits a complex low-temperature magnetic phase diagram with signatures of first-order metamagnetic transitions and strong history dependence. Magnetization and transport measurements reveal that the transition into the lowest temperature phase becomes kinetically hindered, producing anomalous isothermal hysteresis loops, magnetic-memory effects under field-cooled protocols, and devitrification behavior in cooling-and-heating-in-unequal-fields measurements. Together, these results indicate the formation of a magnetic glass state arising from incomplete phase transformation across a disorder-broadened first-order phase transition. The arrested phase fraction can be tuned systematically by field-cooling history, and the electrical resistivity is directly sensitive to this tuning.
\end{abstract}

\maketitle

\section*{Acronyms}
\begingroup
\footnotesize
\noindent \textbf{AFM}, antiferromagnetic; \textbf{CCD}, charge-coupled device; \textbf{CHUF}, cooling and heating in unequal fields; \textbf{DC}, direct current; \textbf{DM}, Dzyaloshinskii--Moriya; \textbf{EDS}, energy-dispersive X-ray spectroscopy; \textbf{ETO}, electrical transport option; \textbf{FC}, field cooling / field cooled; \textbf{FCC}, field-cooled cooling; \textbf{FCW}, field-cooled warming; \textbf{FM}, ferromagnetic; \textbf{HT}, high-temperature; \textbf{KHFOPT}, kinetically hindered first-order phase transition; \textbf{LR}, long-range; \textbf{LT}, low-temperature; \textbf{$\mu$SR}, muon spin rotation/relaxation; \textbf{MPMS}, Magnetic Property Measurement System; \textbf{PPMS}, Physical Property Measurement System; \textbf{QD}, Quantum Design; \textbf{RKKY}, Ruderman--Kittel--Kasuya--Yosida; \textbf{SEM}, scanning electron microscopy / scanning electron microscope; \textbf{SQUID}, superconducting quantum interference device; \textbf{vdP}, van der Pauw; \textbf{VSM}, vibrating sample magnetometer; \textbf{ZFC}, zero-field cooling / zero-field cooled.
\par
\endgroup

\section{Introduction}\label{Introduction}

Traditional (von Neumann) computing architectures separate memory and processing, so substantial energy and time are spent shuttling data between the two. Neuromorphic and in-memory computing concepts therefore motivate device elements that can store and update information locally using multilevel, analog, nonvolatile conductance states \cite{Sebastian2020NatNanoInMemory,Xia2019NatMaterCrossbar,Zhou2024NML2DInMemory}. A wide range of candidate materials systems are being explored because no single platform simultaneously optimizes all relevant figures of merit, such as programmability, retention, device-to-device variability, endurance, and operating conditions \cite{Sebastian2020NatNanoInMemory,Xia2019NatMaterCrossbar,Zhou2024NML2DInMemory}.

A natural place to look for multistate behavior is in glassy magnetic systems. Spin glasses, in particular, are famous for their rugged energy landscape, strong history dependence, and memory effects \cite{Jonason1998PRLMemoryChaos,mydosh1993spin}. Moreover, spin-glass-inspired models (e.g., Hopfield networks) have played an influential role in conceptual links between spin glass magnetism and associative memory \cite{hopfield1982neural,mydosh1993spin}. At the same time, the same complexity that enables many accessible metastable states also makes them difficult to control, a necessity for neuromorphic memory technologies \cite{grollier2020neuromorphic}.

A closely related, but potentially more controllable, route is the magnetic glass state which is expected to emerge from kinetic arrest across a first-order magnetic transition \cite{manekar2001first,chattopadhyay2005kinetic,roy2006evidence,kumar2006relating}. In this scenario, the glassiness does not arise from an intrinsic spin-glass ground state. Instead, it reflects incomplete transformation across a disorder-broadened first-order phase transition: a fraction of the HT (or high-field) phase becomes frozen in on experimental timescales even though the system is cooled well below the supercooling transition. The relevant state variable can then be viewed as an arrested phase fraction that is programmable by well-defined field--temperature protocols.

\begin{figure}[t]
\centering
\includegraphics[width=1\columnwidth]{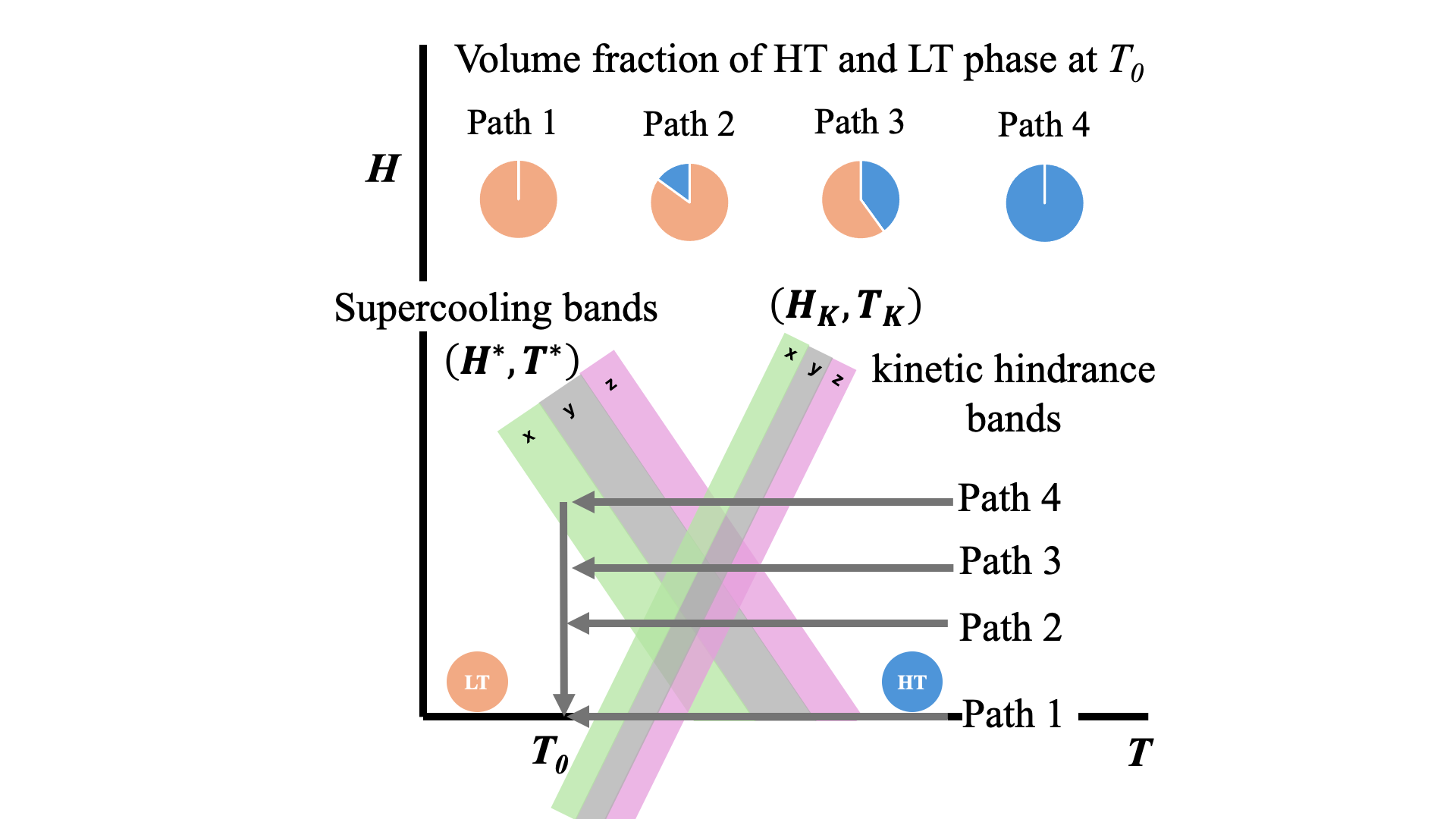}
\caption{\label{fig:cartoon} \textbf{Tunable phase fractions from kinetically hindered first-order phase transitions.} Schematic magnetic field--temperature $H$--$T$ phase diagram for a disorder-broadened first-order phase transition between a high-temperature (HT) phase and a low-temperature (LT) phase, adapted from Ref. \cite{kumar2006relating}. For illustration, the HT phase is taken to be ferromagnetic-like and the LT phase antiferromagnetic-like. The $(H^*,T^*)$ bands denote the supercooling lines, below which an infinitesimal fluctuation drives the system into the LT state. Crossing the $(H_K,T_K)$ band before all $(H^*,T^*)$ bands are crossed kinetically arrests part of the HT phase, leaving it frozen in at temperature $T_0$ well below the supercooling lines. The colored regions x, y, and z represent different spatial regions of the sample that transform at slightly different temperatures because of disorder. As a result of this presumed energy landscape, field-cooling paths 1--4 yield different fractions of frozen HT phase coexisting with the LT phase at $T_0$ as demonstrated by the pie charts above.}
\end{figure}

The basic idea is sketched in the schematic magnetic-field--temperature ($H$--$T$) phase diagram in Fig.~\ref{fig:cartoon}. For concreteness, consider a first-order transition between an HT ferromagnetic (FM) phase and an LT antiferromagnetic (AFM) phase such that applying a magnetic field favors the FM phase \cite{manekar2001first,kumar2006relating}. In an ideal clean system, the first-order character implies supercooling and superheating limits: on cooling, the high-temperature phase can persist metastably down to a supercooling boundary $(H^{*},T^{*})$, while on warming the low-temperature phase can persist up to a superheating boundary $(H^{**},T^{**})$. Between these limits, phase coexistence is natural near the transition. We have only included the $(H^{*},T^{*})$ bands in Fig.~\ref{fig:cartoon} for simplicity. In real materials, quenched disorder and strain can broaden the transition, so that the supercooling/superheating conditions become bands rather than sharp lines, reflecting a distribution of local transition conditions across the sample \cite{imry1979influence,soibel2000imaging,roy2004first,manekar2001first,chattopadhyay2004metastable,Chattopadhyay2005PRBMagneticGlass,kumar2006relating,roy2013first}. Therefore, the different colored bands in Fig.~\ref{fig:cartoon} represent different spatial regions of the sample.

In the thermodynamic picture alone, when cooled below the supercooling band, an infinitesimal fluctuation would drive the system fully into the low-temperature phase. However, it is now established that kinetics can play a decisive role near disorder-broadened first-order transitions \cite{imry1979influence,soibel2000imaging,roy2004first,manekar2001first,chattopadhyay2004metastable,Chattopadhyay2005PRBMagneticGlass,kumar2006relating,roy2013first}. In the kinetic-arrest framework, an additional band $(H_K,T_K)$ separates regions of the phase diagram where transformation is dynamically allowed from regions where kinetics become so slow that the system's high-temperature phase freezes or becomes hindered below $(H_K,T_K)$ \cite{manekar2001first,kumar2006relating}. The shape of the $(H_K,T_K)$ bands relative to the $(H^{*},T^{*})$ bands in Fig.~\ref{fig:cartoon} is drawn such that increasing field favors the HT phase since it is assumed to be FM \cite{kumar2006relating}. A magnetic glass state is created when the system crosses into the kinetic-arrest band while a finite fraction of the high-temperature/high-field phase still remains untransformed. The remaining fraction that has not undergone the thermodynamic phase transitions is then frozen in, producing long-lived phase coexistence at low temperature. In the FM--AFM picture presented, increasing the applied magnetic field in which the sample is cooled would then increase the relative volume fraction of the frozen HT phase as shown schematically in Fig.~\ref{fig:cartoon}, where the pie charts represent the volume fraction of the HT and LT in the sample at $T_0 < T^*$ for different field-cooling paths. ``Devitrification'' corresponds to re-entering a part of the phase diagram where kinetics are sufficiently fast for the arrested fraction to relax toward the equilibrium low-temperature state \cite{banerjee2006coexisting,banerjee2009conversion}.

This picture immediately suggests a route to multilevel memory: if an electrical transport observable is sensitive to the relative phase fractions and/or domain populations, then the arrested fraction set by the field--temperature history can act as a programmable internal state variable. In other words, the same first-order competition that produces phase coexistence also provides a knob for writing and tuning multiple distinct, nonvolatile resistive states \cite{saha2018tunable}, which could be nonvolatile provided the kinetics are slow enough to stabilize them over relevant timescales.

Here, we present evidence for tunable magnetoresistance in the polar, non-centrosymmetric correlated metal \CeCo{} \cite{pecharsky1993unusual}. Its non-centrosymmetric crystal structure and the presence of competing interactions (including Kondo screening and Ruderman--Kittel--Kasuya--Yosida (RKKY) exchange) produce a complex $H$--$T$ magnetic phase diagram for $H \parallel c$ \cite{thamizhavel2005unique}, where it has recently been suggested that the valence electron count specific to CeCoGe$_3$ amplifies the phase competition \cite{dominguez2024role}.

We map this phase diagram and identify a low-temperature first-order boundary that exhibits signatures consistent with kinetic hindrance, including an anomalous virgin curve which lies outside the envelope of the main hysteresis loop in isothermal magnetization and magnetoresistance \cite{banerjee2006coexisting,banerjee2006ferromagnetic,rawat2007anomalous,sengupta2006field,singh2002first}. Using established memory protocols and cooling and heating in unequal fields (CHUF), as well as memory tests, we show that the anomalous hysteresis is attributable to kinetic arrest rather than an intrinsic spin- or cluster-glass ground state \cite{banerjee2006coexisting,banerjee2009conversion,pal2021memorylike}. Anisotropic resistance measurements suggest evidence of a magnetostructural phase transition which may be the origin of the quenched disorder. We further find that the arrested fraction, and the associated domain pinning, can be systematically tuned by field-cooling procedures, and that electrical transport is directly sensitive to this tuning, enabling multilevel resistive states that remain stable over many field cycles, as evidenced in minor loops at low temperature. Thus, using \CeCo{} as an example, we have demonstrated that kinetically hindered first-order phase transitions (KHFOPTs) are an interesting avenue to study in the context of multilevel analog memory technologies. Finally, because \CeCo{} lacks inversion symmetry, it provides a natural platform to explore whether electrically driven spin--orbit torques could ultimately bias domain populations and enable current-driven programming of these metastable states \cite{wadley2016electrical,bodnar2018writing,Zelezny2014}.

\section{Results}

\subsection{Crystal structure}

CeCoGe$_3$ crystallizes in the tetragonal, polar, non-centrosymmetric BaNiSn$_3$-type structure (space group $I4mm$, number 107). The composition was verified with energy-dispersive X-ray spectroscopy (EDS) and the crystal structure was confirmed by single-crystal X-ray diffraction (see Supplementary Note I, Fig. S1-S5, and Tables 1-6) \cite{pecharsky1993unusual}. In this structure, shown in the inset of Fig.~\ref{fig:Basics}a, the Ce atoms (red) occupy the four corners and the body-center of the tetragonal unit cell. The Co (dark blue)--Ge (light blue) connectivity is asymmetric around the body-centered Ce along the crystallographic $c$-axis, breaking inversion symmetry. The lack of an inversion center inherent to the crystal structure introduces the Dzyaloshinskii--Moriya (DM) interaction \cite{dzyaloshinsky1958thermodynamic,moriya1960new}, which competes with the Heisenberg-like RKKY interaction and can induce exotic magnetic domains and a complicated energy landscape in the magnetic field--temperature ($H$--$T$) phase space \cite{fert2013skyrmions,tokura2010multiferroics,parkin2015memory,nagaosa2013topological}. Further complicating the energy landscape is the presence of the Kondo interaction, which acts to screen the Ce local moments \cite{pecharsky1993unusual,smidman2013neutron,li2023photoemission}. The competing interactions result in a complex magnetic field--temperature ($H$--$T$) phase diagram presented in Fig.~\ref{fig:Basics}a which was measured with $H~\parallel~c$.

\subsection{Magnetic phase diagram and evidence for first-order phase transitions}

\begin{figure*}[t]
\centering
\includegraphics[width=0.95\textwidth]{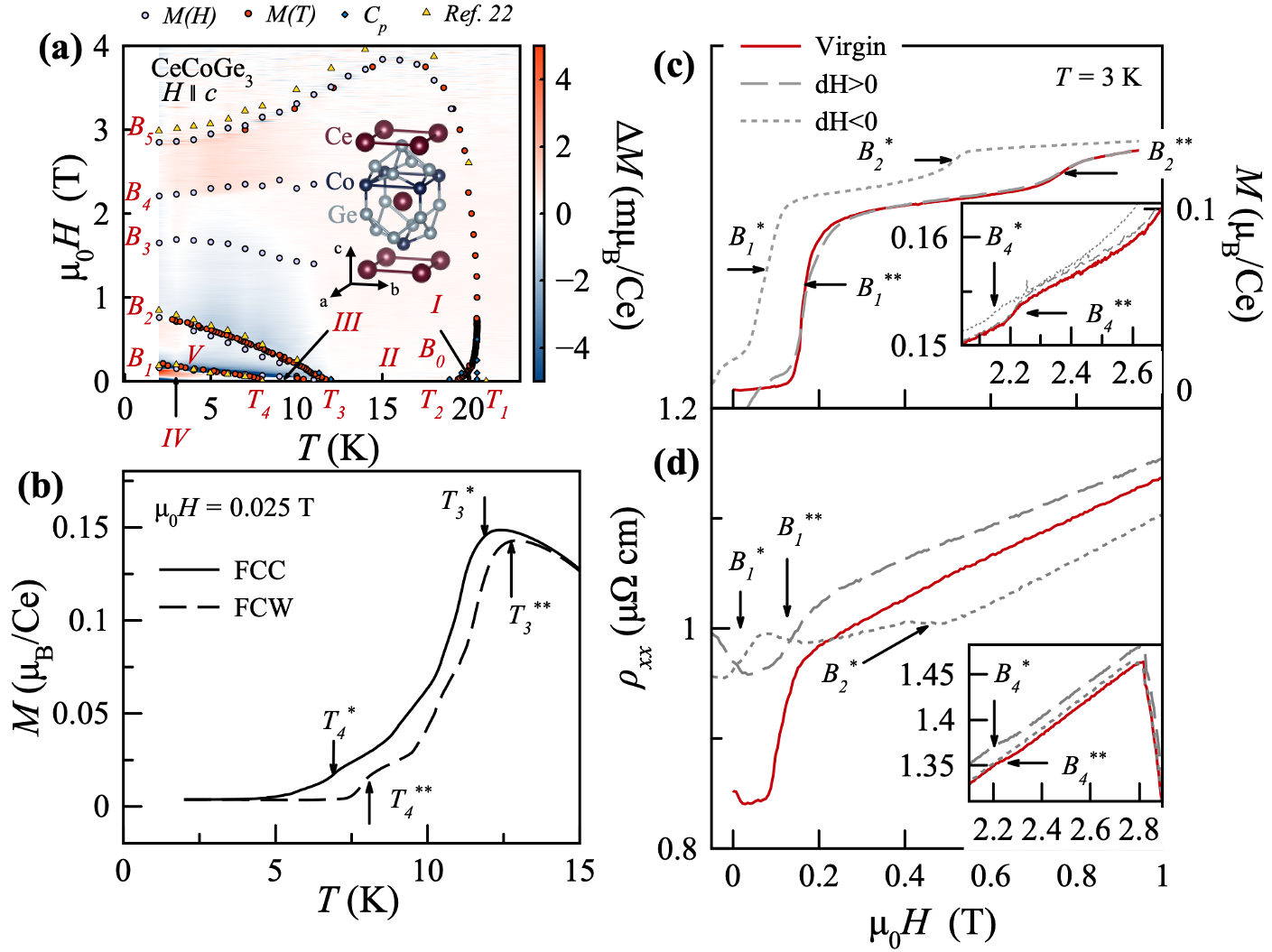}
\caption{\textbf{Phase diagram of CeCoGe$_3$.} \textbf{(a)} Magnetic field--temperature ($H$--$T$) phase diagram of CeCoGe$_3$ for $H \parallel c$, constructed from features identified in isothermal magnetization $M(H)$ (blue circles), temperature-dependent magnetization $M(T)$ (red circles), and temperature-dependent heat capacity $C_P$ (triangles) measurements, and compared with previous results reported in Ref.~\cite{thamizhavel2005unique}. See Supplementary Note 2 for details of how the phase boundaries were assigned. Features observed under near-zero-field conditions are labeled $T_1$--$T_4$ on cooling, while field-induced metamagnetic features are labeled $B_0$--$B_5$. The background color map shows $\Delta M$, defined as the difference between the virgin $M(H)$ curve measured from 0 to 7 T after the sample was zero-field cooled (ZFC) and the corresponding 0 to 7 T segment measured on the return branch of the hysteresis loop. Positive $\Delta M$, shown in red, marks the region of anomalous virgin-curve behavior. The inset shows the crystal structure of CeCoGe$_3$. \textbf{(b)} Low-field $M(T)$ shown below 15 K in $\mu_0H = 0.025$ T using field-cooled cooling (FCC, solid) and field-cooled warming (FCW, dashed) protocols. Clear thermal hysteresis at the $T_3$ and $T_4$ transitions indicates first-order character. The corresponding transition temperatures on cooling and warming are labeled $T_i^*$ and $T_i^{**}$, respectively, to denote supercooling and superheating. \textbf{(c)} Expanded view of low-field $M(H)$ between 0 and 1 T, highlighting the $B_1$ and $B_2$ transitions. \textbf{(d)} Expanded view of the corresponding field-dependent isothermal resistivity $\rho_{xx}$ over the same field range, showing similar low-field transition structure in transport. In both \textbf{(c)} and \textbf{(d)}, the red solid curve is the virgin curve measured after zero-field cooling; the short-dashed curve ($dH<0$) was measured after first increasing the field above 7 T and then sweeping from 7 to $-7$ T; and the long-dashed curve ($dH>0$) was measured on the subsequent sweep from $-7$ to 7 T. The notation $B_i^*$ and $B_i^{**}$ again denotes features associated with supercooling and superheating, respectively. Insets in \textbf{(c)} and \textbf{(d)} highlight the higher-field $B_4$ anomaly. All measurements shown were performed with $H \parallel c$.}
\label{fig:Basics}
\end{figure*}

The $H$--$T$ phase diagram of CeCoGe$_3$ shown in Fig.~\ref{fig:Basics}a was constructed from anomalies in direct-current (DC) isothermal ($M(H)$, blue circles) and temperature-dependent ($M(T)$, red circles) magnetization, as well as temperature-dependent heat capacity ($C_P$, blue diamonds) measurements.
Details of the boundary determinations can be found in Supplementary Note 2 and Figs.~6-8.

In zero-field (Fig.~\ref{fig:Basics}a), four magnetic phases are identified: Phase I ($T_N = 21$ K to $T_2 = 18.5$ K), Phase II ($T_2$ to $T_3 = 12$ K), Phase III ($T_3$ to $T_4 = 8$ K), and Phase IV (below $T_4$). Previous neutron scattering measurements reported magnetic propagation vectors $\mathbf{k} = (0,0,2/3)$, $(0,0,5/8)$, and $(0,0,1/2)$ in Phases II, III, and IV, respectively \cite{smidman2013neutron}. Phases I--III exhibit a ferromagnetic component, while Phase IV is proposed to be an AFM with a two-up two-down spin structure with moments aligned along the $c$-axis \cite{smidman2013neutron}.

Below $T_3$, with application of a magnetic field $H \parallel c$, five features are observed in the virgin curve of the $M(H)$ (Supplementary Fig.~\ref{fig:MH1p8}) marked $B_1$--$B_5$ in Fig.~\ref{fig:Basics}a. $B_1$ and $B_2$ delineate Phases III and IV from Phase V, while $B_5$ marks the transition to the field-polarized state. The other two features, $B_3$ and $B_4$, are more subtle (Supplementary Fig.~\ref{fig:MH1p8}) and the temperature dependence of the $M(H)$ loops indicates that $B_3$ and $B_4$ persist up to $T_3$ without a vertical phase boundary, suggesting a domain effect rather than a true phase transition. Similar features have been attributed to the formation of soliton defects in EuRhAl$_4$Si$_2$ \cite{allen2026atomically}, but we did not investigate this possibility further here. Above $T_2$, we find that a low-field boundary ($B_0$) separates Phases I and II. Compared to previous studies \cite{thamizhavel2005unique} (yellow triangles in Fig.~\ref{fig:Basics}a), our high-resolution measurements resolve all previously reported features while identifying new features at $B_3$ and $B_4$, and separating Phases I and II as well as Phases V and III.

The first indications of a KHFOPT in \CeCo{} appear when we examine the low-temperature behavior in the magnetization and transport measurements: the temperature-dependent magnetization, $M(T)$ (Fig.~\ref{fig:Basics}b); the field-dependent magnetization, $M(H)$ (Fig.~\ref{fig:Basics}c); and the longitudinal resistivity, $\rho_{xx}(H)$ (Fig.~\ref{fig:Basics}d). In $M(T)$ (Fig.~\ref{fig:Basics}b), the transitions at $T_3$ and $T_4$ show clear thermal hysteresis between the FCC and FCW curves. Here, $T^{*}$ and $T^{**}$ mark the cooling and warming transition temperatures, respectively. This hysteresis is consistent with the first-order character of these transitions. In addition, an unusual field-dependent hysteresis is observed at $T=3$~K in the ZFC $M(H)$ loop (Fig.~\ref{fig:Basics}c) and in $\rho_{xx}(H)$ (Fig.~\ref{fig:Basics}d). There is hysteresis between the anomalies on the field-decreasing branch ($\mathrm{d}H<0$), labeled $B_{1(2)}^{*}$, and on the field-increasing branch ($\mathrm{d}H>0$), labeled $B_{1(2)}^{**}$. Moreover, the virgin (initial) field-increasing curve (red solid line) lies outside the main hysteresis loop---most clearly at low fields in the main panels and at high fields in the insets. Taken together, these observations suggest that the transitions associated with $B_1$ ($T_4$), $B_2$ ($T_3$), and $B_5$ are likely first order.

Such behavior, in which the virgin curve lies outside the envelope loop, has been attributed in some materials to kinetic arrest associated with a KHFOPT~\cite{banerjee2006coexisting,banerjee2006ferromagnetic,rawat2007anomalous,sengupta2006field,singh2002first}. In this picture (see Fig.~\ref{fig:cartoon}), the anomalous virgin curve can be understood as follows. After zero-field cooling to the LT phase, none of the sample becomes kinetically arrested in the HT phase because the cooling trajectory crosses the supercooling band $(H^{*},T^{*})$ before encountering the kinetic-arrest band $(H_K,T_K)$. Thus, at 3~K in the ZFC state, away from the first-order transitions at $T_3$ and $T_4$, \CeCo{} is entirely in the LT AFM state (phase IV). If the field is then increased beyond the superheating band $(H^{**},T^{**})$ for the relevant transitions (in particular, beyond the superheating condition for the $B_5$ transition) and subsequently reduced back toward $H=0$ (as in our measurement protocol), then on the decreasing-field branch a portion of the sample can become kinetically arrested in the HT/high-field phase. This occurs because the kinetic-arrest band $(H_K,T_K)$ lies at higher temperature than the supercooling band $(H^{*},T^{*})$, allowing part of the transformed high-field phase to remain ``frozen in'' as the field is reduced. This arrested volume fraction does not fully de-arrest until the field is decreased further, crossing into negative field, as discussed later.

While this anomalous virgin-curve behavior is consistent with a KHFOPT, similar behavior has also been reported in cluster-glass systems, where it is often described as mictomagnetism~\cite{mydosh1978spin,shull1975mictomagnetic,shull1976transition,benka2022interplay}. In addition, related effects can sometimes appear over a narrow temperature window near a first-order FM--AFM transition even without kinetic arrest~\cite{chattopadhyay2008metamagnetic}.

\begin{figure}[t]
\centering
\includegraphics[width=0.8\columnwidth]{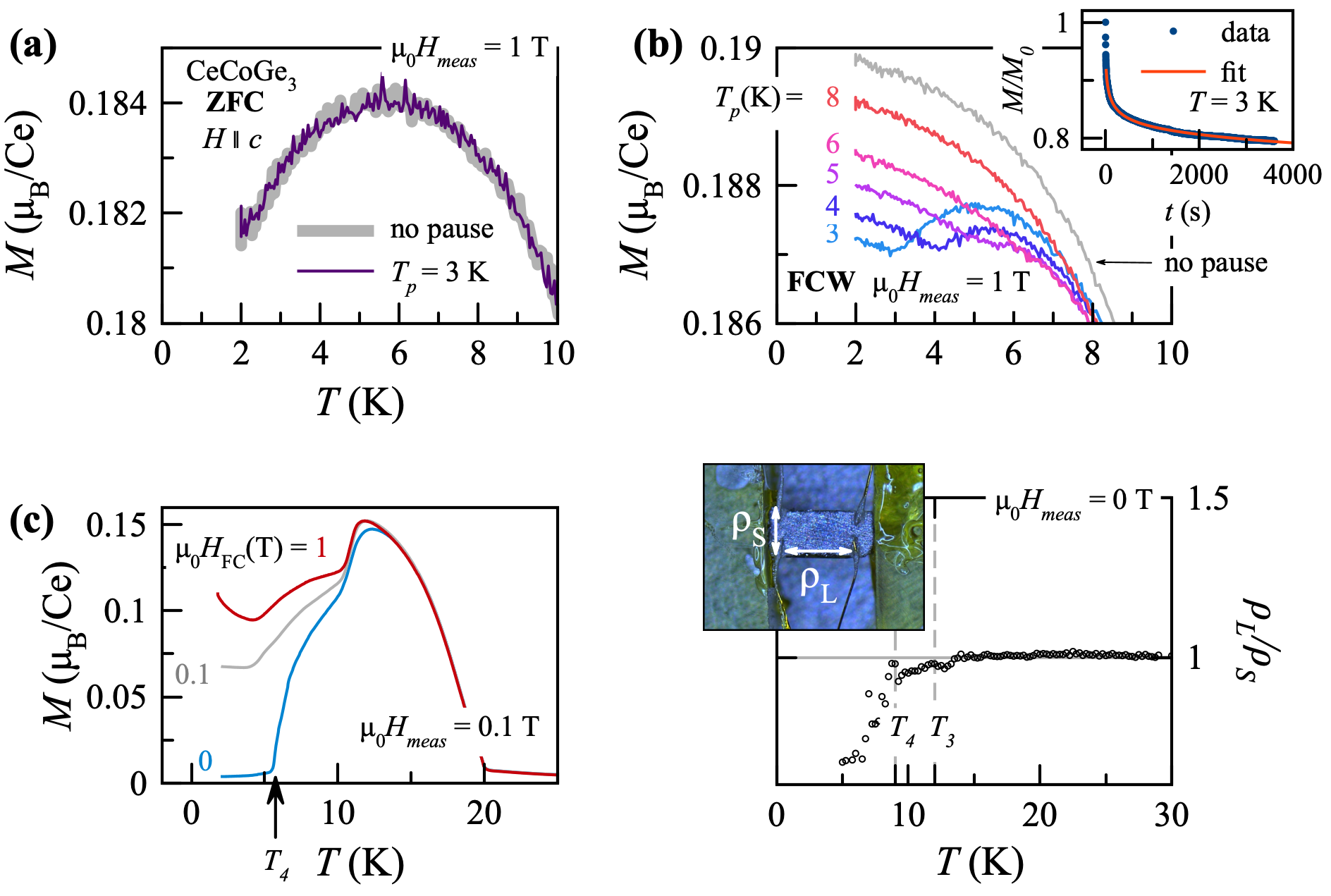}
\caption{\label{fig:memory}
\textbf{Evidence for magnetic glass behavior in CeCoGe$_3$.} \textbf{(a)} Temperature-dependent magnetization $M(T)$ measured below 10 K with $\mu_0H_{\mathrm{meas}} \parallel c = 1$ T using the no-pause protocol (gray) and a pause protocol with a stop at $T_p = 3$ K (purple). No measurable difference is observed, indicating the absence of a memory effect under this protocol. \textbf{(b)} $M(T)$ measured using $\mu_0H_{\mathrm{meas}} \parallel c = 1$ T under the FCW protocol with either no pause (gray) or a pause at selected temperatures $T_p$ between 3 K and 8 K. The memory effect appears as a local minimum in $M(T)$ at $T_p$. \textbf{(Inset)} Time dependence of the remanent magnetization $M$ after the field is switched off at $T_p = 3$ K. The data are plotted as $M/M_0$ (blue circles), where $M_0$ is the magnetization at the first recorded time after the applied field reaches zero. The red line is a fit to a stretched exponential function. \textbf{(c)} Magnetization measured on warming in $\mu_0H_{\mathrm{meas}} \parallel c = 0.1$ T after cooling in $\mu_0H_{\mathrm{FC}} = 1$ T (red), 0.1 T (gray), and 0 T (blue). Devitrification is observed as a rapid decrease of $M$ for the $\mu_0H_{\mathrm{FC}} = 1$ T case toward the equilibrium curve measured after $\mu_0H_{\mathrm{FC}} = 0$ T. \textbf{(d)} Temperature-dependent anisotropic resistivity ratio $\rho_L/\rho_S$, where $\rho_L$ and $\rho_S$ correspond to resistivity measured along the long and short axes of the crystal shown in the inset. These directions correspond to the degenerate $a=b$ crystallographic axes of the tetragonal phase. The ratio $\rho_L/\rho_S$ deviates from unity below $T_3$, indicating a loss of four-fold rotational symmetry.
 }
\end{figure}

To further quantify the anomalous behavior, we note that the $M(H)$ loops were collected using a five-quadrant measurement sequence after zero-field cooling from above $T_N$: $Q\mathrm{I}$ ($0 \to 7$~T), $Q\mathrm{II}$ ($7 \to 0$~T), $Q\mathrm{III}$ ($0 \to -7$~T), $Q\mathrm{IV}$ ($-7 \to 0$~T), and $Q\mathrm{V}$ ($0 \to 7$~T). For each loop, we compute $\Delta M = M(Q\mathrm{V}) - M(Q\mathrm{I})$, which is positive by construction when the virgin curve lies outside the main hysteresis loop. Plotting $\Delta M$ as a color map in Fig.~\ref{fig:Basics}a, we find that the anomalous behavior (red, positive regions) persists across phases IV and V, and also for fields above $B_4$ up to $T_3$. Exemplary curves can be found in Fig.~\ref{fig:micto} of the Supplementary Materials. Because this effect is not confined to a narrow temperature interval and instead increases with decreasing temperature in Phase IV, the behavior likely does not originate from a simple first-order phase transition~\cite{chattopadhyay2008metamagnetic}. We therefore carried out additional measurements to distinguish between the kinetic-arrest and cluster-glass interpretations.

\subsection{Evidence for kinetic hindrance and a magnetic glass state}

In a KHFOPT, arrest of the HT phase can produce a magnetic glass state. In that case, magnetic memory-like effects are expected, similar to those observed in spin-glass or cluster-glass systems \cite{pal2021memorylike}. If the energy landscape resembles Fig.~\ref{fig:cartoon}, cooling across the $(H^*,T^*)$ boundary in a magnetic field creates a magnetic glass state, whereas zero-field cooling produces the thermodynamic LT phase. Under this kinetic-arrest scenario, no memory effect is expected after zero-field cooling. By contrast, if the ground state is intrinsically a spin-glass or cluster-glass--like state, memory effects are expected even after zero-field cooling \cite{bag2018cluster,pal2021memorylike,vincent2022spin}. Memory tests are therefore informative for distinguishing between these scenarios.

A memory test was first performed under ZFC conditions, where we will refer to the following protocols as the ZFC memory protocols. The sample was cooled in zero field to the pause temperature $T_p = 3$~K at a rate of 1~K/min, held for 60~minutes, and then cooled to 1.8~K. At 1.8~K, a 1~T magnetic field was applied isothermally parallel to the $c$-axis. Magnetization $M$ was then measured on warming at 1~K/min, as shown in Fig.~\ref{fig:memory}a (dark line). This curve was compared to a reference ``no pause'' curve measured after uninterrupted cooling (Fig.~\ref{fig:memory}a, light line). No discernible difference is observed between the two curves, indicating no measurable memory effect under these ZFC conditions. Because strong magnetic fields can suppress memory signatures in spin- or cluster-glass systems, analogous measurements were performed in a 0.01 T applied field (Supplementary Fig.\ref{fig:ZFC100Oe}), yielding the same conclusion. These results indicate that the anomalous virgin curve in \CeCo{} is not due to a cluster-glass ground state.

Field-cooled memory protocols were then used to test the magnetic glass scenario, as shown in Fig.~\ref{fig:memory}b. The sample was field-cooled in 1~T at 3~K/min down to $T_p$. At $T_p$, the magnetic field was reduced isothermally to 0~T and the sample was held for one hour. The field was then increased isothermally back to 1~T, cooling resumed to 1.8~K, and $M$ was measured on warming. Measurements were carried out for several $T_p$ below $T_3$ and compared against a field-cooled ``no pause'' reference curve (gray curve in Fig.~\ref{fig:memory}b). Characteristic memory dips are observed as a local minimum in $M(T)$ near $T_p$. The dips are strongest at low temperature, where $\Delta M$ (Fig.~\ref{fig:Basics}a) is largest, and they are absent for $T_p = 8$~K. Additional support for glassy kinetics includes incomplete rejuvenation on warming (Fig.~\ref{fig:memory}b, main panel) and pronounced aging during the pause, demonstrated explicitly for $T_p = 3$~K in the inset of Fig.~\ref{fig:memory}b. Together, these results suggest that, at minimum, the transition into Phase IV is a kinetically hindered first-order phase transition.

The observations in Fig.~\ref{fig:memory}b can be understood as follows. Since phases II and III have an uncompensated magnetic moment \cite{smidman2013neutron}, it is likely that Phase V also carries an uncompensated moment. Field cooling of \CeCo{} in 1~T to 3~K therefore places the system in phase II with a ferrimagnetic moment. When the field is turned off at $T_p$, and assuming the $B_1$ phase boundary behaves similarly to Fig.~\ref{fig:cartoon}, part of the sample is kinetically hindered at zero field. As a result, ferrimagnetic phase II and antiferromagnetic phase IV coexist at 0~T and 3~K. During the dwell, the system relaxes slowly toward the antiferromagnetic ground state, as reflected in the time dependence of the normalized remanent magnetization $M/M_0$, where $M_0$ is the magnetization at $t=0$. The relaxation can be described by the empirical relation of the form $\propto \exp\!\left[-(t/\tau)^{\beta}\right]$ (Fig.~\ref{fig:memory}b inset, red line), with the relaxation time $\tau \sim 20 \times 10^6$~s and exponent $\beta \sim 0.2$ used for other glass formers \cite{bag2018cluster}. Exponents $0<\beta<1$ suggest the system evolves through multiple energy barriers as the system devitrifies and the extremely large $\tau$ is consistent with KHFOPTs \cite{chattopadhyay2005kinetic}. After the field is reapplied to 1~T and cooling proceeds to 1.8~K, the paused curves exhibit a reduced magnetization relative to the ``no pause'' reference, consistent with increased domain-wall pinning (domain stiffening) during aging. On subsequent warming, recovery is delayed until the temperature exceeds $T_p$, where thermal activation partially overcomes pinning. This produces a rapid increase in $M$ near $T_p$, which yields the memory-dip feature centered near the pause temperature.

To further confirm that the transition into Phase IV is a KHFOPT, the CHUF protocol was applied to \CeCo{} \cite{banerjee2006coexisting,banerjee2009conversion}. In CHUF, the temperature dependence of $M$ is measured on warming in a constant measurement field $\mu_0H_{\mathrm{meas}}$ after field cooling in a different field $\mu_0H_{\mathrm{FC}}$. Two unequal-field cases are considered: (i) $\mu_0H_{\mathrm{FC}}<\mu_0H_{\mathrm{meas}}$ and (ii) $\mu_0H_{\mathrm{FC}}>\mu_0H_{\mathrm{meas}}$. A reference curve is also measured with $\mu_0H_{\mathrm{FC}}=\mu_0H_{\mathrm{meas}}$. Figure~\ref{fig:memory}c shows three such curves with $\mu_0H_{\mathrm{meas}}=0.1$~T held fixed and $\mu_0H_{\mathrm{FC}}$ varied: 0~T (ZFC; blue), 0.1~T (reference; gray), and 1~T (red). The ZFC curve ($\mu_0H_{\mathrm{FC}}<\mu_0H_{\mathrm{meas}}$) is consistent with little to no trapped HT phase in the antiferromagnetic phase IV: $M$ remains small at low temperature and increases only gradually on warming until a larger increase occurs near $T_4$. In contrast, the reference curve ($\mu_0H_{\mathrm{FC}}=\mu_0H_{\mathrm{meas}}$) shows a larger low-temperature $M$, consistent with a finite fraction of the HT ferrimagnetic phase being trapped on cooling. When $\mu_0H_{\mathrm{FC}}>\mu_0H_{\mathrm{meas}}$ (1~T $\rightarrow$ 0.1~T), $M$ is further enhanced, consistent with an even larger trapped fraction. On warming, a rapid decrease in $M$ is observed at low temperature before the system recovers at higher temperature, which is naturally understood as devitrification of the arrested ferrimagnetic fraction toward the antiferromagnetic ground state. This behavior is characteristic of field-induced kinetic arrest and the magnetic glass state associated with it \cite{banerjee2006coexisting,banerjee2009conversion}.

While the transition into phase IV shows multiple signatures of a KHFOPT, an important question is why the transition becomes kinetically hindered in the first place. In many systems, kinetic hindrance is tied to quenched disorder associated with a magnetostructural phase transition \cite{roy2013first}. To probe whether symmetry breaking accompanies the magnetic transitions in \CeCo{}, anisotropic in-plane transport was measured on single crystals. In the high-temperature tetragonal phase, the electrical resistivity should be isotropic within the $ab$ plane. To test this, four electrical contacts were placed on the $ab$ face in a van der Pauw (vdP) geometry (inset of Fig.~\ref{fig:memory}d), and the in-plane anisotropic resistivities were measured in zero field. The resistivity ratio between the long and short edges of the crystal, $\rho_L/\rho_S$, was extracted using the Montgomery method \cite{montgomery1971method,dos2011procedure}. The long and short edges were aligned with the nominal $a$ and $b$ directions; in the tetragonal state these directions are equivalent and $\rho_L/\rho_S=1$. Consistent with this expectation, $\rho_L/\rho_S \approx 1$ above $T_3$ (Fig.~\ref{fig:memory}d). Below $T_3$, a small but systematic deviation develops, and below $T_4$ the anisotropy grows strongly away from 1. This deviation is reproducible in additional measurements on independent crystals (Supplementary Fig.~\ref{fig:anisotropy}). The emergence of $\rho_L/\rho_S \neq 1$ implies that four-fold symmetry is broken---possibly already below $T_3$, and definitively below $T_4$. The origin could arise from loss of four-fold symmetry due to the magnetic propagation vector or an accompanying structural distortion. However, since neutron scattering indicates the moments are confined to the crystallographic $c$-axis \cite{smidman2013neutron}, the transport anisotropy points to a loss of tetragonal symmetry that is most naturally associated with a magnetostructural transition. This provides a plausible mechanism for quenched disorder and, in turn, for the kinetic hindrance observed in the transition into phase IV.

\begin{figure*}[t]
\centering
\includegraphics[width=0.8\textwidth]{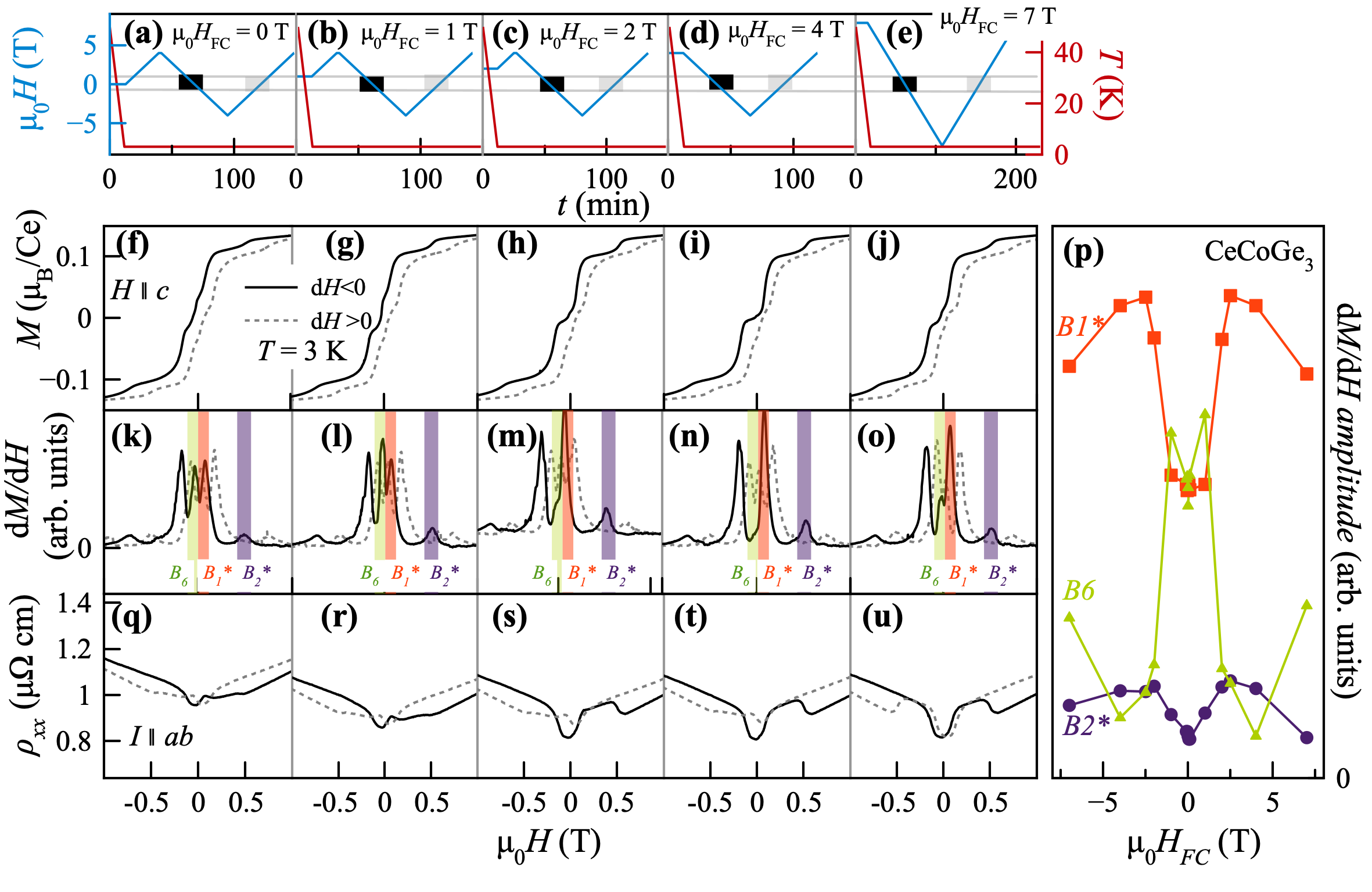}
\caption{\label{fig:FC_loops} \textbf{Field-cooling dependence of low-field properties.} The magnetic field $\mu_0H$ (blue, left axis) and temperature $T$ (red, right axis) histories corresponding to the field-cooling protocols with $\mu_0H_{FC} =$ ({\bf{a}}) 0 T, ({\bf{b}}) 1 T, ({\bf{c}}) 2 T, ({\bf{d}}) 4 T, and ({\bf{e}}) 7 T for which the isothermal magnetization $M$ was measured for $H \parallel c$ at $T~=~3$ K in ({\bf{f--j}}), respectively. The dark (light) squares in ({\bf{a--e}}) highlight the regions of decreasing (increasing) field, $dH<0$ ($dH>0$), for which $M$ is plotted in ({\bf{f--j}}) as solid (dashed) lines. ({\bf{k--o}}) Field-dependent derivative $dM/dH$ for the respective field-cooling procedures. The features associated with $B_2^{*}$, $B_1^{*}$, and $B_6$ are highlighted in purple, orange, and green, respectively. ({\bf{p}}) $\mu_0H_{FC}$ dependence of $B_2^{*}$ (purple circles), $B_1^{*}$ (orange squares), and $B_6$ (green diamonds), extracted from $dM/dH$ in ({\bf{k--o}}) as well as Fig.~\ref{fig:MHFC}. The field-dependent isothermal resistivity $\rho_{xx}$, measured at 3 K with $I \parallel ab$ and $H \parallel c$, is shown for the same field-cooling protocols in ({\bf{q}}) 0 T, ({\bf{r}}) 1 T, ({\bf{s}}) 2 T, ({\bf{t}}) 4 T, and ({\bf{u}}) 7 T.}
\end{figure*}

%The magnetic field $\mu_0H$ (blue, left axis) and temperature $T$ (red, right axis) histories corresponding to the $\mu_0H_{FC}~=$ ({\bf{a}}) 0 T, ({\bf{b}}) 1 T, ({\bf{c}}) 2 T, ({\bf{d}}) 4 T, and ({\bf{e}}) 7 T field-cooling protocols for which the isothermal magnetization $M$ was measured for $H \parallel c$ at $T~=~3$ K in ({\bf{f--j}}), respectively. The dark(light) squares in ({\bf{a--e}}) highlight the region of decreasing(increasing) field [d$H<0$(d$H>0$)] for which $M$ is plotted in ({\bf{f--j}}) as solid(dashed) lines. ({\bf{k--o}}) Field dependent derivative of $M$ d$M$/d$H$ for the respective field-cooling procedures. The peaks corresponding to the $B_2^{*}$, $B_1^{*}$, and $B_6$ are highlighted in purple, orange and green, respectively. ({\bf{p}}) The $\mu_0H_{FC}$ dependence of the $B_2^{*}$ (purple, circles), $B_1^{*}$ (orange, squares), and $B_6$ (green, diamonds) generated from d$M$/d$H$ in ({\bf{k--o}}) as well as \textcolor{red}{Supplementary Fig.~xx}. The field dependent isothermal resistivity $\rho_{xx}$ measured with current $I \parallel ab$ at $H \parallel c$ at 3 K for the same ({\bf{q}}) 0 T, ({\bf{r}}) 1 T, ({\bf{s}}) 2 T, ({\bf{t}}) 4 T, and ({\bf{u}}) 7 T shown in ({\bf{a--e}}) for which $M$ was also measured.

\subsection{Tunable response by field cooling}

Next, we set out to demonstrate that because \CeCo{} hosts a KHFOPT, domain pinning can be systematically tuned by field-cooling procedures, and that the electrical transport is sensitive to this tuning. To do this, we devised a field-cooling protocol in which the sample was cooled from above \(T_N\) with \(H \parallel c\) in selected cooling fields \(\mu_0H_{FC}\). The chosen values were \(|\mu_0H_{FC}|=\) 0, 0.05, 0.1, 1, 2, 2.5, 4, and 7 T. After reaching 3~K, the field was cycled according to the sign of \(\mu_0H_{FC}\): for \(|\mu_0H_{FC}|\le 4\)~T we used \(\mu_0H_{FC}\rightarrow \pm4~\text{T}\rightarrow \mp4~\text{T}\rightarrow \pm4~\text{T}\), while for \(\mu_0H_{FC}=7\)~T we used \(\mu_0H_{FC}\rightarrow \pm7~\text{T}\rightarrow \mp7~\text{T}\rightarrow \pm 7~\text{T}\). During the \(\pm4~\text{T}\rightarrow \mp4~\text{T}\rightarrow \pm4~\text{T}\) (or \(\pm 7\)~T) cycle, \(M(H)\) and \(\rho_{xx}(H)\) hysteresis loops were recorded. Figure~\ref{fig:FC_loops}a--e shows representative field (blue, left axis) and temperature (red, right axis) histories for \(\mu_0H_{FC}=\) 0, 1, 2, 4, and 7~T. The dark shaded regions correspond to the decreasing-field segments (\(dH<0\)) and the light shaded regions correspond to the increasing-field segments (\(dH>0\)), with the comparisons below focused on the window between \(\mu_0H=+1~\text{T}\rightarrow -1~\text{T}\) for \(dH<0\) and \(\mu_0H=-1~\text{T}\rightarrow +1~\text{T}\) for \(dH>0\). The corresponding 3~K data for \(M\) (Fig.~\ref{fig:FC_loops}f--j), \(dM/dH\) (Fig.~\ref{fig:FC_loops}k--o), and \(\rho_{xx}\) (Fig.~\ref{fig:FC_loops}q--u) are shown for those segments. %\textcolor{red}{Insert the field sweep rate here.}

We first focus on the magnetization on the descending branch (solid curves in Fig.~\ref{fig:FC_loops}f--j) and track the features labeled \(B_2^*\), \(B_1^*\), and \(B_6\), which are highlighted most clearly in the derivative \(dM/dH\) (Fig.~\ref{fig:FC_loops}k--o). The \(\mu_0H_{FC}\)-dependence of the \(dM/dH\) peak amplitudes for these features is summarized in Fig.~\ref{fig:FC_loops}p. The \(B_6\) feature (highlighted in green) does not appear on the ZFC virgin curve or in the ZFC phase diagram (Fig.~\ref{fig:Basics}a,c), since it exists only in negative fields for \(dH<0\) (or, equivalently, only in positive fields for \(dH>0\)). We therefore attribute \(B_6\) to unpinning, or devitrification, of higher-field phases that remain trapped by kinetic hindrance. Consistent with this interpretation, the \(B_6\) peak amplitude slightly increases upon field cooling in \(\mu_0H_{FC}=1\)~T (Fig.~\ref{fig:FC_loops}l) compared to \(\mu_0H_{FC}=0\)~T (Fig.~\ref{fig:FC_loops}k), suggesting that the 1~T cooling procedure increases the fraction of the sample pinned in a higher-field state at \(\mu_0H=0\)~T. In contrast, for intermediate cooling fields \(1~\text{T}<\mu_0H_{FC}<7~\text{T}\), the \(B_6\) amplitude is dramatically reduced (Fig.~\ref{fig:FC_loops}m,n), while the \(B_2^*\) and \(B_1^*\) peak amplitudes increase relative to the \(\mu_0H_{FC}=0\)~T case. This indicates that these field-cooling procedures promote a larger volume fraction transforming into the LT phase IV at the thermodynamic \(B_2^*\) and \(B_1^*\) boundaries, which in turn leaves less kinetically hindered high-field phase at \(\mu_0H=0\)~T and therefore suppresses the \(B_6\) devitrification signature. Interestingly, upon further increasing the cooling field to \(\mu_0H_{FC}=7\)~T, the \(B_6\) amplitude grows again (Fig.~\ref{fig:FC_loops}o) while the \(B_2^*\) and \(B_1^*\) amplitudes decrease, implying a re-entrant increase of kinetically trapped phase at \(\mu_0H=0\)~T compared to the \(\mu_0H_{FC}=4\)~T case.

While Fig.~\ref{fig:cartoon} captures many qualitative aspects of the behavior in \CeCo{}, it does not account for the non-monotonic dependence on \(\mu_0H_{FC}\). At face value, Fig.~\ref{fig:cartoon} suggests that increasing \(\mu_0H_{FC}\) should monotonically increase the pinned volume fraction (up to saturation at \(T_0=3\)~K), whereas Fig.~\ref{fig:FC_loops}p instead shows a rise in the trapped HT fraction up to \(\mu_0H_{FC}\approx 1\)~T, a minimum near \(\mu_0H_{FC}\approx 4\)~T, and then an increase again at \(\mu_0H_{FC}=7\)~T. We have corroborated on an independent crystal that this trend is the same when the field was reversed directly after field-cooling rather than first being driven to \(\mu_0H=\pm 4\)~T; the same conclusions are reached (see Supplemental Fig.~\ref{fig:reversal}).

One plausible way to rationalize this non-monotonic trend is to borrow intuition from exchange-bias systems~\cite{meiklejohn1957new}, where pinning is controlled by a competition between interfacial coupling and the Zeeman tendency of the ``softer'' component to rotate in field. In that language, increasing \(\mu_0H_{FC}\) can initially strengthen pinning by increasing the amount of high-field phase that survives to low temperature, which increases the opportunity for interfacial coupling to bias the subsequent reversal. However, once the high-field portion becomes too dominant, the relevant interfaces between coexisting regions can shrink in a relative sense (because the sample becomes more single-phase-like), even while the net moment of the high-field component grows. In that regime, the coupling that produces pinning becomes less effective and the trapped fraction can decrease---naturally producing a minimum near intermediate \(\mu_0H_{FC}\). The re-entrant growth of the trapped fraction at \(\mu_0H_{FC}=7\)~T then suggests that additional kinetics enter at the highest fields, consistent with the possibility that the \(B_5\) transition itself becomes kinetically hindered, as also hinted by the anomalous virgin curve lying outside the main hysteresis loop (inset of Fig.~\ref{fig:Basics}c).

The \(\rho_{xx}\) measurements under the same field-cooling procedures corroborate this domain-pinning picture. For small cooling fields (\(\mu_0H_{FC}=0\) and 1~T), the resulting \(\rho_{xx}(H)\) loops (Fig.~\ref{fig:FC_loops}q,r) show no hysteresis at \(\mu_0H=0\). Compared to the ZFC virgin curve (Fig.~\ref{fig:Basics}d), these loops appear hindered, with less sharply defined signatures of the transitions. Increasing the cooling field to \(\mu_0H_{FC}=2\) and 4~T produces a pronounced low-resistance state on the \(dH<0\) branch (Fig.~\ref{fig:FC_loops}s, solid), which more closely resembles the ZFC virgin curve in Fig.~\ref{fig:Basics}d, again indicating that this branch is less hindered under these cooling conditions than in the \(\mu_0H_{FC}=0\) and 1~T cases. Notably, once the field is swept through \(-B_5\) and returned to \(+4\)~T, the transport on the \(dH>0\) branch becomes re-hindered, and for \(\mu_0H_{FC}\le 4\)~T the \(dH>0\) branches are essentially indistinguishable from one another (Fig.~\ref{fig:FC_loops}q--s, dashed). This creates an apparent zero-applied-field (\(\mu_0H_z\)) hysteresis in \(\rho_{xx}\). By the Onsager--Casimir reciprocal relations, \(\rho_{xx}(B_z)\) must be even in \(B_z\), which forbids intrinsic hysteresis at fixed \(B_z\); the observed zero-\(\mu_0H_z\) hysteresis therefore implies that the system accesses different domain populations on the \(dH<0\) and \(dH>0\) branches, altering \(M_z\) and thus shifting \(B_z=\mu_0(H_z+M_z)\) even when \(\mu_0H_z=0\). We verified this behavior on two independent crystals using both positive and negative field-cooling procedures, eliminating trapped flux as the origin as shown in Figs.~\ref{fig:CH2}--\ref{fig:CH1} of the Supplementary Materials. Finally, for \(\mu_0H_{FC}=7\)~T (Fig.~\ref{fig:FC_loops}t), the zero-field hysteresis in \(\rho_{xx}\) disappears, indicating that the \(dH<0\) and \(dH>0\) states are again equivalent to each other, similar to the ZFC case. The overall shape resembles the less-hindered \(\mu_0H_{FC}=4\)~T \(dH<0\) state (Fig.~\ref{fig:FC_loops}s, solid), but with a slightly higher value at zero field, consistent with the magnetization-based conclusion that the 7~T cooling procedure restores some degree of kinetic trapping at \(\mu_0H=0\).

In summary, Fig.~\ref{fig:FC_loops} shows that field-cooling protocols control how much phase coexistence (and therefore kinetic trapping) remains at zero applied field on the \(dH<0\) branch in both \(M\) and \(\rho_{xx}\), producing a clear memory effect. Driving the field beyond \(-B_5\) and then reversing it largely erases that memory, yielding \(dH>0\) branches that are similar for all protocols in which the maximum field reached was 4~T. The remaining distinctions in the \(dH>0\) response are instead set primarily by the maximum field experienced: the \(\mu_0H_{FC}=7\)~T loop is qualitatively different, which we attribute to the presence of the additional first-order transition at \(B_5\) and the associated kinetics.

\subsection{Multilevel tunable magnetoresistance}

\begin{figure*}[t]
\centering
\includegraphics[width=0.8\textwidth]{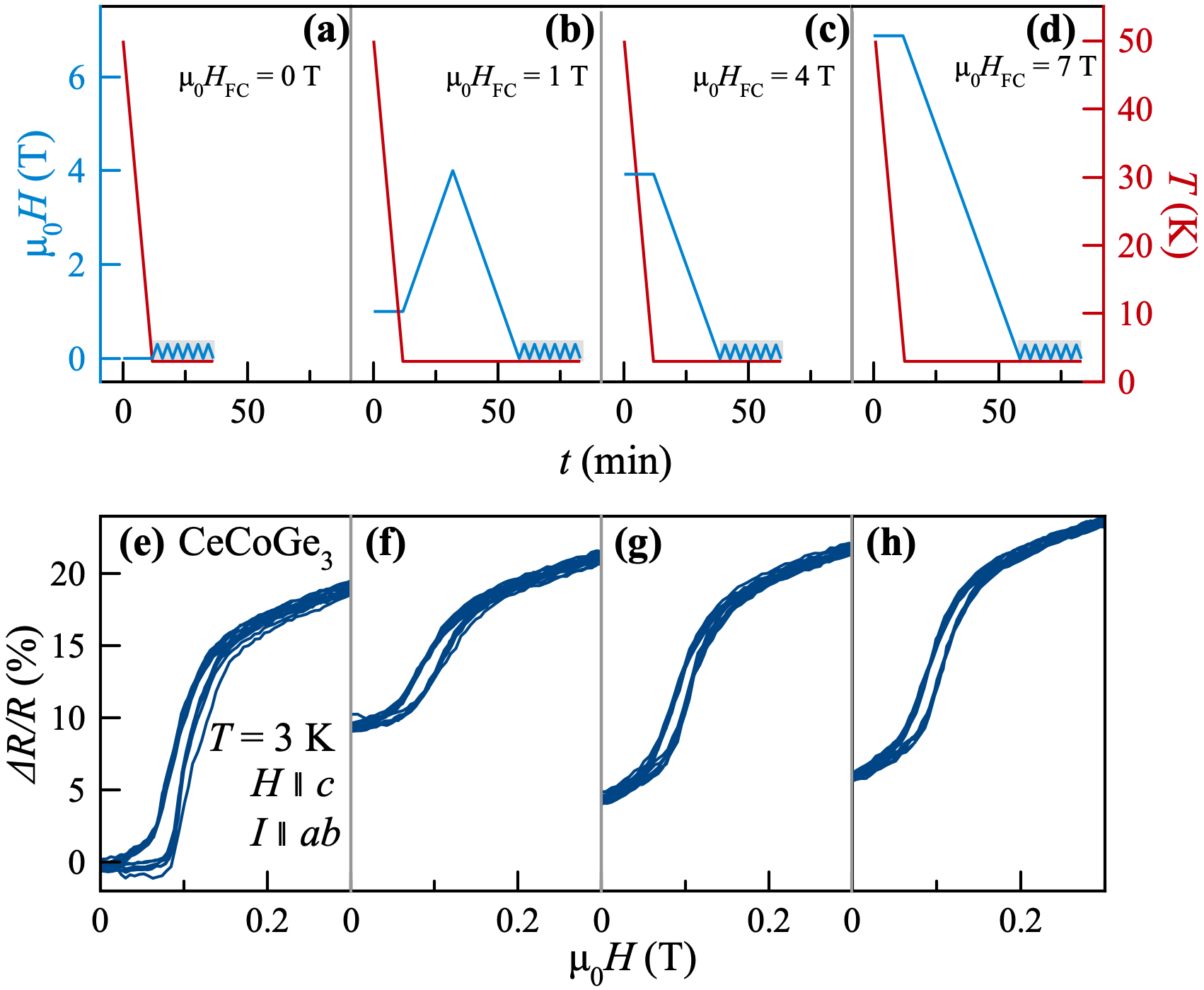}
\caption{\label{fig:minor_loops} \textbf{Tunable electrical transport in CeCoGe$_3$.} The magnetic field $\mu_0H$ (blue, left axis) and temperature $T$ (red, right axis) histories corresponding to the field-cooling protocols with $\mu_0H_{FC} =$ ({\bf{a}}) 0 T, ({\bf{b}}) 1 T, ({\bf{c}}) 4 T, and ({\bf{d}}) 7 T, for which the magnetoresistance, calculated as $\Delta R/R$ (see text for details), was measured in ({\bf{e--h}}). The field-cooling procedure dictates the magnetoresistance. The magnetoresistance was measured at $T = 3$ K for $H \parallel c$ and $I \parallel ab$.}
\end{figure*}

Finally, building on our understanding of phase coexistence arising from the KHFOPT, we show that \CeCo{} supports a predictable and tunable multilevel resistive switching behavior. To demonstrate this, we measured \(\rho_{xx}\) using a set of minor-loop protocols, with the corresponding field and temperature histories shown in Fig.~\ref{fig:minor_loops}a--d. The initial portion of each protocol matches the field-cooling procedures used in Fig.~\ref{fig:FC_loops}. We then chose four representative cooling fields, \(\mu_0H_{FC}=0\), 1, 4, and 7~T, which span the extrema in the \(B_6\) \(dM/dH\) amplitude and therefore bracket the range of kinetically trapped phase fractions at \(\mu_0H=0\). In all cases, the minor loop consists of repeatedly cycling the field between \(\mu_0H=0\) and \(\mu_0H=0.3\)~T at \(T=3\)~K, where 0.3~T lies above \(B_1\) but below \(B_2\), and recording the corresponding magnetoresistance.

We present the data as \(\Delta R/R\), defined as
\begin{equation}
\Delta R/R = \frac{R_{xx}(H)-R^{\mu_0H_{FC}=0~\mathrm{T}}_{xx}(H=0)}{R^{\mu_0H_{FC}=0~\mathrm{T}}_{xx}(H=0)}\times 100\%,
\end{equation}

where \(R^{\mu_0H_{FC}=0~\mathrm{T}}_{xx}(H=0)\) is the zero-field resistance of the ZFC (virgin) state at 3~K. For the ZFC protocol (Fig.~\ref{fig:minor_loops}a), the field was ramped to 0.3~T and then cycled six times between 0.3~T and 0~T; the resulting \(\Delta R/R\) is shown in Fig.~\ref{fig:minor_loops}e. In this case, the system reproducibly toggles between a high- and low-resistance state, with a magnetoresistance of order \(\sim 20\%\) for the crystal shown (typically \(\sim 20\)--40\% across crystals). Importantly, both resistance states are stable under repeated cycling, drifting by less than \(\sim 5\%\) between cycles 1 and 6.

We next use the \(\mu_0H_{FC}=1\)~T protocol (Fig.~\ref{fig:minor_loops}b) to prepare a zero-field state with the largest fraction of kinetically hindered phase. The sample was field-cooled in \(\mu_0H_{FC}=1\)~T to \(T=3\)~K, driven to 4~T, ramped back to 0~T, and then cycled six times between 0 and 0.3~T. The resulting \(\Delta R/R\) is shown in Fig.~\ref{fig:minor_loops}f. Compared to the ZFC case, the zero-field resistance is clearly enhanced, consistent with increased phase coexistence at \(\mu_0H=0\). At the same time, the minor-loop switching amplitude is reduced substantially, with the extrema of \(\Delta R/R\) suppressed by approximately a factor of two.

Figures~\ref{fig:minor_loops}c and d show the corresponding \(H\)--\(T\) histories for the \(\mu_0H_{FC}=4\) and 7~T protocols. As discussed above, field-cooling in 4~T produces a zero-field state with relatively little phase coexistence compared to the 1~T case and is therefore most similar to the ZFC state. Consistent with this, the \(\mu_0H_{FC}=4\)~T minor-loop response exhibits the smallest zero-field offset among the finite-field-cooled protocols. In contrast, the \(\mu_0H_{FC}=7\)~T protocol yields a modestly higher zero-field resistance than the 4~T case, indicating a larger trapped fraction at \(\mu_0H=0\), and it additionally exhibits subtly different kinetics: the resistance near 0.3~T remains elevated by \(\sim 5\%\) relative to the ZFC reference. Taken together, these minor-loop measurements demonstrate a stable multilevel write--read scheme in \CeCo{}, where the zero-field resistive state is programmed by the field-cooling history (i.e., the amount of phase coexistence at \(\mu_0H=0\)) and then reproducibly toggled by cycling through the metamagnetic transition at low field.

\section{Discussion}

Our measurements show that \CeCo{} hosts a kinetically hindered first-order phase transition at the $B_1$ metamagnetic transition (Fig.~\ref{fig:Basics}a) that can be used as a practical control knob: magnetic-field history programs the degree of phase coexistence below the $B_1$ transition, and that programmed state is directly readable in electrical transport (Figs.~\ref{fig:FC_loops} and \ref{fig:minor_loops}). Magnetic memory and CHUF measurements (Fig.~\ref{fig:memory}a-c) demonstrate the arrested state is characteristic of a magnetic glass \cite{banerjee2006coexisting,banerjee2009conversion}.

Within this kinetic-arrest picture, \CeCo{} provides a tunable mixture of the low-temperature antiferromagnetic state (Phase IV in Fig.~\ref{fig:Basics}a) and a metastable ferrimagnetic-like high-field state that can remain frozen in when the field is reduced (all states besides Phase IV). This arrested fraction behaves like a state variable set by the path through the phase diagram. In practice, this leads to an operational ``write--read--erase'' framework that is already embedded in our protocols: the write step is the chosen field history, the read step is the value of \(\rho_{xx}\) at \(\mu_0H=0\), and the erase/reset step can be achieved by warming above $T_N$, or cycling the field above $B_5$ (Fig.~\ref{fig:FC_loops}). The minor-loop measurements (Fig.~\ref{fig:minor_loops}) make this framing explicit: repeated cycling between 0 and 0.3~T toggles reproducibly between resistance states with small drift, while the baseline (zero-field) resistance level is programmable across multiple distinct values by the preparation history.

The non-monotonic dependence of pinning on the cooling field (Fig.~\ref{fig:FC_loops}p) is also informative. If increasing \(\mu_0H_{FC}\) simply increases the amount of trapped high-field phase, one might expect monotonic behavior. Instead, the trapped fraction inferred from the devitrification signature passes through a minimum at intermediate \(\mu_0H_{FC}\), implying that kinetic arrest is controlled by competing effects: the field history changes both how much high-field phase is created and how effectively that phase is stabilized (or destabilized) upon return to low fields. Qualitatively, this is consistent with a picture where pinning depends not only on volume fractions but also on the availability and effectiveness of interfaces and pinning landscapes that bias the return path. The emergence of in-plane transport anisotropy below \(T_3\) (and especially below \(T_4\)) (Fig.~\ref{fig:memory}d) suggests that four-fold symmetry is broken in the low-temperature regime, pointing to magnetostructural coupling and/or texture formation as a plausible source of a rugged energy landscape and strong pinning. In that sense, \CeCo{} may be an especially favorable platform because the non-centrosymmetric crystal structure (and associated DM interaction) already promotes complex magnetic textures, while additional symmetry lowering can supply the disorder/strain accommodation needed to slow transformation kinetics.

More broadly, our results suggest a set of practical design rules for realizing tunable multilevel switching from KHFOPTs in bulk materials. The first ingredient is a first-order boundary between competing magnetic states often involving an antiferromagnetic ground state and a higher-field state with an uncompensated moment. The second is a kinetic bottleneck, arising from magnetostructural coupling, quenched disorder, strain accommodation, or strong domain-wall pinning, that allows a controllable fraction of the high-field phase to remain arrested on returning to low fields. The third is an electronic transport channel (the system should be metallic) that is sensitive to phase fraction and domain population, so that small changes in the arrested fraction translate into large, stable changes in resistance. These ingredients are not unique to \CeCo{}, as KHFOPTs have been reported across diverse material families, including manganites, itinerant magnets, and alloys in which AFM-FM phase competition is coupled to magnetostructural transitions \cite{Roy2008ActaMaterMetamagnetic,chattopadhyay2005kinetic,banerjee2006ferromagnetic,Sharma2007PRBNiMnIn}. It would be interesting to see if similar multilevel memory can be achieved by appropriate field-cycling protocols in these compounds where the huge value of the magnetoresistance makes them particularly interesting \cite{Tokura2006RPP,Dagotto2001PhysRep}.

From an applications standpoint, there are clear next steps and constraints. The switching in \CeCo{} occurs at 3 K, so materials with similar physics but higher operating temperatures are needed. Colossal magnetoresistance manganites provide a natural place to search for such behavior, since related kinetic-arrest phenomena have been reported in systems with competing FM metallic and AFM or charge-ordered insulating phases at substantially higher temperatures \cite{Tokura2006RPP,Dagotto2001PhysRep,gayathri2023interfacial}. On the other hand, \CeCo{} serves as a platform for particularly important near-term experiments to test replacing field-cooling with isothermal programming by electric fields or currents. \CeCo{} is an appealing platform because its lack of inversion symmetry makes electrically driven spin--orbit torques \cite{Manchon2019RMP,Zelezny2017PRB,Edelstein1990SSC,Zelezny2014PRL,Wadley2016Science} a realistic route to bias domain populations, potentially enabling multilevel switching without magnetic field-cooling.

\section{Methods}

\subsubsection{Single-crystal synthesis}
Details of the synthesis conditions can be found in Ref. \cite{moya2025measuring}.

\subsubsection{Single-crystal X-ray diffraction}
 Measurements were performed at 100 K using an APEX2 charge-coupled device (CCD) diffractometer equipped with a Mo K$\alpha$ ($\lambda$ = 0.71073 \AA), sealed-tube X-ray source and a graphite monochromator. Data collection covered a full hemisphere up to a resolution of 0.65 \AA, with indexation and integration carried out using APEX2 software. Run list generation and frame processing were also performed in APEX2. A spherical absorption correction was applied before importing the processed peak list into JANA2020 for further structural refinement \cite{petvrivcek2023jana2020}.

 \subsubsection{Direct-current (DC) magnetization measurements}
DC magnetization measurements up to 7 T were performed in a Quantum Design (QD) Magnetic Property Measurement System (MPMS3) using the vibrating sample magnetometer--superconducting quantum interference device (VSM--SQUID) option. Samples were mounted on quartz braces and inserted onto the brass sample holder. Measurements up to 14 T were performed in a QD DynaCool Physical Property Measurement System (PPMS) cryostat equipped with a VSM option. %The magnetic field was either stabilized or ramped at $\approx 25$ Oe/s for $M(H)$ loops while the temperature was typically ramped at 1.6 K/min for magnetic susceptibility measurements.
%Typical measurement parameters were: peak amplitude 5 mm, averaging time 2 s, and maximum acceleration 38.138 m/sec$^2$.

 \subsubsection{Resistivity measurements}
Longitudinal resistivity measurements were performed in a standard four-point collinear geometry with the applied current $i$ and measured voltage along the $a$ (or by the tetragonal symmetry $b$) axis in a QD DynaCool PPMS equipped with the electrical transport option (ETO). Typical measurement parameters were $i$ = 3--5 mA and $f~=~18.31$ Hz. The data were not symmetrized because of the asymmetry that can appear with respect to $\mu_0H~=~0~$ T.

The anisotropic resistivity measurements were performed using a QD DynaCool PPMS equipped with the D542 vdP Switchbox option. The in-plane resistivity anisotropy was extracted from the vdP geometry using the modified Montgomery \cite{montgomery1971method,dos2011procedure}.

 \subsubsection{Heat capacity measurements}
 Heat capacity measurements were measured in a QD DynaCool equipped with the heat capacity option. The sample was contacted to the sapphire substrate using Apiezon N-grease. Three measurements were performed at each temperature using the semi-adiabatic method and averaged together.

\begin{acknowledgments}
This work is supported by an NSF CAREER grant (DMR-2144295) to LMS, by the Air Force Office of Scientific Research under award number FA9550-25-1-0177 and by NSF through the Princeton Center for Complex Materials, a Materials Research Science and Engineering Center DMR-2011750. This work was further supported by the Gordon and Betty Moore Foundation (EPiQS Synthesis Award) through grant GBMF9064, and the David and Lucile Packard Foundation. SBL and CJP were supported by National Science Foundation Graduate Research Fellowship Program under Grant No. DGE-2039656. GS was supported by the Arnold and Mabel Beckman Foundation via an AOB postdoctoral fellowship (https://doi.org/10.13039/100000997).
\end{acknowledgments}

% \section*{Author contributions}
%  LMS and JMM conceived of the project. LMS supervised the entire project. JMM grew the single crystals, performed thermodynamic and transport measurements, and analyzed the data. SBL, GS, CP, and JMM performed the structural and chemical characterization. SC and NM provided scientific insight. JMM and LMS wrote the manuscript with input from all authors.

\section*{Competing interests}
There are no competing interests to declare.

\section*{Data and materials availability}
All data to support the conclusions of the manuscript are presented in the main text or Supplementary Materials.

\bibliographystyle{apsrev4-2}
\bibliography{science_template}

\begin{thebibliography}{63}%
\makeatletter
\providecommand \@ifxundefined [1]{%
 \@ifx{#1\undefined}
}%
\providecommand \@ifnum [1]{%
 \ifnum #1\expandafter \@firstoftwo
 \else \expandafter \@secondoftwo
 \fi
}%
\providecommand \@ifx [1]{%
 \ifx #1\expandafter \@firstoftwo
 \else \expandafter \@secondoftwo
 \fi
}%
\providecommand \natexlab [1]{#1}%
\providecommand \enquote  [1]{``#1''}%
\providecommand \bibnamefont  [1]{#1}%
\providecommand \bibfnamefont [1]{#1}%
\providecommand \citenamefont [1]{#1}%
\providecommand \href@noop [0]{\@secondoftwo}%
\providecommand \href [0]{\begingroup \@sanitize@url \@href}%
\providecommand \@href[1]{\@@startlink{#1}\@@href}%
\providecommand \@@href[1]{\endgroup#1\@@endlink}%
\providecommand \@sanitize@url [0]{\catcode `\\12\catcode `\$12\catcode `\&12\catcode `\#12\catcode `\^12\catcode `\_12\catcode `\%12\relax}%
\providecommand \@@startlink[1]{}%
\providecommand \@@endlink[0]{}%
\providecommand \url  [0]{\begingroup\@sanitize@url \@url }%
\providecommand \@url [1]{\endgroup\@href {#1}{\urlprefix }}%
\providecommand \urlprefix  [0]{URL }%
\providecommand \Eprint [0]{\href }%
\providecommand \doibase [0]{https://doi.org/}%
\providecommand \selectlanguage [0]{\@gobble}%
\providecommand \bibinfo  [0]{\@secondoftwo}%
\providecommand \bibfield  [0]{\@secondoftwo}%
\providecommand \translation [1]{[#1]}%
\providecommand \BibitemOpen [0]{}%
\providecommand \bibitemStop [0]{}%
\providecommand \bibitemNoStop [0]{.\EOS\space}%
\providecommand \EOS [0]{\spacefactor3000\relax}%
\providecommand \BibitemShut  [1]{\csname bibitem#1\endcsname}%
\let\auto@bib@innerbib\@empty
%</preamble>
\bibitem [{\citenamefont {Sebastian}\ \emph {et~al.}(2020)\citenamefont {Sebastian}, \citenamefont {Le~Gallo}, \citenamefont {Khaddam-Aljameh},\ and\ \citenamefont {Eleftheriou}}]{Sebastian2020NatNanoInMemory}%
  \BibitemOpen
  \bibfield  {author} {\bibinfo {author} {\bibfnamefont {A.}~\bibnamefont {Sebastian}}, \bibinfo {author} {\bibfnamefont {M.}~\bibnamefont {Le~Gallo}}, \bibinfo {author} {\bibfnamefont {R.}~\bibnamefont {Khaddam-Aljameh}},\ and\ \bibinfo {author} {\bibfnamefont {E.}~\bibnamefont {Eleftheriou}},\ }\href {https://doi.org/10.1038/s41565-020-0655-z} {\bibfield  {journal} {\bibinfo  {journal} {Nature Nanotechnology}\ }\textbf {\bibinfo {volume} {15}},\ \bibinfo {pages} {529} (\bibinfo {year} {2020})}\BibitemShut {NoStop}%
\bibitem [{\citenamefont {Xia}\ and\ \citenamefont {Yang}(2019)}]{Xia2019NatMaterCrossbar}%
  \BibitemOpen
  \bibfield  {author} {\bibinfo {author} {\bibfnamefont {Q.}~\bibnamefont {Xia}}\ and\ \bibinfo {author} {\bibfnamefont {J.~J.}\ \bibnamefont {Yang}},\ }\href {https://doi.org/10.1038/s41563-019-0291-x} {\bibfield  {journal} {\bibinfo  {journal} {Nature Materials}\ }\textbf {\bibinfo {volume} {18}},\ \bibinfo {pages} {309} (\bibinfo {year} {2019})}\BibitemShut {NoStop}%
\bibitem [{\citenamefont {Zhou}\ \emph {et~al.}(2024)\citenamefont {Zhou}, \citenamefont {Li}, \citenamefont {Ang},\ and\ \citenamefont {Zhang}}]{Zhou2024NML2DInMemory}%
  \BibitemOpen
  \bibfield  {author} {\bibinfo {author} {\bibfnamefont {H.}~\bibnamefont {Zhou}}, \bibinfo {author} {\bibfnamefont {S.}~\bibnamefont {Li}}, \bibinfo {author} {\bibfnamefont {K.-W.}\ \bibnamefont {Ang}},\ and\ \bibinfo {author} {\bibfnamefont {Y.-W.}\ \bibnamefont {Zhang}},\ }\href {https://doi.org/10.1007/s40820-024-01335-2} {\bibfield  {journal} {\bibinfo  {journal} {Nano-Micro Letters}\ }\textbf {\bibinfo {volume} {16}},\ \bibinfo {pages} {121} (\bibinfo {year} {2024})}\BibitemShut {NoStop}%
\bibitem [{\citenamefont {Jonason}\ \emph {et~al.}(1998)\citenamefont {Jonason}, \citenamefont {Vincent}, \citenamefont {Hammann}, \citenamefont {Bouchaud},\ and\ \citenamefont {Nordblad}}]{Jonason1998PRLMemoryChaos}%
  \BibitemOpen
  \bibfield  {author} {\bibinfo {author} {\bibfnamefont {K.}~\bibnamefont {Jonason}}, \bibinfo {author} {\bibfnamefont {E.}~\bibnamefont {Vincent}}, \bibinfo {author} {\bibfnamefont {J.}~\bibnamefont {Hammann}}, \bibinfo {author} {\bibfnamefont {J.-P.}\ \bibnamefont {Bouchaud}},\ and\ \bibinfo {author} {\bibfnamefont {P.}~\bibnamefont {Nordblad}},\ }\href {https://doi.org/10.1103/PhysRevLett.81.3243} {\bibfield  {journal} {\bibinfo  {journal} {Physical Review Letters}\ }\textbf {\bibinfo {volume} {81}},\ \bibinfo {pages} {3243} (\bibinfo {year} {1998})}\BibitemShut {NoStop}%
\bibitem [{\citenamefont {Mydosh}(1993)}]{mydosh1993spin}%
  \BibitemOpen
  \bibfield  {author} {\bibinfo {author} {\bibfnamefont {J.~A.}\ \bibnamefont {Mydosh}},\ }\href@noop {} {\emph {\bibinfo {title} {Spin glasses: an experimental introduction}}}\ (\bibinfo  {publisher} {CRC Press},\ \bibinfo {year} {1993})\BibitemShut {NoStop}%
\bibitem [{\citenamefont {Hopfield}(1982)}]{hopfield1982neural}%
  \BibitemOpen
  \bibfield  {author} {\bibinfo {author} {\bibfnamefont {J.~J.}\ \bibnamefont {Hopfield}},\ }\href@noop {} {\bibfield  {journal} {\bibinfo  {journal} {Proceedings of the national academy of sciences}\ }\textbf {\bibinfo {volume} {79}},\ \bibinfo {pages} {2554} (\bibinfo {year} {1982})}\BibitemShut {NoStop}%
\bibitem [{\citenamefont {Grollier}\ \emph {et~al.}(2020)\citenamefont {Grollier}, \citenamefont {Querlioz}, \citenamefont {Camsari}, \citenamefont {Everschor-Sitte}, \citenamefont {Fukami},\ and\ \citenamefont {Stiles}}]{grollier2020neuromorphic}%
  \BibitemOpen
  \bibfield  {author} {\bibinfo {author} {\bibfnamefont {J.}~\bibnamefont {Grollier}}, \bibinfo {author} {\bibfnamefont {D.}~\bibnamefont {Querlioz}}, \bibinfo {author} {\bibfnamefont {K.~Y.}\ \bibnamefont {Camsari}}, \bibinfo {author} {\bibfnamefont {K.}~\bibnamefont {Everschor-Sitte}}, \bibinfo {author} {\bibfnamefont {S.}~\bibnamefont {Fukami}},\ and\ \bibinfo {author} {\bibfnamefont {M.~D.}\ \bibnamefont {Stiles}},\ }\href@noop {} {\bibfield  {journal} {\bibinfo  {journal} {Nature electronics}\ }\textbf {\bibinfo {volume} {3}},\ \bibinfo {pages} {360} (\bibinfo {year} {2020})}\BibitemShut {NoStop}%
\bibitem [{\citenamefont {Manekar}\ \emph {et~al.}(2001)\citenamefont {Manekar}, \citenamefont {Chaudhary}, \citenamefont {Chattopadhyay}, \citenamefont {Singh}, \citenamefont {Roy},\ and\ \citenamefont {Chaddah}}]{manekar2001first}%
  \BibitemOpen
  \bibfield  {author} {\bibinfo {author} {\bibfnamefont {M.}~\bibnamefont {Manekar}}, \bibinfo {author} {\bibfnamefont {S.}~\bibnamefont {Chaudhary}}, \bibinfo {author} {\bibfnamefont {M.}~\bibnamefont {Chattopadhyay}}, \bibinfo {author} {\bibfnamefont {K.}~\bibnamefont {Singh}}, \bibinfo {author} {\bibfnamefont {S.}~\bibnamefont {Roy}},\ and\ \bibinfo {author} {\bibfnamefont {P.}~\bibnamefont {Chaddah}},\ }\href@noop {} {\bibfield  {journal} {\bibinfo  {journal} {Physical Review B}\ }\textbf {\bibinfo {volume} {64}},\ \bibinfo {pages} {104416} (\bibinfo {year} {2001})}\BibitemShut {NoStop}%
\bibitem [{\citenamefont {Chattopadhyay}\ \emph {et~al.}(2005{\natexlab{a}})\citenamefont {Chattopadhyay}, \citenamefont {Roy},\ and\ \citenamefont {Chaddah}}]{chattopadhyay2005kinetic}%
  \BibitemOpen
  \bibfield  {author} {\bibinfo {author} {\bibfnamefont {M.}~\bibnamefont {Chattopadhyay}}, \bibinfo {author} {\bibfnamefont {S.}~\bibnamefont {Roy}},\ and\ \bibinfo {author} {\bibfnamefont {P.}~\bibnamefont {Chaddah}},\ }\href@noop {} {\bibfield  {journal} {\bibinfo  {journal} {Physical Review B—Condensed Matter and Materials Physics}\ }\textbf {\bibinfo {volume} {72}},\ \bibinfo {pages} {180401} (\bibinfo {year} {2005}{\natexlab{a}})}\BibitemShut {NoStop}%
\bibitem [{\citenamefont {Roy}\ \emph {et~al.}(2006)\citenamefont {Roy}, \citenamefont {Chattopadhyay}, \citenamefont {Chaddah}, \citenamefont {Moore}, \citenamefont {Perkins}, \citenamefont {Cohen}, \citenamefont {Gschneidner~Jr},\ and\ \citenamefont {Pecharsky}}]{roy2006evidence}%
  \BibitemOpen
  \bibfield  {author} {\bibinfo {author} {\bibfnamefont {S.}~\bibnamefont {Roy}}, \bibinfo {author} {\bibfnamefont {M.}~\bibnamefont {Chattopadhyay}}, \bibinfo {author} {\bibfnamefont {P.}~\bibnamefont {Chaddah}}, \bibinfo {author} {\bibfnamefont {J.}~\bibnamefont {Moore}}, \bibinfo {author} {\bibfnamefont {G.}~\bibnamefont {Perkins}}, \bibinfo {author} {\bibfnamefont {L.}~\bibnamefont {Cohen}}, \bibinfo {author} {\bibfnamefont {K.}~\bibnamefont {Gschneidner~Jr}},\ and\ \bibinfo {author} {\bibfnamefont {V.}~\bibnamefont {Pecharsky}},\ }\href@noop {} {\bibfield  {journal} {\bibinfo  {journal} {Physical Review B—Condensed Matter and Materials Physics}\ }\textbf {\bibinfo {volume} {74}},\ \bibinfo {pages} {012403} (\bibinfo {year} {2006})}\BibitemShut {NoStop}%
\bibitem [{\citenamefont {Kumar}\ \emph {et~al.}(2006)\citenamefont {Kumar}, \citenamefont {Pramanik}, \citenamefont {Banerjee}, \citenamefont {Chaddah}, \citenamefont {Roy}, \citenamefont {Park}, \citenamefont {Zhang},\ and\ \citenamefont {Cheong}}]{kumar2006relating}%
  \BibitemOpen
  \bibfield  {author} {\bibinfo {author} {\bibfnamefont {K.}~\bibnamefont {Kumar}}, \bibinfo {author} {\bibfnamefont {A.}~\bibnamefont {Pramanik}}, \bibinfo {author} {\bibfnamefont {A.}~\bibnamefont {Banerjee}}, \bibinfo {author} {\bibfnamefont {P.}~\bibnamefont {Chaddah}}, \bibinfo {author} {\bibfnamefont {S.}~\bibnamefont {Roy}}, \bibinfo {author} {\bibfnamefont {S.}~\bibnamefont {Park}}, \bibinfo {author} {\bibfnamefont {C.}~\bibnamefont {Zhang}},\ and\ \bibinfo {author} {\bibfnamefont {S.-W.}\ \bibnamefont {Cheong}},\ }\href@noop {} {\bibfield  {journal} {\bibinfo  {journal} {Physical Review B—Condensed Matter and Materials Physics}\ }\textbf {\bibinfo {volume} {73}},\ \bibinfo {pages} {184435} (\bibinfo {year} {2006})}\BibitemShut {NoStop}%
\bibitem [{\citenamefont {Imry}\ and\ \citenamefont {Wortis}(1979)}]{imry1979influence}%
  \BibitemOpen
  \bibfield  {author} {\bibinfo {author} {\bibfnamefont {Y.}~\bibnamefont {Imry}}\ and\ \bibinfo {author} {\bibfnamefont {M.}~\bibnamefont {Wortis}},\ }\href@noop {} {\bibfield  {journal} {\bibinfo  {journal} {Physical Review B}\ }\textbf {\bibinfo {volume} {19}},\ \bibinfo {pages} {3580} (\bibinfo {year} {1979})}\BibitemShut {NoStop}%
\bibitem [{\citenamefont {Soibel}\ \emph {et~al.}(2000)\citenamefont {Soibel}, \citenamefont {Zeldov}, \citenamefont {Rappaport}, \citenamefont {Myasoedov}, \citenamefont {Tamegai}, \citenamefont {Ooi}, \citenamefont {Konczykowski},\ and\ \citenamefont {Geshkenbein}}]{soibel2000imaging}%
  \BibitemOpen
  \bibfield  {author} {\bibinfo {author} {\bibfnamefont {A.}~\bibnamefont {Soibel}}, \bibinfo {author} {\bibfnamefont {E.}~\bibnamefont {Zeldov}}, \bibinfo {author} {\bibfnamefont {M.}~\bibnamefont {Rappaport}}, \bibinfo {author} {\bibfnamefont {Y.}~\bibnamefont {Myasoedov}}, \bibinfo {author} {\bibfnamefont {T.}~\bibnamefont {Tamegai}}, \bibinfo {author} {\bibfnamefont {S.}~\bibnamefont {Ooi}}, \bibinfo {author} {\bibfnamefont {M.}~\bibnamefont {Konczykowski}},\ and\ \bibinfo {author} {\bibfnamefont {V.~B.}\ \bibnamefont {Geshkenbein}},\ }\href@noop {} {\bibfield  {journal} {\bibinfo  {journal} {Nature}\ }\textbf {\bibinfo {volume} {406}},\ \bibinfo {pages} {282} (\bibinfo {year} {2000})}\BibitemShut {NoStop}%
\bibitem [{\citenamefont {Roy}\ \emph {et~al.}(2004)\citenamefont {Roy}, \citenamefont {Perkins}, \citenamefont {Chattopadhyay}, \citenamefont {Nigam}, \citenamefont {Sokhey}, \citenamefont {Chaddah}, \citenamefont {Caplin},\ and\ \citenamefont {Cohen}}]{roy2004first}%
  \BibitemOpen
  \bibfield  {author} {\bibinfo {author} {\bibfnamefont {S.}~\bibnamefont {Roy}}, \bibinfo {author} {\bibfnamefont {G.}~\bibnamefont {Perkins}}, \bibinfo {author} {\bibfnamefont {M.}~\bibnamefont {Chattopadhyay}}, \bibinfo {author} {\bibfnamefont {A.}~\bibnamefont {Nigam}}, \bibinfo {author} {\bibfnamefont {K.}~\bibnamefont {Sokhey}}, \bibinfo {author} {\bibfnamefont {P.}~\bibnamefont {Chaddah}}, \bibinfo {author} {\bibfnamefont {A.}~\bibnamefont {Caplin}},\ and\ \bibinfo {author} {\bibfnamefont {L.}~\bibnamefont {Cohen}},\ }\href@noop {} {\bibfield  {journal} {\bibinfo  {journal} {Physical review letters}\ }\textbf {\bibinfo {volume} {92}},\ \bibinfo {pages} {147203} (\bibinfo {year} {2004})}\BibitemShut {NoStop}%
\bibitem [{\citenamefont {Chattopadhyay}\ \emph {et~al.}(2004)\citenamefont {Chattopadhyay}, \citenamefont {Manekar}, \citenamefont {Pecharsky}, \citenamefont {Pecharsky}, \citenamefont {Gschneidner~Jr}, \citenamefont {Moore}, \citenamefont {Perkins}, \citenamefont {Bugoslavsky}, \citenamefont {Roy}, \citenamefont {Chaddah} \emph {et~al.}}]{chattopadhyay2004metastable}%
  \BibitemOpen
  \bibfield  {author} {\bibinfo {author} {\bibfnamefont {M.}~\bibnamefont {Chattopadhyay}}, \bibinfo {author} {\bibfnamefont {M.}~\bibnamefont {Manekar}}, \bibinfo {author} {\bibfnamefont {A.}~\bibnamefont {Pecharsky}}, \bibinfo {author} {\bibfnamefont {V.}~\bibnamefont {Pecharsky}}, \bibinfo {author} {\bibfnamefont {K.}~\bibnamefont {Gschneidner~Jr}}, \bibinfo {author} {\bibfnamefont {J.}~\bibnamefont {Moore}}, \bibinfo {author} {\bibfnamefont {G.}~\bibnamefont {Perkins}}, \bibinfo {author} {\bibfnamefont {Y.}~\bibnamefont {Bugoslavsky}}, \bibinfo {author} {\bibfnamefont {S.}~\bibnamefont {Roy}}, \bibinfo {author} {\bibfnamefont {P.}~\bibnamefont {Chaddah}}, \emph {et~al.},\ }\href@noop {} {\bibfield  {journal} {\bibinfo  {journal} {Physical Review B—Condensed Matter and Materials Physics}\ }\textbf {\bibinfo {volume} {70}},\ \bibinfo {pages} {214421} (\bibinfo {year} {2004})}\BibitemShut {NoStop}%
\bibitem [{\citenamefont {Chattopadhyay}\ \emph {et~al.}(2005{\natexlab{b}})\citenamefont {Chattopadhyay}, \citenamefont {Roy},\ and\ \citenamefont {Chaddah}}]{Chattopadhyay2005PRBMagneticGlass}%
  \BibitemOpen
  \bibfield  {author} {\bibinfo {author} {\bibfnamefont {M.~K.}\ \bibnamefont {Chattopadhyay}}, \bibinfo {author} {\bibfnamefont {S.~B.}\ \bibnamefont {Roy}},\ and\ \bibinfo {author} {\bibfnamefont {P.}~\bibnamefont {Chaddah}},\ }\href {https://doi.org/10.1103/PhysRevB.72.180401} {\bibfield  {journal} {\bibinfo  {journal} {Physical Review B}\ }\textbf {\bibinfo {volume} {72}},\ \bibinfo {pages} {180401(R)} (\bibinfo {year} {2005}{\natexlab{b}})}\BibitemShut {NoStop}%
\bibitem [{\citenamefont {Roy}(2013)}]{roy2013first}%
  \BibitemOpen
  \bibfield  {author} {\bibinfo {author} {\bibfnamefont {S.~B.}\ \bibnamefont {Roy}},\ }\href@noop {} {\bibfield  {journal} {\bibinfo  {journal} {Journal of Physics: Condensed Matter}\ }\textbf {\bibinfo {volume} {25}},\ \bibinfo {pages} {183201} (\bibinfo {year} {2013})}\BibitemShut {NoStop}%
\bibitem [{\citenamefont {Banerjee}\ \emph {et~al.}(2006{\natexlab{a}})\citenamefont {Banerjee}, \citenamefont {Pramanik}, \citenamefont {Kumar},\ and\ \citenamefont {Chaddah}}]{banerjee2006coexisting}%
  \BibitemOpen
  \bibfield  {author} {\bibinfo {author} {\bibfnamefont {A.}~\bibnamefont {Banerjee}}, \bibinfo {author} {\bibfnamefont {A.}~\bibnamefont {Pramanik}}, \bibinfo {author} {\bibfnamefont {K.}~\bibnamefont {Kumar}},\ and\ \bibinfo {author} {\bibfnamefont {P.}~\bibnamefont {Chaddah}},\ }\href@noop {} {\bibfield  {journal} {\bibinfo  {journal} {Journal of Physics: Condensed Matter}\ }\textbf {\bibinfo {volume} {18}},\ \bibinfo {pages} {L605} (\bibinfo {year} {2006}{\natexlab{a}})}\BibitemShut {NoStop}%
\bibitem [{\citenamefont {Banerjee}\ \emph {et~al.}(2009)\citenamefont {Banerjee}, \citenamefont {Kumar},\ and\ \citenamefont {Chaddah}}]{banerjee2009conversion}%
  \BibitemOpen
  \bibfield  {author} {\bibinfo {author} {\bibfnamefont {A.}~\bibnamefont {Banerjee}}, \bibinfo {author} {\bibfnamefont {K.}~\bibnamefont {Kumar}},\ and\ \bibinfo {author} {\bibfnamefont {P.}~\bibnamefont {Chaddah}},\ }\href@noop {} {\bibfield  {journal} {\bibinfo  {journal} {Journal of Physics: Condensed Matter}\ }\textbf {\bibinfo {volume} {21}},\ \bibinfo {pages} {026002} (\bibinfo {year} {2009})}\BibitemShut {NoStop}%
\bibitem [{\citenamefont {Saha}\ and\ \citenamefont {Rawat}(2018)}]{saha2018tunable}%
  \BibitemOpen
  \bibfield  {author} {\bibinfo {author} {\bibfnamefont {P.}~\bibnamefont {Saha}}\ and\ \bibinfo {author} {\bibfnamefont {R.}~\bibnamefont {Rawat}},\ }\href@noop {} {\bibfield  {journal} {\bibinfo  {journal} {Applied Physics Letters}\ }\textbf {\bibinfo {volume} {112}} (\bibinfo {year} {2018})}\BibitemShut {NoStop}%
\bibitem [{\citenamefont {Pecharsky}\ \emph {et~al.}(1993)\citenamefont {Pecharsky}, \citenamefont {Hyun},\ and\ \citenamefont {Gschneidner~Jr}}]{pecharsky1993unusual}%
  \BibitemOpen
  \bibfield  {author} {\bibinfo {author} {\bibfnamefont {V.}~\bibnamefont {Pecharsky}}, \bibinfo {author} {\bibfnamefont {O.-B.}\ \bibnamefont {Hyun}},\ and\ \bibinfo {author} {\bibfnamefont {K.}~\bibnamefont {Gschneidner~Jr}},\ }\href@noop {} {\bibfield  {journal} {\bibinfo  {journal} {Physical Review B}\ }\textbf {\bibinfo {volume} {47}},\ \bibinfo {pages} {11839} (\bibinfo {year} {1993})}\BibitemShut {NoStop}%
\bibitem [{\citenamefont {Thamizhavel}\ \emph {et~al.}(2005)\citenamefont {Thamizhavel}, \citenamefont {Takeuchi}, \citenamefont {D~Matsuda}, \citenamefont {Haga}, \citenamefont {Sugiyama}, \citenamefont {Settai},\ and\ \citenamefont {{\=O}nuki}}]{thamizhavel2005unique}%
  \BibitemOpen
  \bibfield  {author} {\bibinfo {author} {\bibfnamefont {A.}~\bibnamefont {Thamizhavel}}, \bibinfo {author} {\bibfnamefont {T.}~\bibnamefont {Takeuchi}}, \bibinfo {author} {\bibfnamefont {T.}~\bibnamefont {D~Matsuda}}, \bibinfo {author} {\bibfnamefont {Y.}~\bibnamefont {Haga}}, \bibinfo {author} {\bibfnamefont {K.}~\bibnamefont {Sugiyama}}, \bibinfo {author} {\bibfnamefont {R.}~\bibnamefont {Settai}},\ and\ \bibinfo {author} {\bibfnamefont {Y.}~\bibnamefont {{\=O}nuki}},\ }\href@noop {} {\bibfield  {journal} {\bibinfo  {journal} {Journal of the Physical Society of Japan}\ }\textbf {\bibinfo {volume} {74}},\ \bibinfo {pages} {1858} (\bibinfo {year} {2005})}\BibitemShut {NoStop}%
\bibitem [{\citenamefont {Dominguez~Montero}\ \emph {et~al.}(2024)\citenamefont {Dominguez~Montero}, \citenamefont {Stoian}, \citenamefont {McCandless}, \citenamefont {Baumbach},\ and\ \citenamefont {Chan}}]{dominguez2024role}%
  \BibitemOpen
  \bibfield  {author} {\bibinfo {author} {\bibfnamefont {A.}~\bibnamefont {Dominguez~Montero}}, \bibinfo {author} {\bibfnamefont {S.~A.}\ \bibnamefont {Stoian}}, \bibinfo {author} {\bibfnamefont {G.~T.}\ \bibnamefont {McCandless}}, \bibinfo {author} {\bibfnamefont {R.~E.}\ \bibnamefont {Baumbach}},\ and\ \bibinfo {author} {\bibfnamefont {J.~Y.}\ \bibnamefont {Chan}},\ }\href@noop {} {\bibfield  {journal} {\bibinfo  {journal} {Chemistry of Materials}\ }\textbf {\bibinfo {volume} {36}},\ \bibinfo {pages} {8534} (\bibinfo {year} {2024})}\BibitemShut {NoStop}%
\bibitem [{\citenamefont {Banerjee}\ \emph {et~al.}(2006{\natexlab{b}})\citenamefont {Banerjee}, \citenamefont {Mukherjee}, \citenamefont {Kumar},\ and\ \citenamefont {Chaddah}}]{banerjee2006ferromagnetic}%
  \BibitemOpen
  \bibfield  {author} {\bibinfo {author} {\bibfnamefont {A.}~\bibnamefont {Banerjee}}, \bibinfo {author} {\bibfnamefont {K.}~\bibnamefont {Mukherjee}}, \bibinfo {author} {\bibfnamefont {K.}~\bibnamefont {Kumar}},\ and\ \bibinfo {author} {\bibfnamefont {P.}~\bibnamefont {Chaddah}},\ }\href@noop {} {\bibfield  {journal} {\bibinfo  {journal} {Physical Review B—Condensed Matter and Materials Physics}\ }\textbf {\bibinfo {volume} {74}},\ \bibinfo {pages} {224445} (\bibinfo {year} {2006}{\natexlab{b}})}\BibitemShut {NoStop}%
\bibitem [{\citenamefont {Rawat}\ \emph {et~al.}(2007)\citenamefont {Rawat}, \citenamefont {Mukherjee}, \citenamefont {Kumar}, \citenamefont {Banerjee},\ and\ \citenamefont {Chaddah}}]{rawat2007anomalous}%
  \BibitemOpen
  \bibfield  {author} {\bibinfo {author} {\bibfnamefont {R.}~\bibnamefont {Rawat}}, \bibinfo {author} {\bibfnamefont {K.}~\bibnamefont {Mukherjee}}, \bibinfo {author} {\bibfnamefont {K.}~\bibnamefont {Kumar}}, \bibinfo {author} {\bibfnamefont {A.}~\bibnamefont {Banerjee}},\ and\ \bibinfo {author} {\bibfnamefont {P.}~\bibnamefont {Chaddah}},\ }\href@noop {} {\bibfield  {journal} {\bibinfo  {journal} {Journal of Physics: Condensed Matter}\ }\textbf {\bibinfo {volume} {19}},\ \bibinfo {pages} {256211} (\bibinfo {year} {2007})}\BibitemShut {NoStop}%
\bibitem [{\citenamefont {Sengupta}\ and\ \citenamefont {Sampathkumaran}(2006)}]{sengupta2006field}%
  \BibitemOpen
  \bibfield  {author} {\bibinfo {author} {\bibfnamefont {K.}~\bibnamefont {Sengupta}}\ and\ \bibinfo {author} {\bibfnamefont {E.}~\bibnamefont {Sampathkumaran}},\ }\href@noop {} {\bibfield  {journal} {\bibinfo  {journal} {Physical Review B—Condensed Matter and Materials Physics}\ }\textbf {\bibinfo {volume} {73}},\ \bibinfo {pages} {020406} (\bibinfo {year} {2006})}\BibitemShut {NoStop}%
\bibitem [{\citenamefont {Singh}\ \emph {et~al.}(2002)\citenamefont {Singh}, \citenamefont {Chaudhary}, \citenamefont {Chattopadhyay}, \citenamefont {Manekar}, \citenamefont {Roy},\ and\ \citenamefont {Chaddah}}]{singh2002first}%
  \BibitemOpen
  \bibfield  {author} {\bibinfo {author} {\bibfnamefont {K.~J.}\ \bibnamefont {Singh}}, \bibinfo {author} {\bibfnamefont {S.}~\bibnamefont {Chaudhary}}, \bibinfo {author} {\bibfnamefont {M.}~\bibnamefont {Chattopadhyay}}, \bibinfo {author} {\bibfnamefont {M.}~\bibnamefont {Manekar}}, \bibinfo {author} {\bibfnamefont {S.}~\bibnamefont {Roy}},\ and\ \bibinfo {author} {\bibfnamefont {P.}~\bibnamefont {Chaddah}},\ }\href@noop {} {\bibfield  {journal} {\bibinfo  {journal} {Physical Review B}\ }\textbf {\bibinfo {volume} {65}},\ \bibinfo {pages} {094419} (\bibinfo {year} {2002})}\BibitemShut {NoStop}%
\bibitem [{\citenamefont {Pal}\ \emph {et~al.}(2021)\citenamefont {Pal}, \citenamefont {Kumar},\ and\ \citenamefont {Banerjee}}]{pal2021memorylike}%
  \BibitemOpen
  \bibfield  {author} {\bibinfo {author} {\bibfnamefont {S.}~\bibnamefont {Pal}}, \bibinfo {author} {\bibfnamefont {K.}~\bibnamefont {Kumar}},\ and\ \bibinfo {author} {\bibfnamefont {A.}~\bibnamefont {Banerjee}},\ }\href@noop {} {\bibfield  {journal} {\bibinfo  {journal} {Physical Review B}\ }\textbf {\bibinfo {volume} {103}},\ \bibinfo {pages} {144434} (\bibinfo {year} {2021})}\BibitemShut {NoStop}%
\bibitem [{\citenamefont {Wadley}\ \emph {et~al.}(2016{\natexlab{a}})\citenamefont {Wadley}, \citenamefont {Howells}, \citenamefont {{\v{Z}}elezn{\`y}}, \citenamefont {Andrews}, \citenamefont {Hills}, \citenamefont {Campion}, \citenamefont {Nov{\'a}k}, \citenamefont {Olejn{\'\i}k}, \citenamefont {Maccherozzi}, \citenamefont {Dhesi} \emph {et~al.}}]{wadley2016electrical}%
  \BibitemOpen
  \bibfield  {author} {\bibinfo {author} {\bibfnamefont {P.}~\bibnamefont {Wadley}}, \bibinfo {author} {\bibfnamefont {B.}~\bibnamefont {Howells}}, \bibinfo {author} {\bibfnamefont {J.}~\bibnamefont {{\v{Z}}elezn{\`y}}}, \bibinfo {author} {\bibfnamefont {C.}~\bibnamefont {Andrews}}, \bibinfo {author} {\bibfnamefont {V.}~\bibnamefont {Hills}}, \bibinfo {author} {\bibfnamefont {R.~P.}\ \bibnamefont {Campion}}, \bibinfo {author} {\bibfnamefont {V.}~\bibnamefont {Nov{\'a}k}}, \bibinfo {author} {\bibfnamefont {K.}~\bibnamefont {Olejn{\'\i}k}}, \bibinfo {author} {\bibfnamefont {F.}~\bibnamefont {Maccherozzi}}, \bibinfo {author} {\bibfnamefont {S.}~\bibnamefont {Dhesi}}, \emph {et~al.},\ }\href@noop {} {\bibfield  {journal} {\bibinfo  {journal} {Science}\ }\textbf {\bibinfo {volume} {351}},\ \bibinfo {pages} {587} (\bibinfo {year} {2016}{\natexlab{a}})}\BibitemShut {NoStop}%
\bibitem [{\citenamefont {Bodnar}\ \emph {et~al.}(2018)\citenamefont {Bodnar}, \citenamefont {{\v{S}}mejkal}, \citenamefont {Turek}, \citenamefont {Jungwirth}, \citenamefont {Gomonay}, \citenamefont {Sinova}, \citenamefont {Sapozhnik}, \citenamefont {Elmers}, \citenamefont {Kl{\"a}ui},\ and\ \citenamefont {Jourdan}}]{bodnar2018writing}%
  \BibitemOpen
  \bibfield  {author} {\bibinfo {author} {\bibfnamefont {S.~Y.}\ \bibnamefont {Bodnar}}, \bibinfo {author} {\bibfnamefont {L.}~\bibnamefont {{\v{S}}mejkal}}, \bibinfo {author} {\bibfnamefont {I.}~\bibnamefont {Turek}}, \bibinfo {author} {\bibfnamefont {T.}~\bibnamefont {Jungwirth}}, \bibinfo {author} {\bibfnamefont {O.}~\bibnamefont {Gomonay}}, \bibinfo {author} {\bibfnamefont {J.}~\bibnamefont {Sinova}}, \bibinfo {author} {\bibfnamefont {A.}~\bibnamefont {Sapozhnik}}, \bibinfo {author} {\bibfnamefont {H.-J.}\ \bibnamefont {Elmers}}, \bibinfo {author} {\bibfnamefont {M.}~\bibnamefont {Kl{\"a}ui}},\ and\ \bibinfo {author} {\bibfnamefont {M.}~\bibnamefont {Jourdan}},\ }\href@noop {} {\bibfield  {journal} {\bibinfo  {journal} {Nature communications}\ }\textbf {\bibinfo {volume} {9}},\ \bibinfo {pages} {348} (\bibinfo {year} {2018})}\BibitemShut {NoStop}%
\bibitem [{\citenamefont {\ifmmode~\check{Z}\else \v{Z}\fi{}elezn\'y}\ \emph {et~al.}(2014)\citenamefont {\ifmmode~\check{Z}\else \v{Z}\fi{}elezn\'y}, \citenamefont {Gao}, \citenamefont {V\'yborn\'y}, \citenamefont {Zemen}, \citenamefont {Ma\ifmmode~\check{s}\else \v{s}\fi{}ek}, \citenamefont {Manchon}, \citenamefont {Wunderlich}, \citenamefont {Sinova},\ and\ \citenamefont {Jungwirth}}]{Zelezny2014}%
  \BibitemOpen
  \bibfield  {author} {\bibinfo {author} {\bibfnamefont {J.}~\bibnamefont {\ifmmode~\check{Z}\else \v{Z}\fi{}elezn\'y}}, \bibinfo {author} {\bibfnamefont {H.}~\bibnamefont {Gao}}, \bibinfo {author} {\bibfnamefont {K.}~\bibnamefont {V\'yborn\'y}}, \bibinfo {author} {\bibfnamefont {J.}~\bibnamefont {Zemen}}, \bibinfo {author} {\bibfnamefont {J.}~\bibnamefont {Ma\ifmmode~\check{s}\else \v{s}\fi{}ek}}, \bibinfo {author} {\bibfnamefont {A.}~\bibnamefont {Manchon}}, \bibinfo {author} {\bibfnamefont {J.}~\bibnamefont {Wunderlich}}, \bibinfo {author} {\bibfnamefont {J.}~\bibnamefont {Sinova}},\ and\ \bibinfo {author} {\bibfnamefont {T.}~\bibnamefont {Jungwirth}},\ }\href {https://doi.org/10.1103/PhysRevLett.113.157201} {\bibfield  {journal} {\bibinfo  {journal} {Phys. Rev. Lett.}\ }\textbf {\bibinfo {volume} {113}},\ \bibinfo {pages} {157201} (\bibinfo {year} {2014})}\BibitemShut {NoStop}%
\bibitem [{\citenamefont {Dzyaloshinsky}(1958)}]{dzyaloshinsky1958thermodynamic}%
  \BibitemOpen
  \bibfield  {author} {\bibinfo {author} {\bibfnamefont {I.}~\bibnamefont {Dzyaloshinsky}},\ }\href@noop {} {\bibfield  {journal} {\bibinfo  {journal} {Journal of Physics and Chemistry of Solids}\ }\textbf {\bibinfo {volume} {4}},\ \bibinfo {pages} {241} (\bibinfo {year} {1958})}\BibitemShut {NoStop}%
\bibitem [{\citenamefont {Moriya}(1960)}]{moriya1960new}%
  \BibitemOpen
  \bibfield  {author} {\bibinfo {author} {\bibfnamefont {T.}~\bibnamefont {Moriya}},\ }\href@noop {} {\bibfield  {journal} {\bibinfo  {journal} {Physical Review Letters}\ }\textbf {\bibinfo {volume} {4}},\ \bibinfo {pages} {228} (\bibinfo {year} {1960})}\BibitemShut {NoStop}%
\bibitem [{\citenamefont {Fert}\ \emph {et~al.}(2013)\citenamefont {Fert}, \citenamefont {Cros},\ and\ \citenamefont {Sampaio}}]{fert2013skyrmions}%
  \BibitemOpen
  \bibfield  {author} {\bibinfo {author} {\bibfnamefont {A.}~\bibnamefont {Fert}}, \bibinfo {author} {\bibfnamefont {V.}~\bibnamefont {Cros}},\ and\ \bibinfo {author} {\bibfnamefont {J.}~\bibnamefont {Sampaio}},\ }\href@noop {} {\bibfield  {journal} {\bibinfo  {journal} {Nature nanotechnology}\ }\textbf {\bibinfo {volume} {8}},\ \bibinfo {pages} {152} (\bibinfo {year} {2013})}\BibitemShut {NoStop}%
\bibitem [{\citenamefont {Tokura}\ and\ \citenamefont {Seki}(2010)}]{tokura2010multiferroics}%
  \BibitemOpen
  \bibfield  {author} {\bibinfo {author} {\bibfnamefont {Y.}~\bibnamefont {Tokura}}\ and\ \bibinfo {author} {\bibfnamefont {S.}~\bibnamefont {Seki}},\ }\href@noop {} {\bibfield  {journal} {\bibinfo  {journal} {Advanced materials}\ }\textbf {\bibinfo {volume} {22}},\ \bibinfo {pages} {1554} (\bibinfo {year} {2010})}\BibitemShut {NoStop}%
\bibitem [{\citenamefont {Parkin}\ and\ \citenamefont {Yang}(2015)}]{parkin2015memory}%
  \BibitemOpen
  \bibfield  {author} {\bibinfo {author} {\bibfnamefont {S.}~\bibnamefont {Parkin}}\ and\ \bibinfo {author} {\bibfnamefont {S.-H.}\ \bibnamefont {Yang}},\ }\href@noop {} {\bibfield  {journal} {\bibinfo  {journal} {Nature nanotechnology}\ }\textbf {\bibinfo {volume} {10}},\ \bibinfo {pages} {195} (\bibinfo {year} {2015})}\BibitemShut {NoStop}%
\bibitem [{\citenamefont {Nagaosa}\ and\ \citenamefont {Tokura}(2013)}]{nagaosa2013topological}%
  \BibitemOpen
  \bibfield  {author} {\bibinfo {author} {\bibfnamefont {N.}~\bibnamefont {Nagaosa}}\ and\ \bibinfo {author} {\bibfnamefont {Y.}~\bibnamefont {Tokura}},\ }\href@noop {} {\bibfield  {journal} {\bibinfo  {journal} {Nature nanotechnology}\ }\textbf {\bibinfo {volume} {8}},\ \bibinfo {pages} {899} (\bibinfo {year} {2013})}\BibitemShut {NoStop}%
\bibitem [{\citenamefont {Smidman}\ \emph {et~al.}(2013)\citenamefont {Smidman}, \citenamefont {Adroja}, \citenamefont {Hillier}, \citenamefont {Chapon}, \citenamefont {Taylor}, \citenamefont {Anand}, \citenamefont {Singh}, \citenamefont {Lees}, \citenamefont {Goremychkin}, \citenamefont {Koza} \emph {et~al.}}]{smidman2013neutron}%
  \BibitemOpen
  \bibfield  {author} {\bibinfo {author} {\bibfnamefont {M.}~\bibnamefont {Smidman}}, \bibinfo {author} {\bibfnamefont {D.}~\bibnamefont {Adroja}}, \bibinfo {author} {\bibfnamefont {A.~D.}\ \bibnamefont {Hillier}}, \bibinfo {author} {\bibfnamefont {L.}~\bibnamefont {Chapon}}, \bibinfo {author} {\bibfnamefont {J.}~\bibnamefont {Taylor}}, \bibinfo {author} {\bibfnamefont {V.}~\bibnamefont {Anand}}, \bibinfo {author} {\bibfnamefont {R.~P.}\ \bibnamefont {Singh}}, \bibinfo {author} {\bibfnamefont {M.~R.}\ \bibnamefont {Lees}}, \bibinfo {author} {\bibfnamefont {E.}~\bibnamefont {Goremychkin}}, \bibinfo {author} {\bibfnamefont {M.}~\bibnamefont {Koza}}, \emph {et~al.},\ }\href@noop {} {\bibfield  {journal} {\bibinfo  {journal} {Physical Review B—Condensed Matter and Materials Physics}\ }\textbf {\bibinfo {volume} {88}},\ \bibinfo {pages} {134416} (\bibinfo {year} {2013})}\BibitemShut {NoStop}%
\bibitem [{\citenamefont {Li}\ \emph {et~al.}(2023)\citenamefont {Li}, \citenamefont {Ye}, \citenamefont {Hu}, \citenamefont {Fang}, \citenamefont {Xiao}, \citenamefont {Wu}, \citenamefont {Shan}, \citenamefont {Singh}, \citenamefont {Balakrishnan}, \citenamefont {Shen} \emph {et~al.}}]{li2023photoemission}%
  \BibitemOpen
  \bibfield  {author} {\bibinfo {author} {\bibfnamefont {P.}~\bibnamefont {Li}}, \bibinfo {author} {\bibfnamefont {H.}~\bibnamefont {Ye}}, \bibinfo {author} {\bibfnamefont {Y.}~\bibnamefont {Hu}}, \bibinfo {author} {\bibfnamefont {Y.}~\bibnamefont {Fang}}, \bibinfo {author} {\bibfnamefont {Z.}~\bibnamefont {Xiao}}, \bibinfo {author} {\bibfnamefont {Z.}~\bibnamefont {Wu}}, \bibinfo {author} {\bibfnamefont {Z.}~\bibnamefont {Shan}}, \bibinfo {author} {\bibfnamefont {R.~P.}\ \bibnamefont {Singh}}, \bibinfo {author} {\bibfnamefont {G.}~\bibnamefont {Balakrishnan}}, \bibinfo {author} {\bibfnamefont {D.}~\bibnamefont {Shen}}, \emph {et~al.},\ }\href@noop {} {\bibfield  {journal} {\bibinfo  {journal} {Physical Review B}\ }\textbf {\bibinfo {volume} {107}},\ \bibinfo {pages} {L201104} (\bibinfo {year} {2023})}\BibitemShut {NoStop}%
\bibitem [{\citenamefont {Allen}\ \emph {et~al.}(2026)\citenamefont {Allen}, \citenamefont {Bouaziz}, \citenamefont {Zhang}, \citenamefont {Du}, \citenamefont {Mishra}, \citenamefont {Bihlmayer}, \citenamefont {Hao}, \citenamefont {Ukleev}, \citenamefont {Luo}, \citenamefont {Radu} \emph {et~al.}}]{allen2026atomically}%
  \BibitemOpen
  \bibfield  {author} {\bibinfo {author} {\bibfnamefont {K.}~\bibnamefont {Allen}}, \bibinfo {author} {\bibfnamefont {J.}~\bibnamefont {Bouaziz}}, \bibinfo {author} {\bibfnamefont {Y.}~\bibnamefont {Zhang}}, \bibinfo {author} {\bibfnamefont {K.}~\bibnamefont {Du}}, \bibinfo {author} {\bibfnamefont {S.}~\bibnamefont {Mishra}}, \bibinfo {author} {\bibfnamefont {G.}~\bibnamefont {Bihlmayer}}, \bibinfo {author} {\bibfnamefont {Y.}~\bibnamefont {Hao}}, \bibinfo {author} {\bibfnamefont {V.}~\bibnamefont {Ukleev}}, \bibinfo {author} {\bibfnamefont {C.}~\bibnamefont {Luo}}, \bibinfo {author} {\bibfnamefont {F.}~\bibnamefont {Radu}}, \emph {et~al.},\ }\href@noop {} {\bibfield  {journal} {\bibinfo  {journal} {arXiv preprint arXiv:2602.10281}\ } (\bibinfo {year} {2026})}\BibitemShut {NoStop}%
\bibitem [{\citenamefont {Mydosh}(1978)}]{mydosh1978spin}%
  \BibitemOpen
  \bibfield  {author} {\bibinfo {author} {\bibfnamefont {J.}~\bibnamefont {Mydosh}},\ }\href@noop {} {\bibfield  {journal} {\bibinfo  {journal} {Journal of Magnetism and Magnetic Materials}\ }\textbf {\bibinfo {volume} {7}},\ \bibinfo {pages} {237} (\bibinfo {year} {1978})}\BibitemShut {NoStop}%
\bibitem [{\citenamefont {Shull}\ and\ \citenamefont {Beck}(1975)}]{shull1975mictomagnetic}%
  \BibitemOpen
  \bibfield  {author} {\bibinfo {author} {\bibfnamefont {R.}~\bibnamefont {Shull}}\ and\ \bibinfo {author} {\bibfnamefont {P.~A.}\ \bibnamefont {Beck}},\ }in\ \href@noop {} {\emph {\bibinfo {booktitle} {AIP Conference Proceedings}}},\ Vol.~\bibinfo {volume} {24}\ (\bibinfo {organization} {American Institute of Physics},\ \bibinfo {year} {1975})\ pp.\ \bibinfo {pages} {95--96}\BibitemShut {NoStop}%
\bibitem [{\citenamefont {Shull}\ \emph {et~al.}(1976)\citenamefont {Shull}, \citenamefont {Okamoto},\ and\ \citenamefont {Beck}}]{shull1976transition}%
  \BibitemOpen
  \bibfield  {author} {\bibinfo {author} {\bibfnamefont {R.}~\bibnamefont {Shull}}, \bibinfo {author} {\bibfnamefont {H.}~\bibnamefont {Okamoto}},\ and\ \bibinfo {author} {\bibfnamefont {P.}~\bibnamefont {Beck}},\ }\href@noop {} {\bibfield  {journal} {\bibinfo  {journal} {Solid State Communications}\ }\textbf {\bibinfo {volume} {20}},\ \bibinfo {pages} {863} (\bibinfo {year} {1976})}\BibitemShut {NoStop}%
\bibitem [{\citenamefont {Benka}\ \emph {et~al.}(2022)\citenamefont {Benka}, \citenamefont {Bauer}, \citenamefont {Schmakat}, \citenamefont {S{\"a}ubert}, \citenamefont {Seifert}, \citenamefont {Jorba},\ and\ \citenamefont {Pfleiderer}}]{benka2022interplay}%
  \BibitemOpen
  \bibfield  {author} {\bibinfo {author} {\bibfnamefont {G.}~\bibnamefont {Benka}}, \bibinfo {author} {\bibfnamefont {A.}~\bibnamefont {Bauer}}, \bibinfo {author} {\bibfnamefont {P.}~\bibnamefont {Schmakat}}, \bibinfo {author} {\bibfnamefont {S.}~\bibnamefont {S{\"a}ubert}}, \bibinfo {author} {\bibfnamefont {M.}~\bibnamefont {Seifert}}, \bibinfo {author} {\bibfnamefont {P.}~\bibnamefont {Jorba}},\ and\ \bibinfo {author} {\bibfnamefont {C.}~\bibnamefont {Pfleiderer}},\ }\href@noop {} {\bibfield  {journal} {\bibinfo  {journal} {Physical Review Materials}\ }\textbf {\bibinfo {volume} {6}},\ \bibinfo {pages} {044407} (\bibinfo {year} {2022})}\BibitemShut {NoStop}%
\bibitem [{\citenamefont {Chattopadhyay}\ and\ \citenamefont {Roy}(2008)}]{chattopadhyay2008metamagnetic}%
  \BibitemOpen
  \bibfield  {author} {\bibinfo {author} {\bibfnamefont {M.}~\bibnamefont {Chattopadhyay}}\ and\ \bibinfo {author} {\bibfnamefont {S.}~\bibnamefont {Roy}},\ }\href@noop {} {\bibfield  {journal} {\bibinfo  {journal} {Journal of Physics: Condensed Matter}\ }\textbf {\bibinfo {volume} {20}},\ \bibinfo {pages} {025209} (\bibinfo {year} {2008})}\BibitemShut {NoStop}%
\bibitem [{\citenamefont {Bag}\ \emph {et~al.}(2018)\citenamefont {Bag}, \citenamefont {Baral},\ and\ \citenamefont {Nath}}]{bag2018cluster}%
  \BibitemOpen
  \bibfield  {author} {\bibinfo {author} {\bibfnamefont {P.}~\bibnamefont {Bag}}, \bibinfo {author} {\bibfnamefont {P.}~\bibnamefont {Baral}},\ and\ \bibinfo {author} {\bibfnamefont {R.}~\bibnamefont {Nath}},\ }\href@noop {} {\bibfield  {journal} {\bibinfo  {journal} {arXiv preprint arXiv:1810.03890}\ } (\bibinfo {year} {2018})}\BibitemShut {NoStop}%
\bibitem [{\citenamefont {Vincent}(2022)}]{vincent2022spin}%
  \BibitemOpen
  \bibfield  {author} {\bibinfo {author} {\bibfnamefont {E.}~\bibnamefont {Vincent}},\ }\href@noop {} {\bibfield  {journal} {\bibinfo  {journal} {arXiv preprint arXiv:2208.00981}\ } (\bibinfo {year} {2022})}\BibitemShut {NoStop}%
\bibitem [{\citenamefont {Montgomery}(1971)}]{montgomery1971method}%
  \BibitemOpen
  \bibfield  {author} {\bibinfo {author} {\bibfnamefont {H.}~\bibnamefont {Montgomery}},\ }\href@noop {} {\bibfield  {journal} {\bibinfo  {journal} {Journal of applied physics}\ }\textbf {\bibinfo {volume} {42}},\ \bibinfo {pages} {2971} (\bibinfo {year} {1971})}\BibitemShut {NoStop}%
\bibitem [{\citenamefont {Dos~Santos}\ \emph {et~al.}(2011)\citenamefont {Dos~Santos}, \citenamefont {De~Campos}, \citenamefont {Da~Luz}, \citenamefont {White}, \citenamefont {Neumeier}, \citenamefont {De~Lima},\ and\ \citenamefont {Shigue}}]{dos2011procedure}%
  \BibitemOpen
  \bibfield  {author} {\bibinfo {author} {\bibfnamefont {C.}~\bibnamefont {Dos~Santos}}, \bibinfo {author} {\bibfnamefont {A.}~\bibnamefont {De~Campos}}, \bibinfo {author} {\bibfnamefont {M.}~\bibnamefont {Da~Luz}}, \bibinfo {author} {\bibfnamefont {B.}~\bibnamefont {White}}, \bibinfo {author} {\bibfnamefont {J.}~\bibnamefont {Neumeier}}, \bibinfo {author} {\bibfnamefont {B.}~\bibnamefont {De~Lima}},\ and\ \bibinfo {author} {\bibfnamefont {C.}~\bibnamefont {Shigue}},\ }\href@noop {} {\bibfield  {journal} {\bibinfo  {journal} {Journal of Applied Physics}\ }\textbf {\bibinfo {volume} {110}} (\bibinfo {year} {2011})}\BibitemShut {NoStop}%
\bibitem [{\citenamefont {Meiklejohn}\ and\ \citenamefont {Bean}(1957)}]{meiklejohn1957new}%
  \BibitemOpen
  \bibfield  {author} {\bibinfo {author} {\bibfnamefont {W.~H.}\ \bibnamefont {Meiklejohn}}\ and\ \bibinfo {author} {\bibfnamefont {C.~P.}\ \bibnamefont {Bean}},\ }\href@noop {} {\bibfield  {journal} {\bibinfo  {journal} {Physical Review}\ }\textbf {\bibinfo {volume} {105}},\ \bibinfo {pages} {904} (\bibinfo {year} {1957})}\BibitemShut {NoStop}%
\bibitem [{\citenamefont {Roy}\ \emph {et~al.}(2008)\citenamefont {Roy}, \citenamefont {Chaddah}, \citenamefont {Pecharsky},\ and\ \citenamefont {Gschneidner}}]{Roy2008ActaMaterMetamagnetic}%
  \BibitemOpen
  \bibfield  {author} {\bibinfo {author} {\bibfnamefont {S.~B.}\ \bibnamefont {Roy}}, \bibinfo {author} {\bibfnamefont {P.}~\bibnamefont {Chaddah}}, \bibinfo {author} {\bibfnamefont {V.~K.}\ \bibnamefont {Pecharsky}},\ and\ \bibinfo {author} {\bibfnamefont {K.~A.~J.}\ \bibnamefont {Gschneidner}},\ }\href {https://doi.org/10.1016/j.actamat.2008.08.040} {\bibfield  {journal} {\bibinfo  {journal} {Acta Materialia}\ }\textbf {\bibinfo {volume} {56}},\ \bibinfo {pages} {5895} (\bibinfo {year} {2008})}\BibitemShut {NoStop}%
\bibitem [{\citenamefont {Sharma}\ \emph {et~al.}(2007)\citenamefont {Sharma}, \citenamefont {Chattopadhyay},\ and\ \citenamefont {Roy}}]{Sharma2007PRBNiMnIn}%
  \BibitemOpen
  \bibfield  {author} {\bibinfo {author} {\bibfnamefont {V.~K.}\ \bibnamefont {Sharma}}, \bibinfo {author} {\bibfnamefont {M.~K.}\ \bibnamefont {Chattopadhyay}},\ and\ \bibinfo {author} {\bibfnamefont {S.~B.}\ \bibnamefont {Roy}},\ }\href {https://doi.org/10.1103/PhysRevB.76.140401} {\bibfield  {journal} {\bibinfo  {journal} {Physical Review B}\ }\textbf {\bibinfo {volume} {76}},\ \bibinfo {pages} {140401(R)} (\bibinfo {year} {2007})}\BibitemShut {NoStop}%
\bibitem [{\citenamefont {Tokura}(2006)}]{Tokura2006RPP}%
  \BibitemOpen
  \bibfield  {author} {\bibinfo {author} {\bibfnamefont {Y.}~\bibnamefont {Tokura}},\ }\href {https://doi.org/10.1088/0034-4885/69/3/R06} {\bibfield  {journal} {\bibinfo  {journal} {Reports on Progress in Physics}\ }\textbf {\bibinfo {volume} {69}},\ \bibinfo {pages} {797} (\bibinfo {year} {2006})}\BibitemShut {NoStop}%
\bibitem [{\citenamefont {Dagotto}\ \emph {et~al.}(2001)\citenamefont {Dagotto}, \citenamefont {Hotta},\ and\ \citenamefont {Moreo}}]{Dagotto2001PhysRep}%
  \BibitemOpen
  \bibfield  {author} {\bibinfo {author} {\bibfnamefont {E.}~\bibnamefont {Dagotto}}, \bibinfo {author} {\bibfnamefont {T.}~\bibnamefont {Hotta}},\ and\ \bibinfo {author} {\bibfnamefont {A.}~\bibnamefont {Moreo}},\ }\href {https://doi.org/10.1016/S0370-1573(00)00121-6} {\bibfield  {journal} {\bibinfo  {journal} {Physics Reports}\ }\textbf {\bibinfo {volume} {344}},\ \bibinfo {pages} {1} (\bibinfo {year} {2001})}\BibitemShut {NoStop}%
\bibitem [{\citenamefont {Gayathri}\ \emph {et~al.}(2023)\citenamefont {Gayathri}, \citenamefont {Amaladass}, \citenamefont {Sathyanarayana}, \citenamefont {Geetha~Kumary}, \citenamefont {Pandian}, \citenamefont {Gupta}, \citenamefont {Rai},\ and\ \citenamefont {Mani}}]{gayathri2023interfacial}%
  \BibitemOpen
  \bibfield  {author} {\bibinfo {author} {\bibfnamefont {V.}~\bibnamefont {Gayathri}}, \bibinfo {author} {\bibfnamefont {E.}~\bibnamefont {Amaladass}}, \bibinfo {author} {\bibfnamefont {A.}~\bibnamefont {Sathyanarayana}}, \bibinfo {author} {\bibfnamefont {T.}~\bibnamefont {Geetha~Kumary}}, \bibinfo {author} {\bibfnamefont {R.}~\bibnamefont {Pandian}}, \bibinfo {author} {\bibfnamefont {P.}~\bibnamefont {Gupta}}, \bibinfo {author} {\bibfnamefont {S.~K.}\ \bibnamefont {Rai}},\ and\ \bibinfo {author} {\bibfnamefont {A.}~\bibnamefont {Mani}},\ }\href@noop {} {\bibfield  {journal} {\bibinfo  {journal} {Scientific Reports}\ }\textbf {\bibinfo {volume} {13}},\ \bibinfo {pages} {2315} (\bibinfo {year} {2023})}\BibitemShut {NoStop}%
\bibitem [{\citenamefont {Manchon}\ \emph {et~al.}(2019)\citenamefont {Manchon}, \citenamefont {\v{Z}elezn\'y}, \citenamefont {Miron}, \citenamefont {Jungwirth}, \citenamefont {Sinova}, \citenamefont {Thiaville}, \citenamefont {Garello},\ and\ \citenamefont {Gambardella}}]{Manchon2019RMP}%
  \BibitemOpen
  \bibfield  {author} {\bibinfo {author} {\bibfnamefont {A.}~\bibnamefont {Manchon}}, \bibinfo {author} {\bibfnamefont {J.}~\bibnamefont {\v{Z}elezn\'y}}, \bibinfo {author} {\bibfnamefont {I.~M.}\ \bibnamefont {Miron}}, \bibinfo {author} {\bibfnamefont {T.}~\bibnamefont {Jungwirth}}, \bibinfo {author} {\bibfnamefont {J.}~\bibnamefont {Sinova}}, \bibinfo {author} {\bibfnamefont {A.}~\bibnamefont {Thiaville}}, \bibinfo {author} {\bibfnamefont {K.}~\bibnamefont {Garello}},\ and\ \bibinfo {author} {\bibfnamefont {P.}~\bibnamefont {Gambardella}},\ }\href {https://doi.org/10.1103/RevModPhys.91.035004} {\bibfield  {journal} {\bibinfo  {journal} {Reviews of Modern Physics}\ }\textbf {\bibinfo {volume} {91}},\ \bibinfo {pages} {035004} (\bibinfo {year} {2019})}\BibitemShut {NoStop}%
\bibitem [{\citenamefont {\v{Z}elezn\'y}\ \emph {et~al.}(2017)\citenamefont {\v{Z}elezn\'y}, \citenamefont {Gao}, \citenamefont {Manchon}, \citenamefont {Freimuth}, \citenamefont {Mokrousov}, \citenamefont {Zemen}, \citenamefont {Ma\v{s}ek}, \citenamefont {Sinova},\ and\ \citenamefont {Jungwirth}}]{Zelezny2017PRB}%
  \BibitemOpen
  \bibfield  {author} {\bibinfo {author} {\bibfnamefont {J.}~\bibnamefont {\v{Z}elezn\'y}}, \bibinfo {author} {\bibfnamefont {H.}~\bibnamefont {Gao}}, \bibinfo {author} {\bibfnamefont {A.}~\bibnamefont {Manchon}}, \bibinfo {author} {\bibfnamefont {F.}~\bibnamefont {Freimuth}}, \bibinfo {author} {\bibfnamefont {Y.}~\bibnamefont {Mokrousov}}, \bibinfo {author} {\bibfnamefont {J.}~\bibnamefont {Zemen}}, \bibinfo {author} {\bibfnamefont {J.}~\bibnamefont {Ma\v{s}ek}}, \bibinfo {author} {\bibfnamefont {J.}~\bibnamefont {Sinova}},\ and\ \bibinfo {author} {\bibfnamefont {T.}~\bibnamefont {Jungwirth}},\ }\href {https://doi.org/10.1103/PhysRevB.95.014403} {\bibfield  {journal} {\bibinfo  {journal} {Physical Review B}\ }\textbf {\bibinfo {volume} {95}},\ \bibinfo {pages} {014403} (\bibinfo {year} {2017})}\BibitemShut {NoStop}%
\bibitem [{\citenamefont {Edelstein}(1990)}]{Edelstein1990SSC}%
  \BibitemOpen
  \bibfield  {author} {\bibinfo {author} {\bibfnamefont {V.~M.}\ \bibnamefont {Edelstein}},\ }\href {https://doi.org/10.1016/0038-1098(90)90963-C} {\bibfield  {journal} {\bibinfo  {journal} {Solid State Communications}\ }\textbf {\bibinfo {volume} {73}},\ \bibinfo {pages} {233} (\bibinfo {year} {1990})}\BibitemShut {NoStop}%
\bibitem [{\citenamefont {\v{Z}elezn\'y}\ \emph {et~al.}(2014)\citenamefont {\v{Z}elezn\'y}, \citenamefont {Gao}, \citenamefont {V\'yborn\'y}, \citenamefont {Zemen}, \citenamefont {Ma\v{s}ek}, \citenamefont {Manchon}, \citenamefont {Wunderlich}, \citenamefont {Sinova},\ and\ \citenamefont {Jungwirth}}]{Zelezny2014PRL}%
  \BibitemOpen
  \bibfield  {author} {\bibinfo {author} {\bibfnamefont {J.}~\bibnamefont {\v{Z}elezn\'y}}, \bibinfo {author} {\bibfnamefont {H.}~\bibnamefont {Gao}}, \bibinfo {author} {\bibfnamefont {K.}~\bibnamefont {V\'yborn\'y}}, \bibinfo {author} {\bibfnamefont {J.}~\bibnamefont {Zemen}}, \bibinfo {author} {\bibfnamefont {J.}~\bibnamefont {Ma\v{s}ek}}, \bibinfo {author} {\bibfnamefont {A.}~\bibnamefont {Manchon}}, \bibinfo {author} {\bibfnamefont {J.}~\bibnamefont {Wunderlich}}, \bibinfo {author} {\bibfnamefont {J.}~\bibnamefont {Sinova}},\ and\ \bibinfo {author} {\bibfnamefont {T.}~\bibnamefont {Jungwirth}},\ }\href {https://doi.org/10.1103/PhysRevLett.113.157201} {\bibfield  {journal} {\bibinfo  {journal} {Physical Review Letters}\ }\textbf {\bibinfo {volume} {113}},\ \bibinfo {pages} {157201} (\bibinfo {year} {2014})}\BibitemShut {NoStop}%
\bibitem [{\citenamefont {Wadley}\ \emph {et~al.}(2016{\natexlab{b}})\citenamefont {Wadley}, \citenamefont {Howells}, \citenamefont {\v{Z}elezn\'y}, \citenamefont {Andrews}, \citenamefont {Hills}, \citenamefont {Campion}, \citenamefont {Nov\'ak}, \citenamefont {Olejn\'ik}, \citenamefont {Maccherozzi}, \citenamefont {Dhesi}, \citenamefont {Martin}, \citenamefont {Wagner}, \citenamefont {Wunderlich}, \citenamefont {Freimuth}, \citenamefont {Mokrousov}, \citenamefont {Kune\v{s}}, \citenamefont {Chauhan}, \citenamefont {Grzybowski}, \citenamefont {Rushforth}, \citenamefont {Edmonds}, \citenamefont {Gallagher},\ and\ \citenamefont {Jungwirth}}]{Wadley2016Science}%
  \BibitemOpen
  \bibfield  {author} {\bibinfo {author} {\bibfnamefont {P.}~\bibnamefont {Wadley}}, \bibinfo {author} {\bibfnamefont {B.}~\bibnamefont {Howells}}, \bibinfo {author} {\bibfnamefont {J.}~\bibnamefont {\v{Z}elezn\'y}}, \bibinfo {author} {\bibfnamefont {C.}~\bibnamefont {Andrews}}, \bibinfo {author} {\bibfnamefont {V.}~\bibnamefont {Hills}}, \bibinfo {author} {\bibfnamefont {R.~P.}\ \bibnamefont {Campion}}, \bibinfo {author} {\bibfnamefont {V.}~\bibnamefont {Nov\'ak}}, \bibinfo {author} {\bibfnamefont {K.}~\bibnamefont {Olejn\'ik}}, \bibinfo {author} {\bibfnamefont {F.}~\bibnamefont {Maccherozzi}}, \bibinfo {author} {\bibfnamefont {S.~S.}\ \bibnamefont {Dhesi}}, \bibinfo {author} {\bibfnamefont {S.~Y.}\ \bibnamefont {Martin}}, \bibinfo {author} {\bibfnamefont {T.}~\bibnamefont {Wagner}}, \bibinfo {author} {\bibfnamefont {J.}~\bibnamefont {Wunderlich}}, \bibinfo {author} {\bibfnamefont {F.}~\bibnamefont {Freimuth}}, \bibinfo {author} {\bibfnamefont {Y.}~\bibnamefont {Mokrousov}}, \bibinfo {author} {\bibfnamefont
  {J.}~\bibnamefont {Kune\v{s}}}, \bibinfo {author} {\bibfnamefont {J.~S.}\ \bibnamefont {Chauhan}}, \bibinfo {author} {\bibfnamefont {M.~J.}\ \bibnamefont {Grzybowski}}, \bibinfo {author} {\bibfnamefont {A.~W.}\ \bibnamefont {Rushforth}}, \bibinfo {author} {\bibfnamefont {K.~W.}\ \bibnamefont {Edmonds}}, \bibinfo {author} {\bibfnamefont {B.~L.}\ \bibnamefont {Gallagher}},\ and\ \bibinfo {author} {\bibfnamefont {T.}~\bibnamefont {Jungwirth}},\ }\href {https://doi.org/10.1126/science.aab1031} {\bibfield  {journal} {\bibinfo  {journal} {Science}\ }\textbf {\bibinfo {volume} {351}},\ \bibinfo {pages} {587} (\bibinfo {year} {2016}{\natexlab{b}})}\BibitemShut {NoStop}%
\bibitem [{\citenamefont {Moya}\ \emph {et~al.}(2025)\citenamefont {Moya}, \citenamefont {Voyemant}, \citenamefont {Chatterjee}, \citenamefont {Lee}, \citenamefont {Skorupskii}, \citenamefont {Pollak},\ and\ \citenamefont {Schoop}}]{moya2025measuring}%
  \BibitemOpen
  \bibfield  {author} {\bibinfo {author} {\bibfnamefont {J.~M.}\ \bibnamefont {Moya}}, \bibinfo {author} {\bibfnamefont {A.}~\bibnamefont {Voyemant}}, \bibinfo {author} {\bibfnamefont {S.}~\bibnamefont {Chatterjee}}, \bibinfo {author} {\bibfnamefont {S.~B.}\ \bibnamefont {Lee}}, \bibinfo {author} {\bibfnamefont {G.}~\bibnamefont {Skorupskii}}, \bibinfo {author} {\bibfnamefont {C.~J.}\ \bibnamefont {Pollak}},\ and\ \bibinfo {author} {\bibfnamefont {L.~M.}\ \bibnamefont {Schoop}},\ }\href@noop {} {\bibfield  {journal} {\bibinfo  {journal} {arXiv preprint arXiv:2512.19427}\ } (\bibinfo {year} {2025})}\BibitemShut {NoStop}%
\bibitem [{\citenamefont {Pet{\v{r}}{\'\i}{\v{c}}ek}\ \emph {et~al.}(2023)\citenamefont {Pet{\v{r}}{\'\i}{\v{c}}ek}, \citenamefont {Palatinus}, \citenamefont {Pl{\'a}{\v{s}}il},\ and\ \citenamefont {Du{\v{s}}ek}}]{petvrivcek2023jana2020}%
  \BibitemOpen
  \bibfield  {author} {\bibinfo {author} {\bibfnamefont {V.}~\bibnamefont {Pet{\v{r}}{\'\i}{\v{c}}ek}}, \bibinfo {author} {\bibfnamefont {L.}~\bibnamefont {Palatinus}}, \bibinfo {author} {\bibfnamefont {J.}~\bibnamefont {Pl{\'a}{\v{s}}il}},\ and\ \bibinfo {author} {\bibfnamefont {M.}~\bibnamefont {Du{\v{s}}ek}},\ }\href@noop {} {\bibfield  {journal} {\bibinfo  {journal} {Zeitschrift f{\"u}r Kristallographie-Crystalline Materials}\ }\textbf {\bibinfo {volume} {238}},\ \bibinfo {pages} {271} (\bibinfo {year} {2023})}\BibitemShut {NoStop}%
\bibitem [{\citenamefont {Kirschbaum}\ \emph {et~al.}(1997)\citenamefont {Kirschbaum}, \citenamefont {Martin},\ and\ \citenamefont {Pinkerton}}]{kirschbaum1997lambda}%
  \BibitemOpen
  \bibfield  {author} {\bibinfo {author} {\bibfnamefont {K.}~\bibnamefont {Kirschbaum}}, \bibinfo {author} {\bibfnamefont {A.}~\bibnamefont {Martin}},\ and\ \bibinfo {author} {\bibfnamefont {A.~A.}\ \bibnamefont {Pinkerton}},\ }\href@noop {} {\bibfield  {journal} {\bibinfo  {journal} {Applied Crystallography}\ }\textbf {\bibinfo {volume} {30}},\ \bibinfo {pages} {514} (\bibinfo {year} {1997})}\BibitemShut {NoStop}%
\end{thebibliography}%

\clearpage
\onecolumngrid
%%%%%%%%%%%%%%%% START OF SUPPLEMENT %%%%%%%%%%%%%%%

% Figures, tables, equations and pages in the supplement are numbered S1, S2 etc.
\renewcommand{\thefigure}{S\arabic{figure}}
\renewcommand{\thetable}{S\arabic{table}}
\renewcommand{\theequation}{S\arabic{equation}}
\setcounter{figure}{0}
\setcounter{table}{0}
\setcounter{equation}{0}
\setcounter{section}{0}% not 0 as \newpage already started a supplementary page
% References continue the numbering from the main text.

%%%%%%%%%%%%%%%% SUPPLEMENT TITLE PAGE %%%%%%%%%%%%%%%

% \begin{center}
% \textbf{\large Supplemental Materials:\\ Incommensurate magnetic orders and topological Hall effect in the square-net centrosymmetric EuGa$_2$Al$_2$ system}

% \end{center}

\section*{Tunable magnetotransport through kinetically hindered first-order phase transitions in an antiferromagnetic metal}
\begin{center}
% Author list for the supplement
% Indicate the corresponding authors, but do NOT include institutions here
% It would be nice if the template auto-generated this, but doing so is complicated...
Jaime M. Moya,
 Scott B. Lee,
 Sudipta Chatterjee,
 Nitish Mathur,
 Grigorii Skorupskii,
 Connor J. Pollak,
 and Leslie M. Schoop

\end{center}

\section{Note 1: Elemental and structural analysis}

\begin{figure} % Do NOT use \begin{figure*}
	\centering
	\includegraphics[width=1\textwidth]{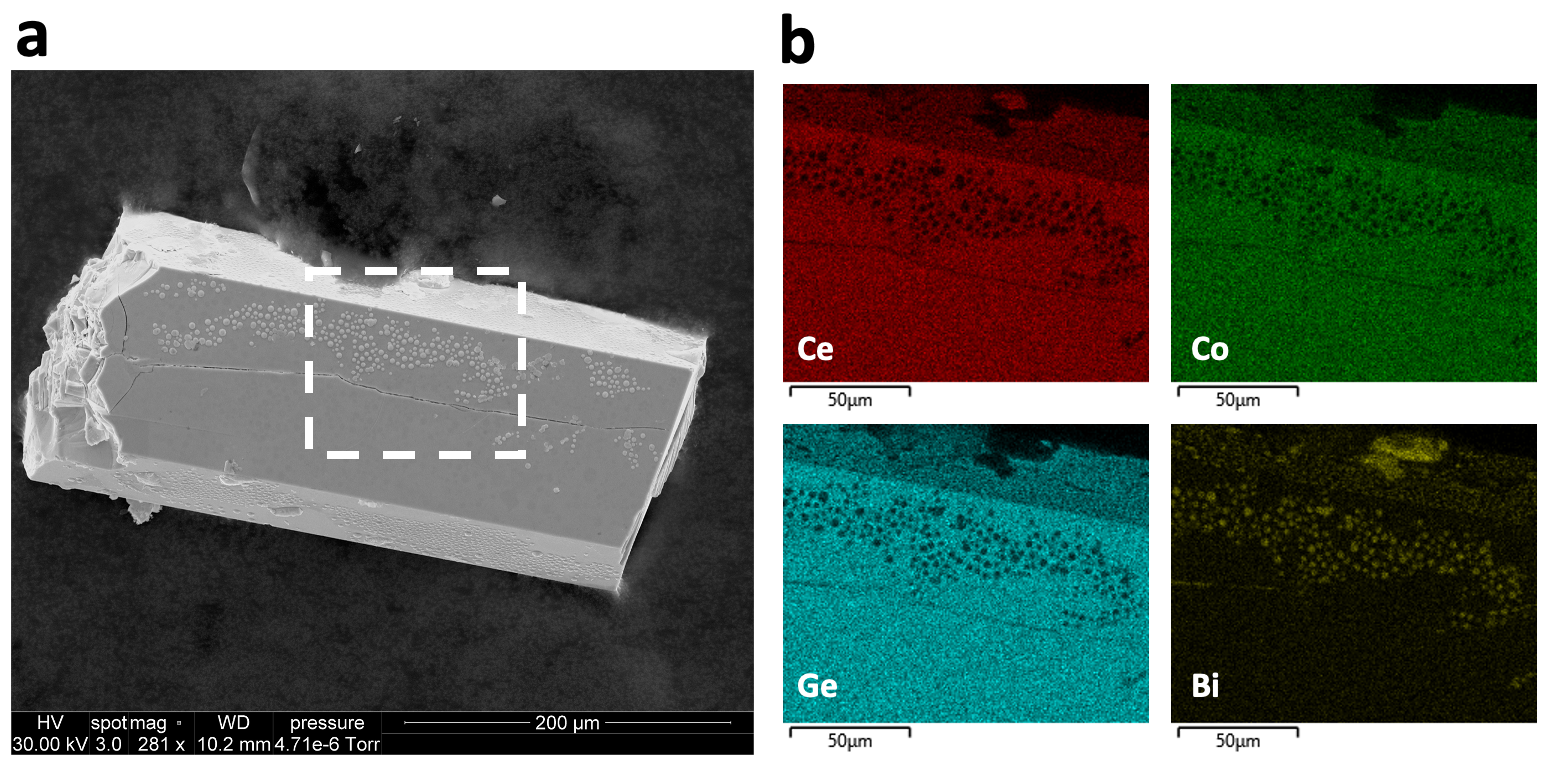}
	\caption{\textbf{a} SEM image of a crystal from the Batch 1 synthesis of CeCoGe$_3$. \textbf{b} EDS maps corresponding to Ce (red), Co (green), Ge (blue), and Bi (yellow)}
	\label{fig:EDS}
\end{figure}

EDS was used to confirm the chemical homogeneity and stoichiometry of the CeCoGe$_3$ samples. Fig.~\ref{fig:EDS}a shows a typical scanning electron microscope (SEM) image of a single crystal of CeCoGe$_3$ together with EDS maps corresponding to Ce (red), Co (green), Ge (blue), and Bi (yellow) in Fig.~\ref{fig:EDS}b. The EDS maps demonstrate that the crystal surfaces are homogeneous and that the Bi flux does not incorporate into the crystal, but rather is isolated on the surface. Furthermore, a point-scan analysis of EDS measurements from four crystals across two growth batches reveals batch-to-batch consistency as well as near-stoichiometric conditions within the expected errors of EDS.

\begin{table}[t]\label{EDS}
\centering
\caption{EDS composition analysis for selected \CeCo\ crystals. Atomic percentages are listed together with compositions normalized (norm) to Ge = 3.}
\label{tab:eds}
\begin{tabular}{llcccccc}
\hline
Crystal & Statistic & Ce (at.\ \%) & Co (at.\ \%) & Ge (at.\ \%) & Ce (norm) & Co (norm) & Ge (norm) \\
\hline
Batch 1, crystal 1 (5 pts) & Average & 19.5 & 20.4 & 60.1 & 0.972 & 1.02 & 3.000 \\
 & St.\ Dev. & 0.1 & 0.2 & 0.1 & 0.007 & 0.01 & 0.005 \\
\hline
Batch 1, crystal 2 (8 pts) & Average & 19.6 & 20.4 & 60.0 & 0.980 & 1.02 & 3.000 \\
 & St.\ Dev. & 0.1 & 0.2 & 0.1 & 0.007 & 0.01 & 0.006 \\
\hline
Batch 1, crystal 3 (4 pts) & Average & 19.23 & 20.09 & 60.7 & 0.951 & 0.993 & 3.000 \\
 & St.\ Dev. & 0.09 & 0.05 & 0.1 & 0.005 & 0.003 & 0.005 \\
\hline
Batch 2, crystal 1 (4 pts) & Average & 19.43 & 20.30 & 60.3 & 0.967 & 1.010 & 3.000 \\
 & St.\ Dev. & 0.02 & 0.11 & 0.1 & 0.002 & 0.006 & 0.006 \\
\hline
All crystals & Average & 19.4 & 20.3 & 60.3 & 0.97 & 1.01 & 3.000 \\
 & St.\ Dev. & 0.1 & 0.1 & 0.3 & 0.01 & 0.01 & --- \\
\hline
\end{tabular}
\end{table}

To confirm the crystal structure of CeCoGe$_3$, single-crystal X-ray diffraction measurements were performed and refined; the results, shown in Tables 2--5, agree with previous work \cite{pecharsky1993unusual}. Interestingly, forbidden reflections are observed, however, as we explain in the following, these are likely due to \(\lambda/2\) contamination \cite{kirschbaum1997lambda}.

\begin{figure}[h] % Do NOT use \begin{figure*}
	\centering
	\includegraphics[width=1\textwidth]{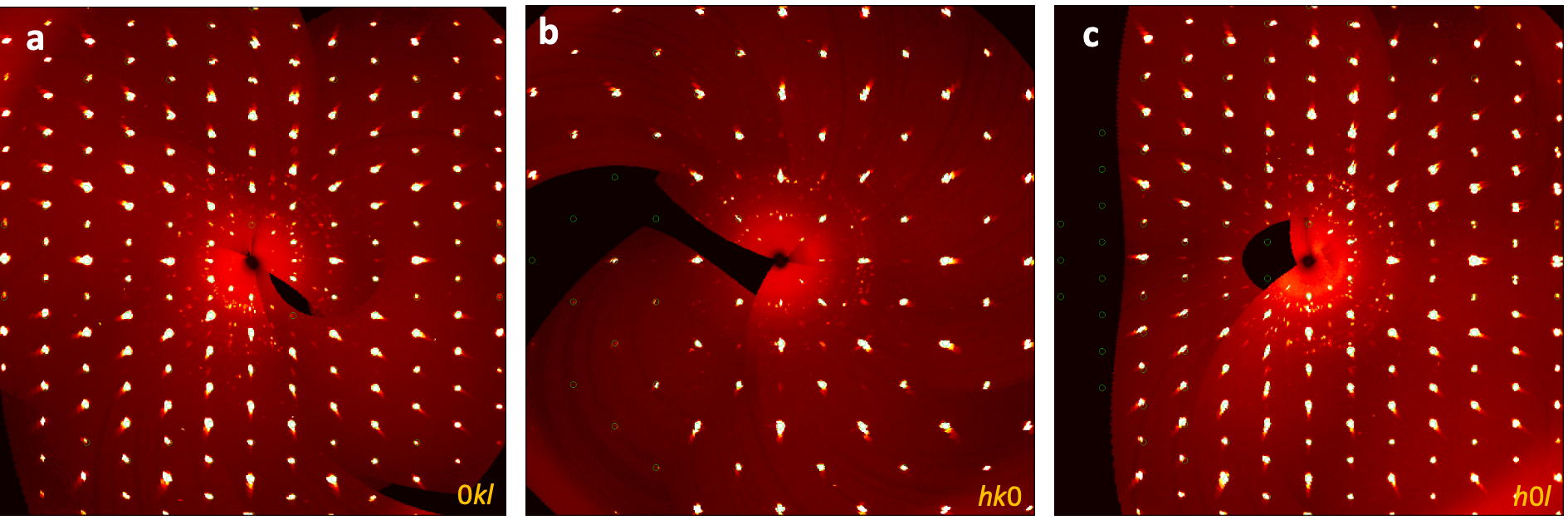}
	\caption{Precession images of the \textbf{(a)} 0\textit{kl}, \textbf{(b)} \textit{hk}0, and \textbf{(c)} \textit{h}0\textit{l} of the crystal used for the single-crystal X-ray diffraction refinement.}
	\label{fig:Precession_old_xtal1}
\end{figure}

\begin{figure}[h] % Do NOT use \begin{figure*}
	\centering
	\includegraphics[width=1\textwidth]{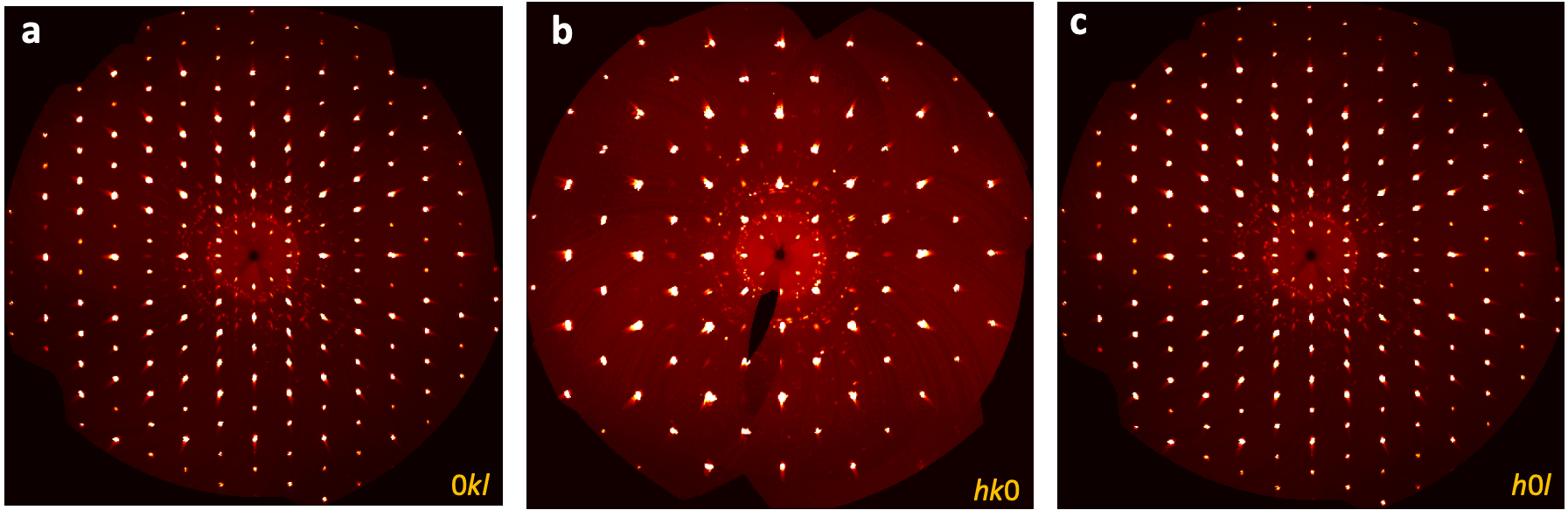}
	\caption{Precession images of the \textbf{(a)} 0\textit{kl}, \textbf{(b)} \textit{hk}0, and \textbf{(c)} \textit{h}0\textit{l} of an independent crystal showing the reproducibility of the forbidden reflections.}
	\label{fig:Precession_old_xtal2}
\end{figure}

\begin{figure}[h] % Do NOT use \begin{figure*}
	\centering
	\includegraphics[width=0.6\textwidth]{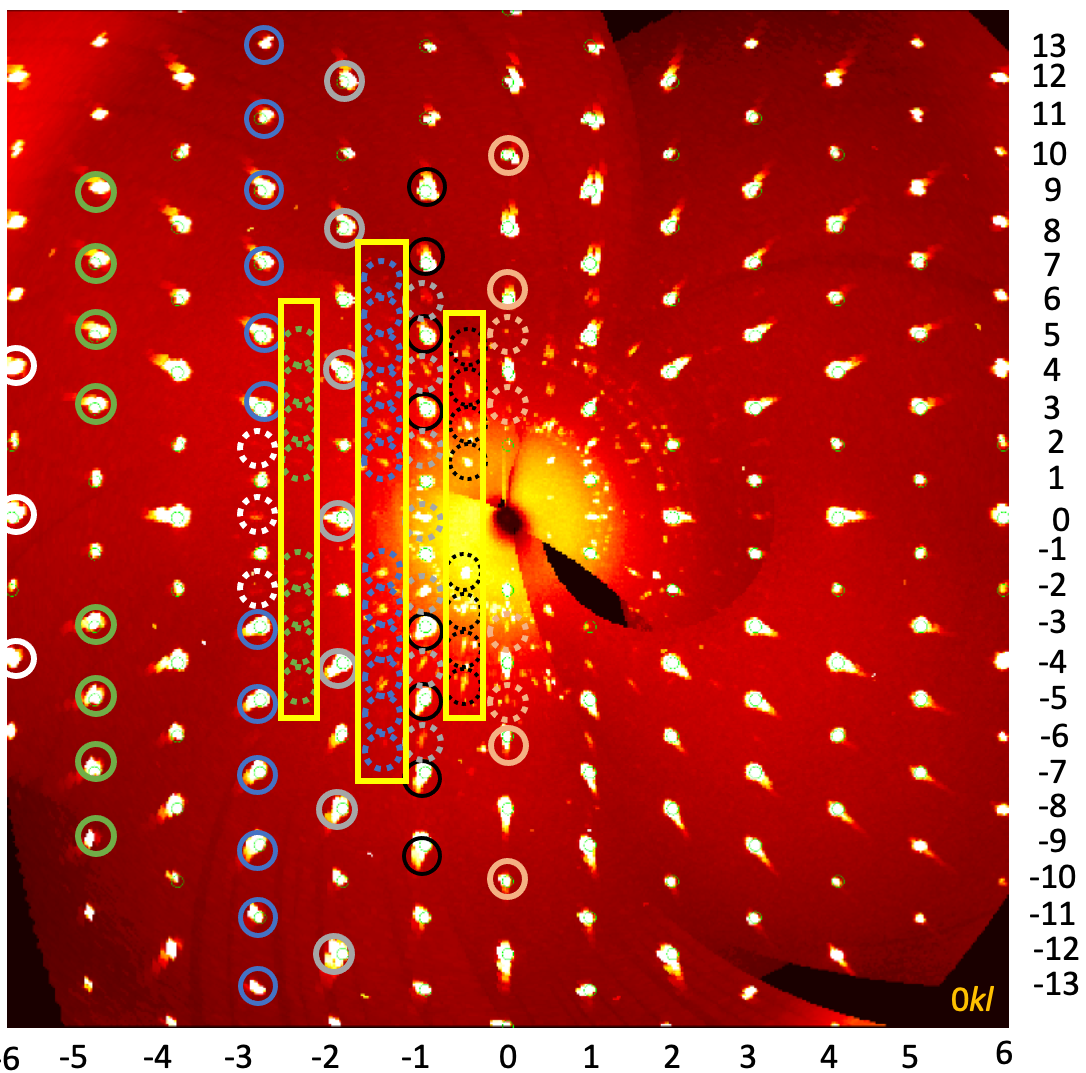}
	\caption{Enlarged view of the 0\textit{kl} precession image presented in Fig.~\ref{fig:Precession_old_xtal1}a. Dashed symbols highlight $\lambda/2$ peaks corresponding to the like-colored solid symbols. See text for details regarding the yellow box.}
	\label{fig:0kl}
\end{figure}

\begin{figure}[h] % Do NOT use \begin{figure*}
	\centering
	\includegraphics[width=0.6\textwidth]{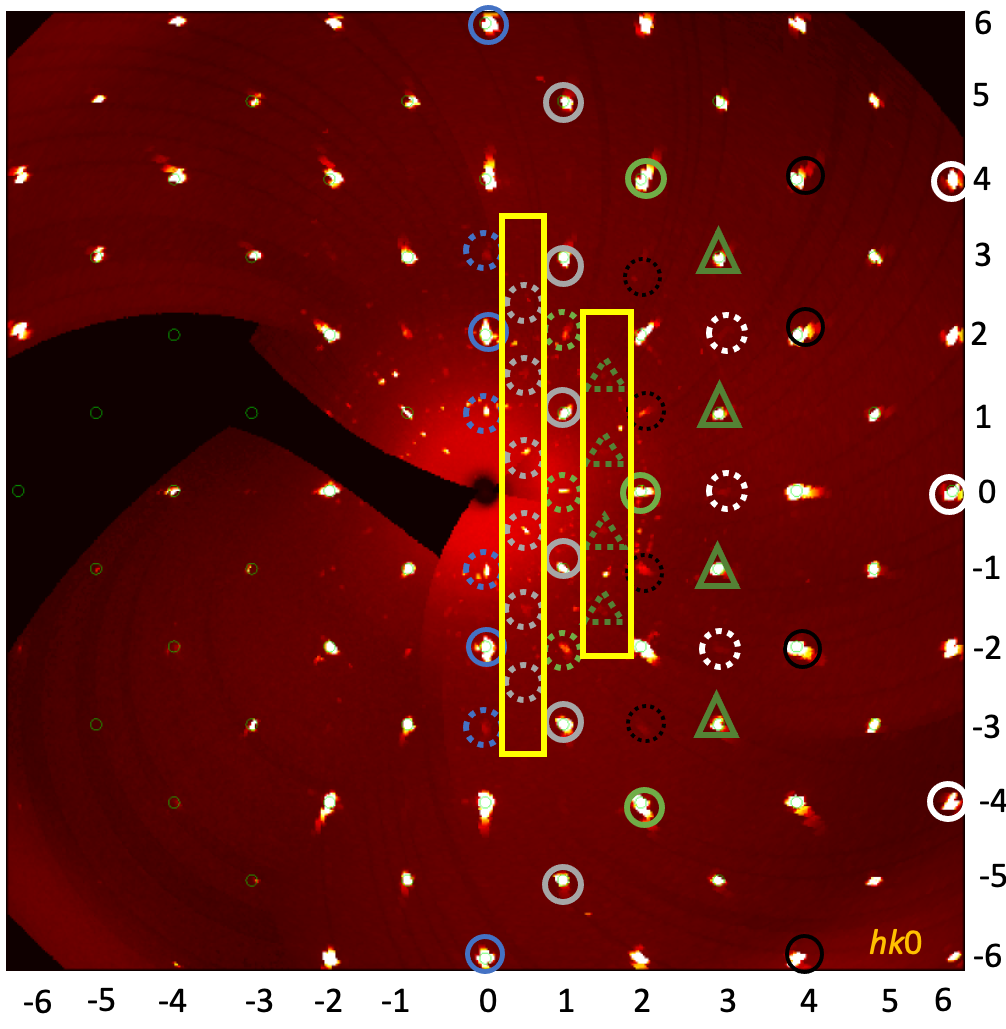}
	\caption{Enlarged view of the \textit{hk}0 precession image presented in Fig.~\ref{fig:Precession_old_xtal1}b. Dashed symbols highlight $\lambda/2$ peaks corresponding to the like-colored solid symbols. See text for details regarding the yellow box.}
	\label{fig:hk0}
\end{figure}

Fig.~\ref{fig:Precession_old_xtal1}a--c shows precession images for the \(0kl\), \(hk0\), and \(h0l\) planes, synthesized from a 100~K dataset. For a body-centered lattice, reflections with \(h+k+l=2n\) are allowed, whereas those with \(h+k+l=2n+1\) are forbidden \((n\in\mathbb{Z})\). Weak but forbidden reflections are present and reproducible, as confirmed on an independent crystal (Fig.~\ref{fig:Precession_old_xtal2}). Both measurements used long exposures (60~s per scan). In Figs.~\ref{fig:0kl} and \ref{fig:hk0} we enlarge the \(0kl\) and \(hk0\) images and highlight the forbidden spots as dashed symbols.

% %%% Crystallographic Refinement Information
\begin{center}
\begin{table}\label{ref1}
\caption{Crystallographic Information for CeCoGe$_{3}$.}
 \centering
 \begin{tabular}{l c c}
\hline
Nominal Composition & & {CeCoGe$_{3}$} \\
Crystal Dimension (mm) & & 0.171$\times$0.293$\times$0.376\\
Radiation Source, $\lambda$ (\AA) & & Mo K$_{\alpha}$, 0.71073 \\
Absorption Correction & & multi-scan \\
Data Collection Temp (K) & & 100 \\
Space Group & & $I4mm$ \\
$a$ (\AA) & & 4.3169(2) \\
$c$ (\AA) & & 9.8293(5) \\
Cell Volume (\AA$^{3}$) & & 183.175(15) \\
Absorption Coefficient (mm$^{-1}$) & & 40.671 \\
$\theta_{min}$, $\theta_{max}$ (deg) & & 4.15, 30.41 \\
Refinement Method & & F$^{2}$ \\
R$_{int}$(I\textgreater{}3$\sigma$, all) & & 4.39, 4.39 \\
Number of Reflections & & 5073 \\
Number of Parameters & & 16 \\
Unique Reflections (I \textgreater 3$\sigma$, all) & & 203, 203 \\
R(I\textgreater{}$3\sigma$), R$_{w}$(I\textgreater{}$3\sigma$) & & 2.20, 5.14 \\
R(all), R$_{w}$(all) & & 2.20, 5.14 \\
S(I\textgreater{}$3\sigma$), S(all) & & 2.09, 2.09 \\
$\Delta\rho_{max}$, $\Delta\rho_{min}$ (e \AA $^{-3}$) & & 2.95, -0.86 \\
\hline
\label{Table_2}
\end{tabular}
\end{table}
\end{center}
%%%
%%% Bi mono Crystallographic Atomic Coordinates
%%%
\begin{center}
\begin{table}\label{ref2}
\renewcommand{\tablename}{Table S}
\caption{Refined atomic coordinates for {CeCoGe$_{3}$}.}
\begin{tabular}{@{}lcccccc@{}}
\hline
Site & Wyckoff Position & x & y & z & Occupancy & \\
\hline
Ce1 & 2a & 0 & 0 & 0.53182(6) & 1 & \\
Co1 & 2a & 1/2 & 1/2 & 0. 6973(2) & 1 & \\
Ge1 & 2a & 1/2 & 1/2 & 0.46337(12) & 1 & \\
Ge2 & 4b & 1/2 & 0 & 0.29137(11) & 1 & \\
\hline
\label{Table_3}
\end{tabular}
\end{table}
\end{center}
\begin{center}
\begin{table}\label{ref3}

\caption{Refined anisotropic displacement parameters for {CeCoGe$_{3}$}.}
\begin{tabular}{lccccccc}
\hline
Site & U$_{11}$ & U$_{22}$ & U$_{33}$ & U$_{12}$ & U$_{13}$ & U$_{23}$ & \\
\hline
Ce1 & 0.0146(3) & 0.0146(3) & 0.0120(3) & 0 & 0 & 0 & \\
Co1 & 0.0141(6) & 0.0141(6) & 0.0124(8) & 0 & 0 & 0 & \\
Ge1 & 0.0152(3) & 0.0152(3) & 0.0103(9) & 0 & 0 & 0 & \\
Ge2 & 0.0127(5) & 0.0186(6) & 0.0113(6) & 0 & 0 & 0 & \\
\hline
\label{Table_4}
\end{tabular}
\end{table}
\end{center}

\begin{center}
\begin{table}\label{ref4}
\caption{Selected interatomic distances for {CeCoGe$_3$}.}
\begin{tabular}{llclclclcl}
\hline
 Site && Neighbor && Multiplicity && Distance (\AA) && \\
\hline
 Ce1 && Ge1 && 4 && 3.1258(4) && \\
 Ge1 && Co1 && 1 && 2.299(2) && \\
 Ge2 && Co1 && 2 && 2.3483(9) && \\
\hline
\label{Table_1}
\end{tabular}
\end{table}
\end{center}

The forbidden spots are observed at \((h+\tfrac{1}{2},\,0,\,l+\tfrac{1}{2})\), and \((0,\,k+\tfrac{1}{2},\,l+\tfrac{1}{2})\). If these were due to a crystallographic distortion, they would imply a doubling of the unit cell along all three directions.
However, we consider it unlikely that these peaks---or the other weak forbidden peaks discussed below---arise from a structural modulation. Instead, we can understand them as consistent with \(\lambda/2\) contamination, a known issue for long exposures with Mo K$_\alpha$ anodes on CCD area detectors even with a graphite monochromator \cite{kirschbaum1997lambda}.

Our reasoning is two-fold. (i) The half-order peaks occur only at low Miller indices, and their intensities track those of the corresponding allowed ``parent'' reflections at doubled indices, \((H,K,L)=2(h+\tfrac{1}{2},k+\tfrac{1}{2},l+\tfrac{1}{2})\), provided \(H+K+L\) is even (the body-centering condition) with $H,K,L$ all odd or 0 to obtain half-integer reflections. For example, the fundamental \((0,-3,l)\) intensities (\(|l|\ge 3\)) decrease with increasing \(|l|\) (blue solid circles), and the forbidden \((0,-\tfrac{3}{2},\,\tfrac{l}{2})\) spots (blue dashed circles) in Fig.~\ref{fig:0kl} show the same trend because they are coming from the same peak with the same structure factor, but the \((0,-\tfrac{3}{2},\,\tfrac{l}{2})\) peak results from the $\lambda/2$ radiation which has much less flux. Notably, \((0,-3,\pm1)\) are comparatively weak, and the corresponding \((0,-\tfrac{3}{2},\,\pm\tfrac{1}{2})\) \(\lambda/2\) peaks are not observed for these exposure times.

(ii) If half-order peaks are \(\lambda/2\) artifacts, then integer symmetry-forbidden spots with \(h+k+l=2n+1\) should also appear at low angle, arising from \(\lambda/2\) contamination of allowed parents at \((H,K,L)=2(h,k,l)\) (with $H,K,L$ (all even)). After further inspection, such spots are indeed present---marked with dashed symbols without yellow boxes in Figs.~\ref{fig:0kl} and \ref{fig:hk0}---and their intensities scale with those of the corresponding allowed reflections (same color, solid symbols). While symmetry-forbidden reflections at \(h+k+l=2n+1\) could, in principle, indicate loss of body centering, here they are confined to low angles, intensity-scaled to parent peaks, and co-occur with half-order features---all pointing to \(\lambda/2\) contamination rather than true symmetry lowering.

In summary, the forbidden reflections we observe are (i) consistent with and indexable as \(\lambda/2\) contamination made visible by long CCD exposures, (ii) reproducible, and (iii) inconsistent with static disorder.

With regard to the refinement, previous work shows that \(\lambda/2\) contamination has negligible effects on refined parameters but can complicate space-group assignment \cite{kirschbaum1997lambda}. Because every forbidden spot maps to an allowed parent at doubled indices, we remain confident in the \(I4mm\) assignment.

To further test whether it is reasonable for a 2$\times$2$\times$2 supercell, some site occupancy disorder, Bi doping, or mixed site occupancy may occur, we repeated our refinements of the published structural solution with these four possibilities in mind.

Refinements in a 2$\times$2$\times$2 supercell did not converge, confirming that these weak reflections are too weak to properly handle. This is further shown by our original structural solution finding 0 forbidden reflections \textemdash while we may be able to judge visually that we see these extra reflections in our reconstructed precession images, they are statistically unobserved.

To test whether the kinetic hindrance in CeCoGe$_3$ and the associated quenched disorder arises from crystallographic disorder, we checked for site occupancy disorder in the 100 K refinements, where we systematically attempted refinements for the occupancy of every site individually, both Ge sites together, and Co/Ge mixed sites.

The results of these tests are summarized in Table~\ref{occupancy}. The results of these refinements are mostly unphysical, with Ce1, Ge1, Ge2, and Ge1/Ge2 refinements all refining to occupancies greater than full occupation. In our refinement of Co1 occupancies, a modest decrease in occupation ($\sim$ 0.07) is seen; however, this is not consistent with EDS data which suggests, if anything, the compound is Ce deficient.

\begin{center}
\begin{table} [ht!]\label{occupancy}
\caption{\textbf{Summary of occupancy refinement tests.}}
 \centering
 \begin{tabular}{c c c c}
\hline
Site refined & Refined Composition & R(I 3$\sigma$) & S(I 3$\sigma$) \\
\hline
none & CeCoGe$_3$ & 2.20 & 2.09 \\
Ce1 & Ce$_{1.047}$CoGe$_3$ & 2.01 & 1.98 \\
Co1 & CeCo$_{0.931}$Ge$_3$ & 1.90 & 1.80 \\
Ge1 & CeCoGe$_{3.008}$ & 2.19 & 2.09 \\
Ge2 & CeCoGe$_{3.148}$ & 2.13 & 1.97 \\
Ge1/Ge2 & CeCoGe$_{3.149}$ & 2.13 & 1.97 \\
\hline
\label{Site_disorder}
\end{tabular}
\end{table}
\end{center}

Another significant outlier in our occupancy test was the Ge2 occupancy refinement, where a modest increase in occupancy was observed. This suggested to us the possibility that Bi was substituting specifically on this crystallographic position. However, a model with Bi1/Ge2 mixed sites did not converge, even when removing the anisotropic refinement parameters for all atoms.

Bi could also potentially substitute onto the Ce1 position, given that our elemental analysis suggests a Ce deficiency. Our model shows that substitution on this site is not likely, converging to a composition of Ce$_{0.993}$Bi$_{0.007}$CoGe$_3$ with a worse R (2.81) and goodness of fit (2.82).

We attempted one more option \textemdash a mixed site occupancy of Ge1/Co1 sites. This was done by adding a split Co1' position onto the Ge1 site, a split Ge1' position on the Co1 site, and adding appropriate constraints to make sure the sum of all occupancies resulted in a 1:1:3 stoichiometry. This situation could possibly retain the observed stoichiometry within our EDS measurements. This model appears to put more electron density on the Ge2 site, and could create very weak reflections in systematic absent positions due to the structure factors of the body-centered position between (Ge1/Co1) and Ge2 positions not equaling exactly 0. However, refinement in this model consistently drove the occupation of the substituted atom (Co1', Ge1') to negative values while driving the occupancy of the original atom (Ge1, Co1) to above full occupancy.

With these results in mind, we firmly believe the weak reflections observed in our precession images are \(\lambda/2\) artifacts, and we have no evidence that they arise from any deviation from our published model. Furthermore, the low-temperature phase transitions mentioned in the text likely arise not from quenched disorder in the as-grown crystals, but rather from a low-temperature magnetostructural phase transition, as discussed in the main text.

\subsection{Supplementary Note 2: Magnetic field-temperature phase diagram for $H \parallel c$}
\label{Note2}

\begin{figure} % Do not use \begin{figure*}
	\centering
	\includegraphics[width=0.6\textwidth]{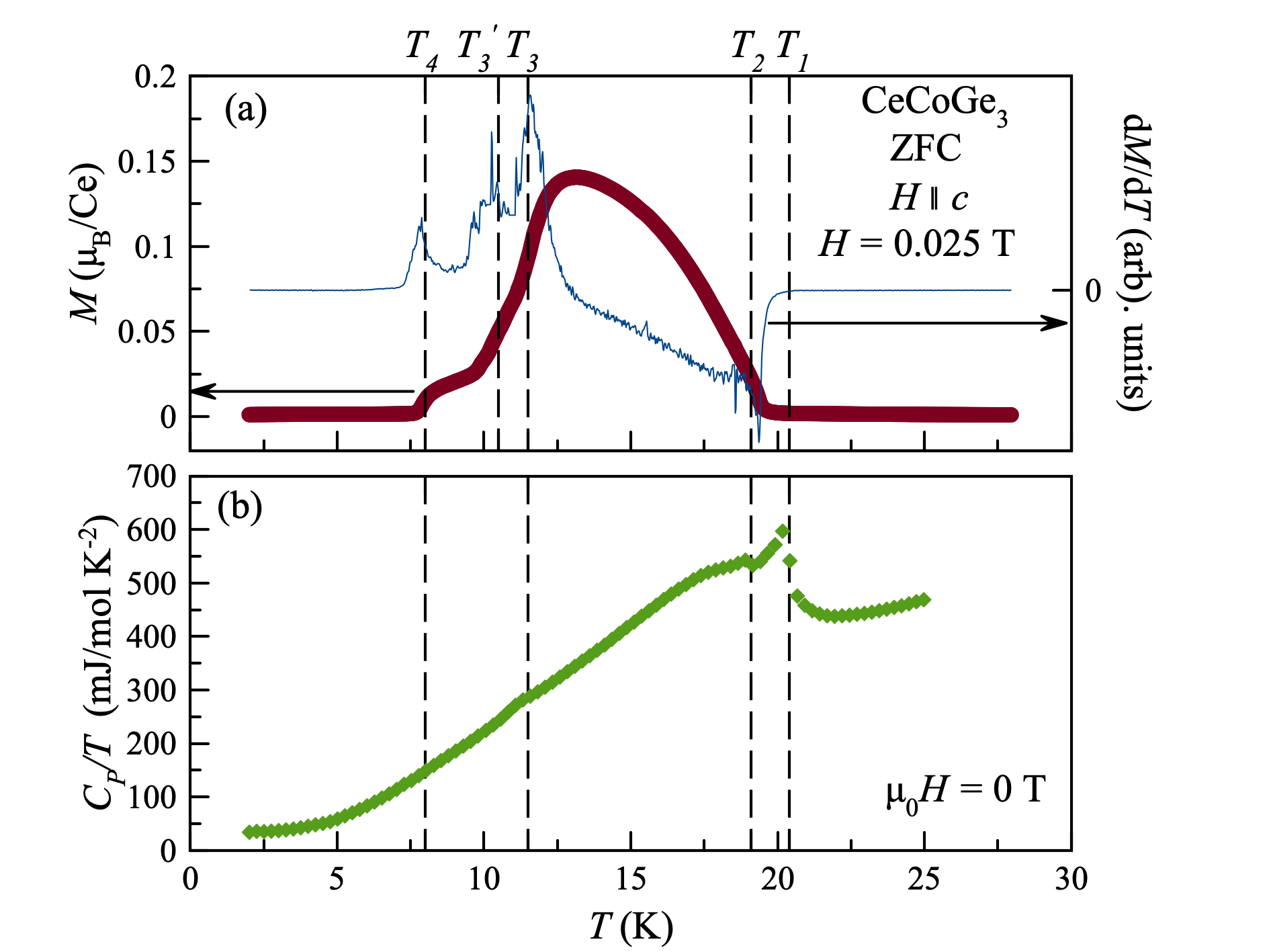} % for an image file named example_figure.*
	% Pick an appropriate width for the size of the image

	% Captions go below figures
	\caption{\textbf{Near zero-field properties of CeCoGe$_3$} The low-temperature ZFC temperature-dependent (\textbf{a}, left axis, open symbols) magnetization $M$ measured with $\mu_0H~=~0.025$ T for $H \parallel c$ together with the derivative (right axis, closed symbols) $dM/dT$, compared to (\textbf{b}) the zero-field heat capacity scaled by temperature $C_P/T$. The dashed lines mark $T_1$, $T_2$, $T_3$, and $T_4$ while $T_3'$ is additionally marked in (a).}
	\label{fig:sup_MTCP} % give each figure a logical label name
\end{figure}

In this Note, we describe measurements used to construct the $H$--$T$ phase diagram in Fig.~\ref{fig:Basics}a of the main text. Focusing on the near-zero-field region, the ZFC temperature-dependent magnetization $M$ measured with $H \parallel c$ for $\mu_0H~=~0.025$ T between 1.8 K and 28 K is presented in Fig.~\ref{fig:sup_MTCP}a (red, left axis). The corresponding derivative $dM/dT$ is shown in Fig.~\ref{fig:sup_MTCP}a (blue, right axis), and compared to the specific heat scaled by temperature $C_P/T$ measured in $\mu_0H~=~0$ T in Fig.~\ref{fig:sup_MTCP}b.

The N\'eel temperature $T_1~\approx~$ 21 K was approximated as the halfway point of the peak in $C_P/T$ (Fig.~\ref{fig:sup_MTCP}b), consistent with previous measurements \cite{thamizhavel2005unique}. Upon further cooling, another anomaly in $C_P/T$ is observed at $T_2~\approx$ 18.5 K, followed by a broad feature at $T_3~\approx$ 12 K. Comparatively, $dM/dT$ (Fig.~\ref{fig:sup_MTCP}a, blue, right axis) captures $T_2$ as a minimum, followed by local maxima at $T_3$, $T_3' \sim 11$ K, and $T_4 \sim 8$ K.

Notably, no feature in $dM/dT$ is resolved at $T_1$ for $\mu_0H~=~0.025$ T, and no feature at $T_4$ and $T_3'$ is resolved in $C_P/T$. Naively, the $T_1$ could be attributed to a structural phase transition, however, both zero-field neutron diffraction and $\mu$SR measurements confirm long-range (LR) order at $T_1$, as well as a magnetic phase transition at $T_4$ \cite{smidman2013neutron}. The feature at $T_3'$ may appear in $dM/dT$ and is therefore likely field-induced. We therefore deduce that CeCoGe$_3$ has four magnetically ordered phases above 2 K in zero field: Phase I from $T_1$ to $T_2$, Phase II from $T_2$ to $T_3$, Phase III from $T_3$ to $T_4$, and Phase IV below $T_4$. We note $T_2$ was not reported in previous $C_P$ measurements \cite{thamizhavel2005unique}, though a weak feature can be seen at $T_2$ that was likely missed due to the lack of temperature resolution.

\begin{figure} % Do not use \begin{figure*}
	\centering
	\includegraphics[width=1\textwidth]{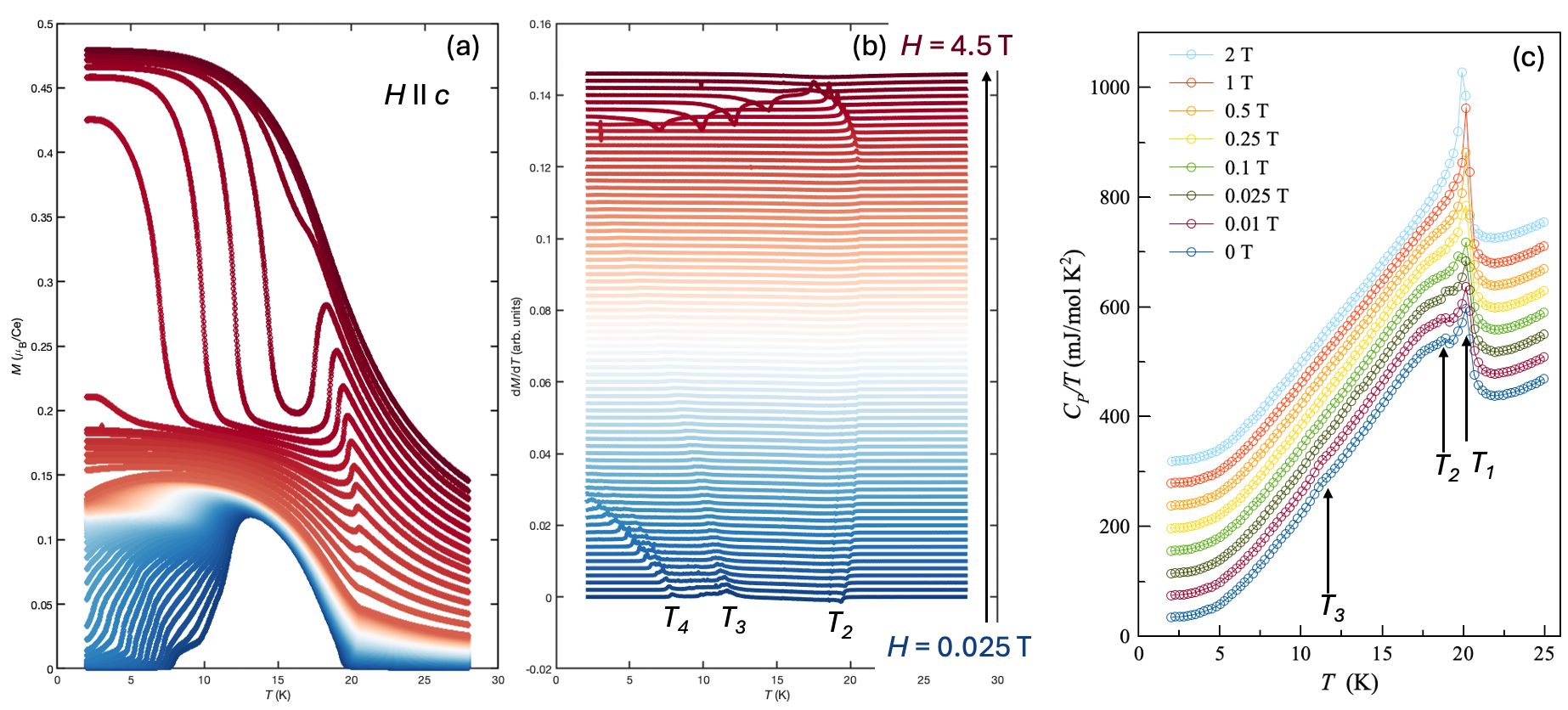} % for an image file named example_figure.*
	% Pick an appropriate width for the size of the image

	% Captions go below figures
	\caption{\textbf{Temperature-dependent thermodynamic measurements}
	 The temperature-dependent (\textbf{a}) magnetization $M$ measured with $H \parallel c$ for select fields from $\mu_0H$ = 0.025 T to 4.5 T and the corresponding (\textbf{b}) derivative $dM/dT$ with $T_2$, $T_3$, and $T_4$ highlighted. The data in (\textbf{b}) are offset for clarity. (\textbf{c}) The temperature-dependent specific heat scaled by temperature $C_P/T$ measured in select fields for $H \parallel c$ showing the low-field evolution of $T_1$, $T_2$, and $T_3$ marked with arrows. All data except the zero-field data are offset for clarity.}
	\label{fig:sup_MT_field} % give each figure a logical label name
\end{figure}

\begin{figure} % Do not use \begin{figure*}
	\centering
	\includegraphics[width=0.6\textwidth]{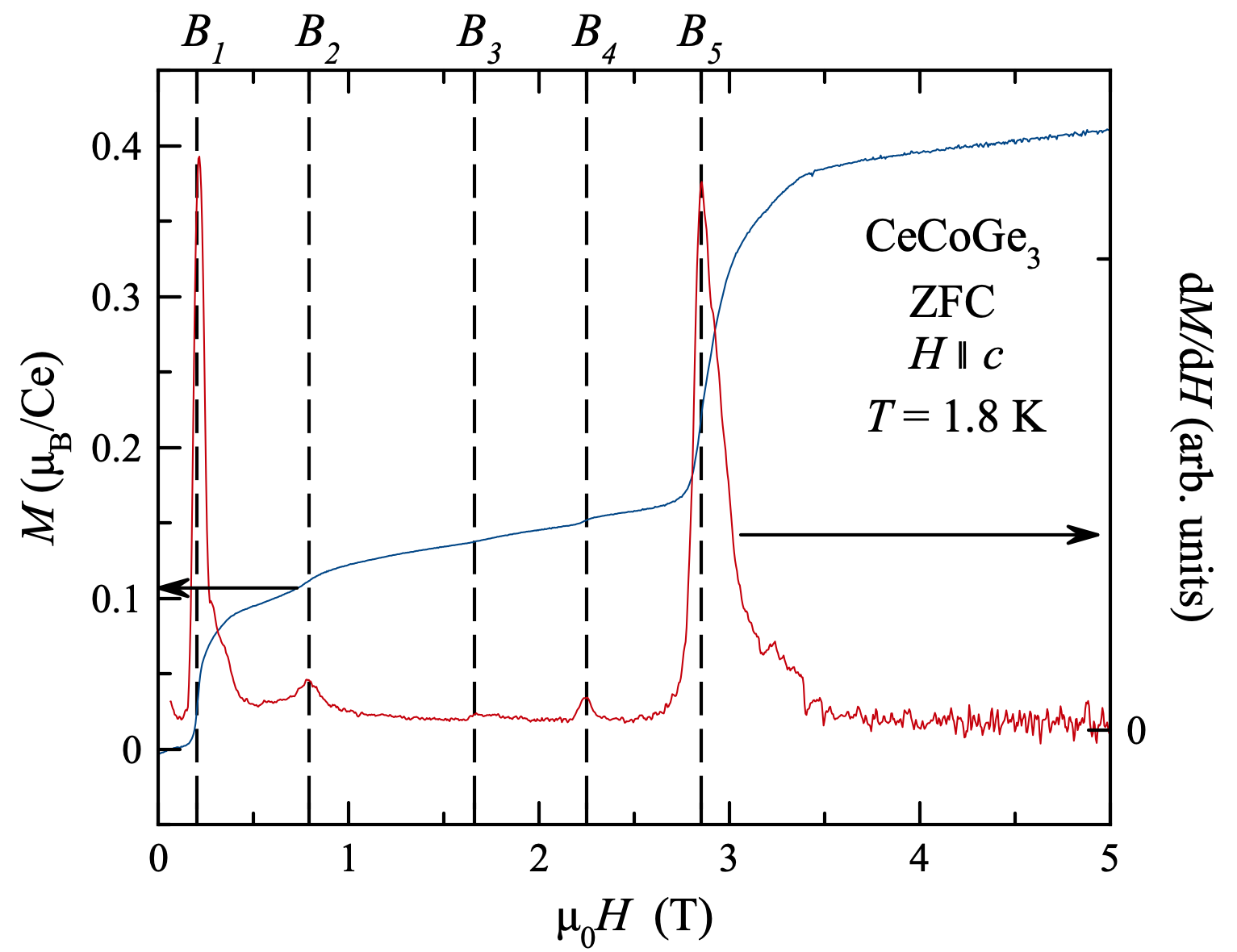} % for an image file named example_figure.*
	% Pick an appropriate width for the size of the image
	% Captions go below figures
	\caption{
	 Isothermal magnetization $M(H)$ (blue, left axis) and the derivative $dM/dH$ (red, right axis) measured at 1.8 K with $H \parallel c$ after zero-field cooling from above $T_1$. Anomalies are marked as local maxima in $dM/dH$ and are labeled $B_1$--$B_5$ on increasing field.}
	\label{fig:MH1p8} % give each figure a logical label name
\end{figure}

To map the isofield dependence of the $H$--$T$ phase diagram, a series of magnetization measurements were performed with increasing field from 0.025 T to 4.5 T for $H \parallel c$, as shown in Fig.~\ref{fig:sup_MT_field}a; the corresponding derivative $dM/dT$ is shown in Fig.~\ref{fig:sup_MT_field}b. The features at $T_2$, $T_3$, and $T_4$ in $dM/dT$ are tracked as a function of the field and are marked as red circles in Fig.~\ref{fig:Basics}a of the main text. Notably, $T_2$ moves up in temperature with increasing field up to $\mu_0H \approx$ 0.3~T until it merges with the $T_1$ phase boundary established by $C_P/T$ measurements, creating the boundary $B_0$ separating Phases I and II with application of field as shown in Fig.~\ref{fig:Basics}a. The boundary $B_0$ is corroborated by $C_P/T$ shown in Fig.~\ref{fig:sup_MT_field}c where $C_P$ was measured at select fields up to 2 T where it is apparent that $T_1$ and $T_2$ merge with increasing field. The boundaries captured by $C_P/T$ are plotted as diamonds in Fig.~\ref{fig:Basics}a of the main text.
To complete the $H$--$T$ phase diagram in Fig.~\ref{fig:Basics}a of the main text, isothermal magnetization $M(H)$ loops were measured at various temperatures after ZFC from above $T_1$. An example of such a loop measured with $H \parallel c$ at $T = 1.8$~K is shown in Fig.~\ref{fig:MH1p8} (blue, left axis). The phase boundaries are captured as local maxima in the derivative $dM/dH$ (Fig.~\ref{fig:MH1p8}) (red, right axis). With increasing field, prominent peaks in $dM/dH$ are registered at $B_1$, $B_2$, and $B_5$, with more subtle features at intermediate fields $B_3$ and $B_4$.

\begin{figure} % Do not use \begin{figure*}
	\centering
	\includegraphics[width=0.6\textwidth]{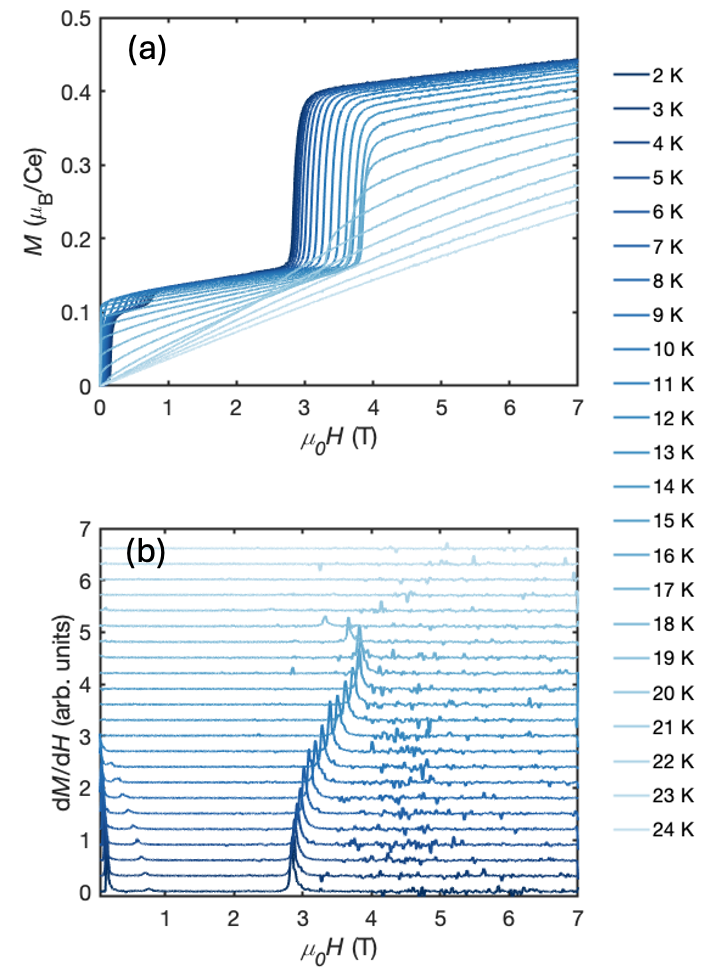} % for an image file named example_figure.*
	% Pick an appropriate width for the size of the image
	% Captions go below figures
	\caption{\textbf{Temperature dependence of isothermal magnetization}
		(\textbf{a}) ZFC isothermal magnetization measured from 2 K to 24 K in 1 K steps with $H \parallel c$. (\textbf{b}) The corresponding derivatives $dM/dH$. Data in (\textbf{b}) are offset for clarity.}
	\label{fig:MHT} % give each figure a logical label name
\end{figure}

The temperature dependence of a subset of the $M(H)$ curves used to map the $H$--$T$ phase diagram is shown in Fig.~\ref{fig:MHT}a; the corresponding $dM/dH$ is shown in Fig.~\ref{fig:MHT}b. Together, all anomalies observed in $dM/dH$ are summarized in Fig.~\ref{fig:Basics}a of the main text.

\begin{figure} % Do not use \begin{figure*}
	\centering
	\includegraphics[width=1\textwidth]{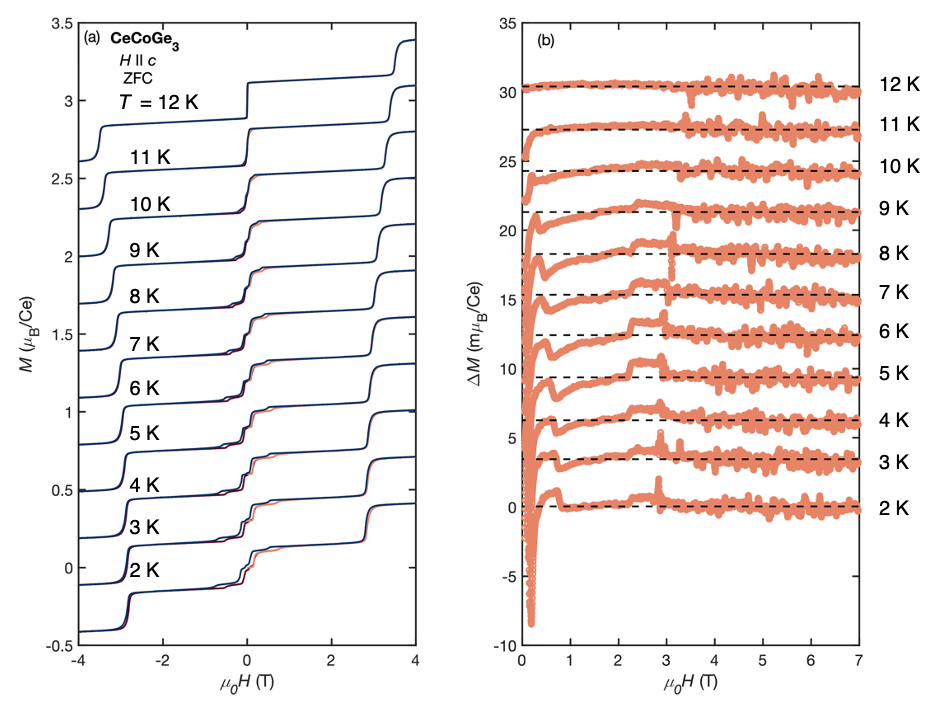} % for an image file named example_figure.*
	% Pick an appropriate width for the size of the image
	% Captions go below figures
	\caption{\textbf{Anomalous virgin curve}
	(\textbf{a}) ZFC isothermal magnetization hysteresis loops measured from 2 K to 12 K in 1 K steps with $H \parallel c$. (\textbf{b}) $\Delta M = M(Q\mathrm{V}) - M(Q\mathrm{I})$ calculated from the loops in (\textbf{a}).}
	\label{fig:micto} % give each figure a logical label name
\end{figure}

\begin{figure} % Do not use \begin{figure*}
	\centering
	\includegraphics[width=0.6\textwidth]{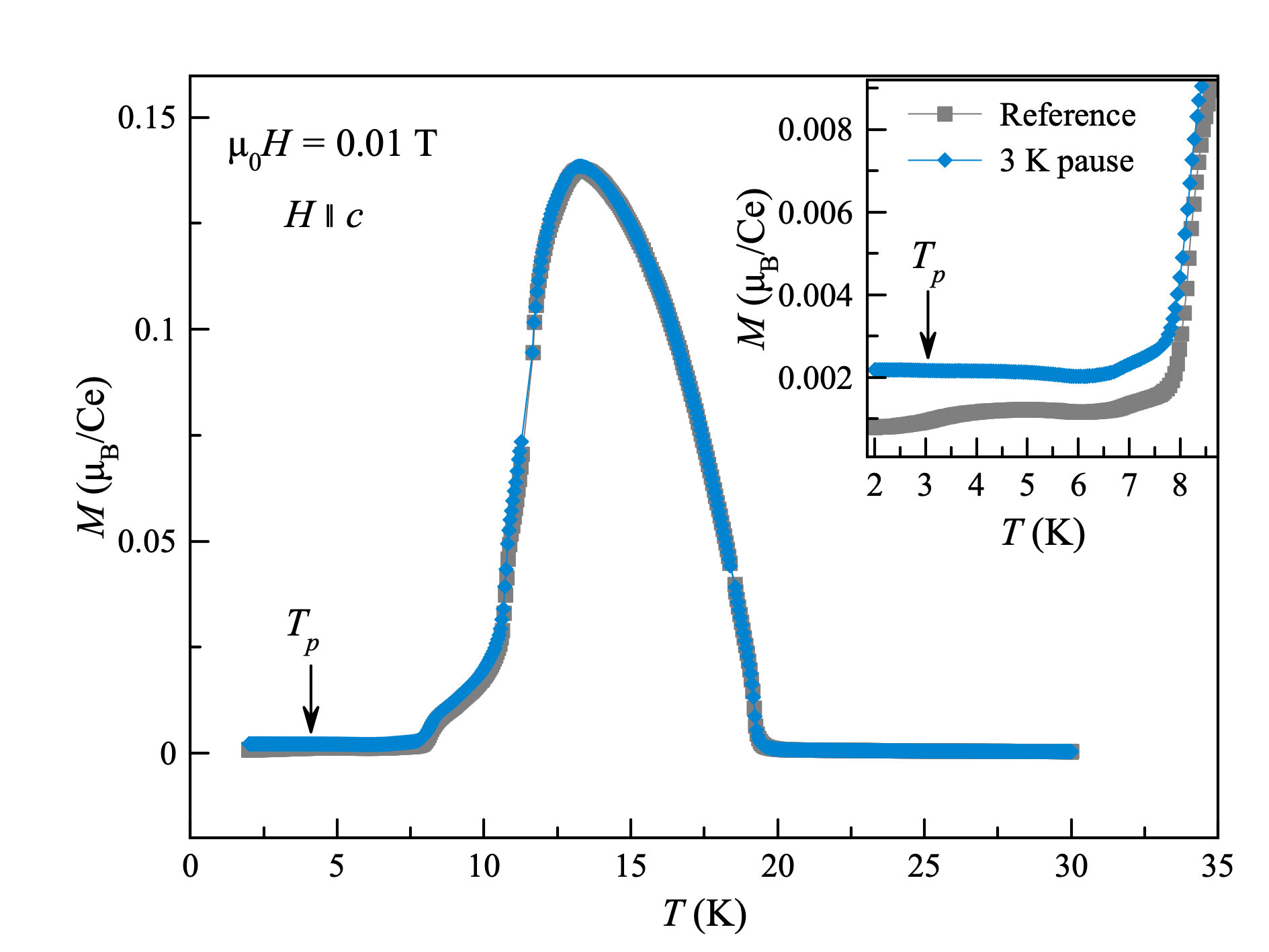} % for an image file named example_figure.*
	% Pick an appropriate width for the size of the image
	% Captions go below figures
	\caption{\textbf{0.01 T ZFC memory test}
	The temperature-dependent magnetization $M$ measured with $H\parallel c$ and $\mu_0H=0.01$ T on warming after (gray) continuous zero-field cooling and (blue) zero-field cooling with a one-hour pause at $T_p$ = 3 K. The inset is the same data zoomed in, demonstrating no memory dip in the $T_p$ = 3 K data.}
	\label{fig:ZFC100Oe} % give each figure a logical label name
\end{figure}

\begin{figure} % Do not use \begin{figure*}
	\centering
	\includegraphics[width=0.6\textwidth]{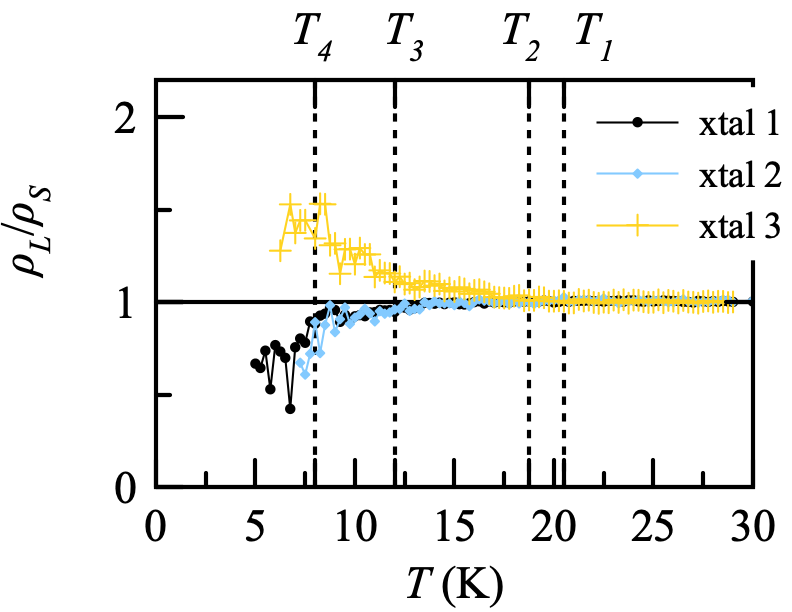} % for an image file named example_figure.*
	% Pick an appropriate width for the size of the image
	% Captions go below figures
	\caption{\textbf{Anisotropic resistance in CeCoGe$_3$}
	Temperature-dependent anisotropic resistance $\rho_L/\rho_S$ measured on three separate crystals labeled crystals 1--3. Crystal 1 is the same data as in Fig.~\ref{fig:memory}d of the main text. Since the system is tetragonal at high temperature, whether $\rho_L/\rho_S$ is less than or greater than one is random. Nonetheless, the four-fold rotational symmetry is absent at low temperatures.}
	\label{fig:anisotropy} % give each figure a logical label name
\end{figure}

\begin{figure} % Do not use \begin{figure*}
	\centering
	\includegraphics[width=1\textwidth]{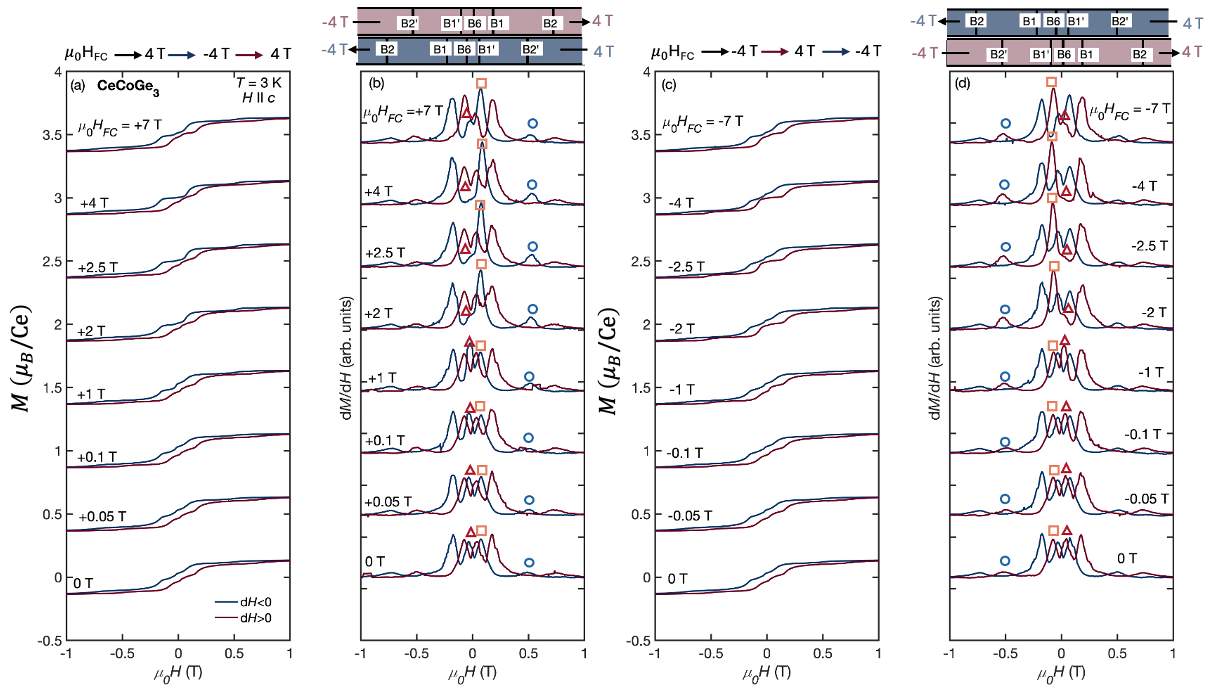} % for an image file named example_figure.*
	% Pick an appropriate width for the size of the image
	% Captions go below figures
	\caption{\textbf{Domain pinning }
		 (\textbf{a,(c)}) Magnetic-field dependence of the field-cooled isothermal magnetization $M$ measured with $H \parallel c$ plotted for fields $-1~\text{T} \leq\mu_0H\leq$1~\text{T} at temperature $T~=~3$~K. All data with the exception of the $\mu_0H_{FC} = 0$ T data are offset for clarity. For each measurement, the sample was field-cooled from 50 K in selected magnetic fields $\mu_0H_{FC}$. The field was then ramped to 4 T, and then a full hysteresis loop was recorded between $\mu_0H~=~+(-)~4~\text{T and } ~-(+)~4~\text{T}$ in (\textbf{a,(c)}). Data collected for increasing field $\text{d}H>0$ are colored blue, while data collected for decreasing field $\text{d}H<0$ are colored red. Domain pinning is tracked by the $\mu_0H_{FC}$ evolution of the peaks in the derivative of $\text{d}M/\text{d}H$ near $B_6$, $B_1^*$, and $B_2^*$ in Fig.~\ref{fig:FC_loops}p of the main text. }
	\label{fig:MHFC} % give each figure a logical label name
\end{figure}

\begin{figure} % Do not use \begin{figure*}
	\centering
	\includegraphics[width=1\textwidth]{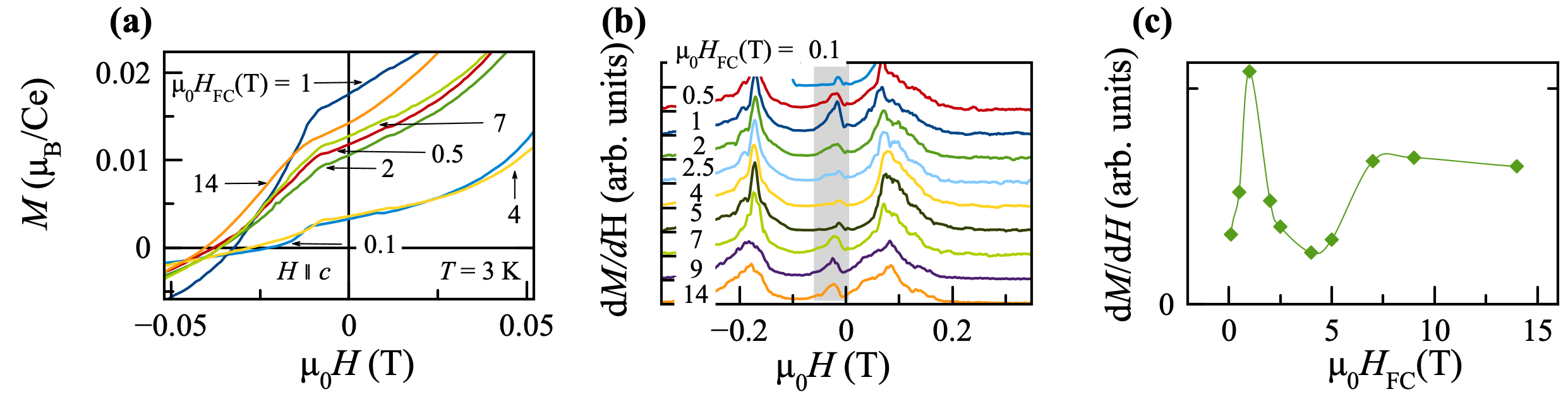} % for an image file named example_figure.*
	% Pick an appropriate width for the size of the image
	% Captions go below figures
	\caption{\textbf{Field-cooling experiments}
		 (\textbf{a}) Magnetic-field dependence of the field-cooled isothermal magnetization $M$ measured with $H \parallel c$ plotted for fields $-0.5~\text{T} \leq\mu_0H\leq$0.5~\text{T} at temperature $T~=~3$~K. For each measurement, the sample was field-cooled from above $T_1$ in selected magnetic fields $\mu_0H_{FC}$. Then $M$ was measured while sweeping from $+\mu_0H_{FC}$ to $-\mu_0H_{FC}$. A subset of the measurements is shown for clarity. (\textbf{b}) The corresponding derivative $dM/dH$. Data are offset for clarity and the black dashed lines represent the zero on the y-axis for each curve. The $B_6$ peak is highlighted in gray. (\textbf{c}) The $\mu_0H_{FC}$ dependence of $dM/dH$ shows non-monotonic behavior consistent with Fig.~\ref{fig:FC_loops}p in the main text. }
	\label{fig:reversal} % give each figure a logical label name
\end{figure}

\begin{figure} % Do not use \begin{figure*}
	\centering
	\includegraphics[width=0.8\textwidth]{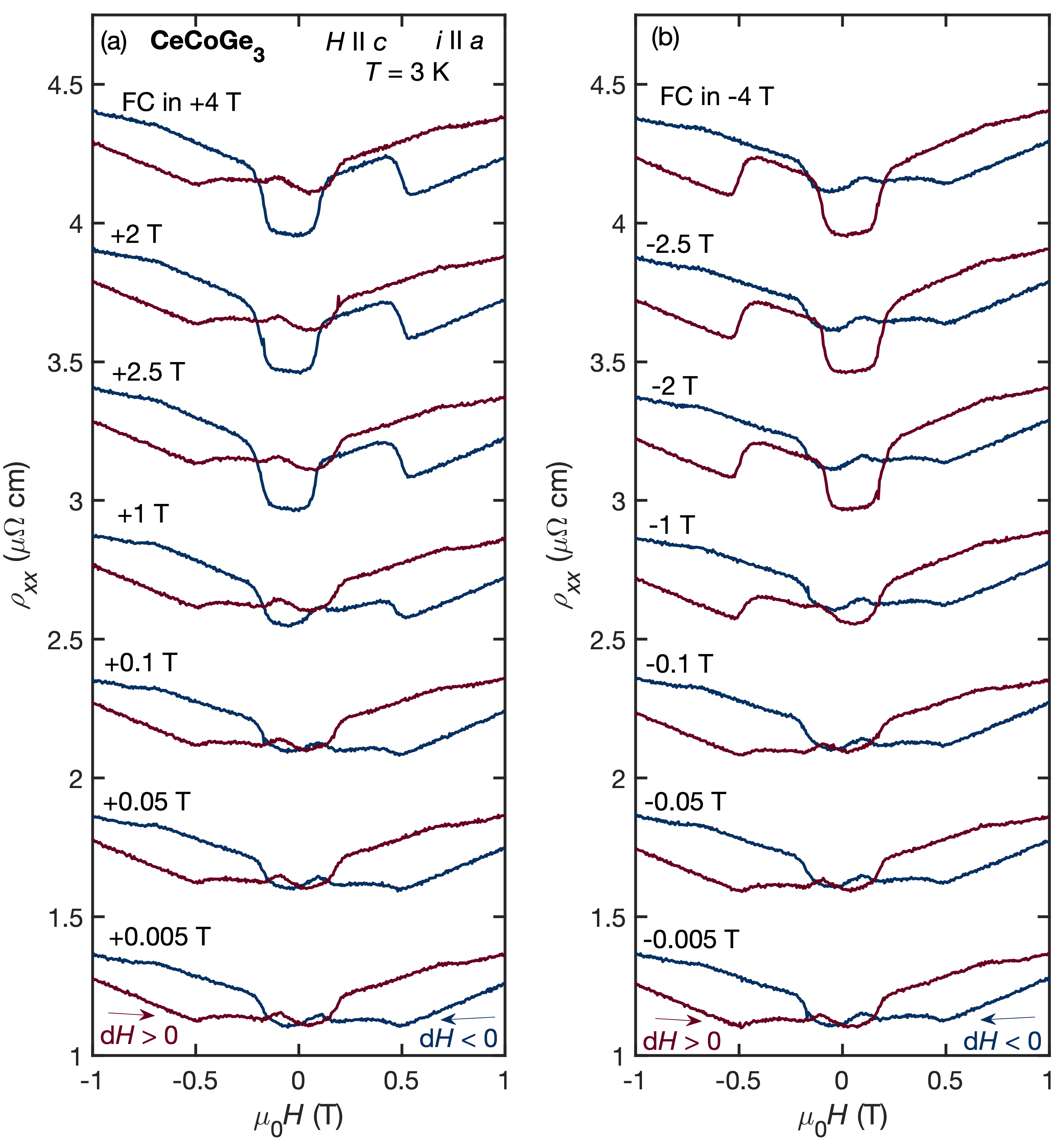} % for an image file named example_figure.*
	% Pick an appropriate width for the size of the image
	% Captions go below figures
	\caption{
		 \textbf{Field-cooling dependence of the longitudinal resistivity}
		 Magnetic field dependence of the longitudinal resistivity $\rho_{xx}$ at temperature $T~=~3$ K with $H \parallel c$ and current $i \parallel a=b$ measured using the same field-cooling protocol described in Fig.~\ref{fig:FC_loops}a--c at various $\mu_0H_{FC}$ measured on a crystal from Batch 1. Data for positive (negative) $\mu_0H_{FC}$ are shown in (\textbf{a,b}), respectively. Data collected for increasing field $\text{d}H>0$ are colored blue, while data collected for decreasing field $\text{d}H<0$ are colored red. The schematic at the top shows where anomalies are present in $M(H)$ loops. All data except for data with $\mu_0H_{FC}~=~0$ T are offset for clarity.}
	\label{fig:CH2} % give each figure a logical label name
\end{figure}

\begin{figure} % Do not use \begin{figure*}
	\centering
	\includegraphics[width=0.8\textwidth]{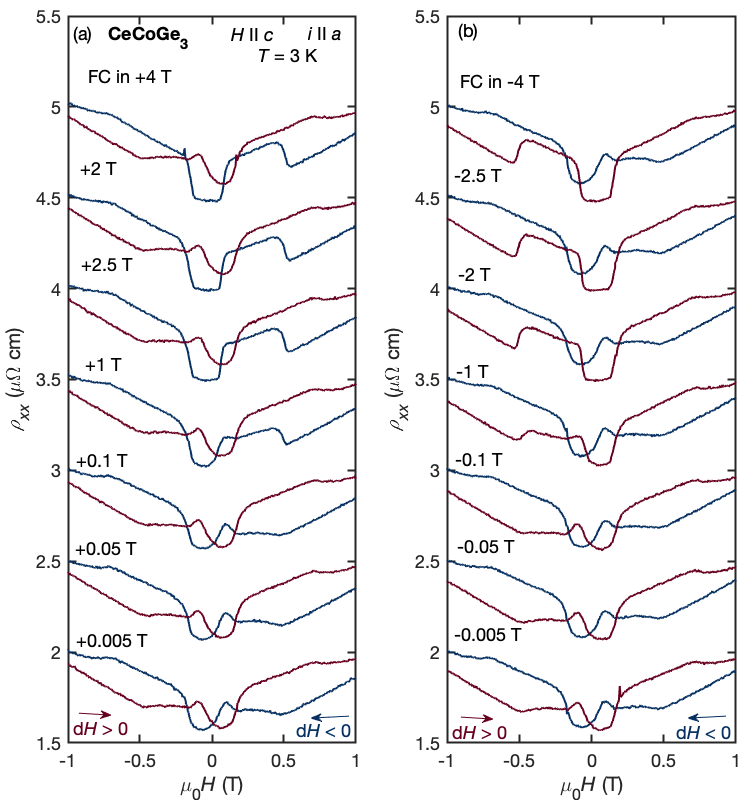} % for an image file named example_figure.*
	% Pick an appropriate width for the size of the image
	% Captions go below figures
	\caption{
		 \textbf{Field-cooling dependence of the longitudinal resistivity}
		 The same as Fig.~\ref{fig:CH2} measured on a different crystal from Batch 2.}
	\label{fig:CH1} % give each figure a logical label name
\end{figure}

\end{document}